\documentclass[english, 11pt,a4paper]{article}

\usepackage{graphicx}
\usepackage{epstopdf}
\usepackage{amsmath,amsfonts,bbm}
\let\OLDthebibliography\thebibliography
\renewcommand\thebibliography[1]{
  \OLDthebibliography{#1}
  \setlength{\parskip}{0pt}
  \setlength{\itemsep}{0pt plus 0.3ex}
}

\usepackage{amsmath, amsthm,mathtools, amssymb,upgreek,slashed}

\begin{document}


\font\msytw=msbm10 scaled\magstep1
\font\msytww=msbm7 scaled\magstep1
\font\msytwww=msbm5 scaled\magstep1
\font\cs=cmcsc10
\font\ottorm=cmr8

\let\a=\alpha \let\b=\beta  \let\g=\gamma  \let\d=\delta \let\e=\varepsilon
\let\z=\zeta  \let\h=\eta   \let\th=\theta \let\k=\kappa \let\l=\lambda
\let\m=\mu    \let\n=\nu    \let\x=\xi     \let\p=\pi    \let\r=r
\let\s=\sigma \let\t=\tau   \let\f=\varphi \let\ph=\varphi\let\c=\chi
\let\ps=\psi  \let\y=\upsilon \let\o=\omega\let\si=\varsigma
\let\G=\Gamma \let\D=\Delta  \let\Th=\Theta\let\L=\Lambda \let\X=\Xi
\let\P=\Pi    \let\Si=\Sigma \let\F=\Phi    \let\Ps=\Psi
\let\O=\Omega \let\Y=\Upsilon

\def\ins#1#2#3{\vbox to0pt{\kern-#2 \hbox{\kern#1 #3}\vss}\nointerlineskip}

\newdimen\xshift \newdimen\xwidth \newdimen\yshift

\def\vdd{{\vec d}}\def\vee{{\vec e}}\def\vkk{{\bk}}\def\vii{{\vec i}}
\def\vmm{{\vec m}}\def\vnn{{\vec n}}\def\vpp{{\vec p}}\def\vqq{{\vec q}}
\def\vxxi{{\vec \xi}}\def\vrr{{\vec r}}\def\vtt{{\vec t}}
\def\vuu{{\vec u}}\def\vvv{{\vec v}}
\def\vxx{{\xx}}\def\vyy{{\vec y}}\def\vzz{{\vec z}}
\def\un{{\underline n}} \def\ux{{\underline x}} \def\uk{{\underline k}}
\def\xxx{{\underline\xx}}\def\vxx{{\xx}} \def\vxxx{{\underline\vxx}}
\def\kkk{{\underline\kk}} \def\vkkk{{\underline\vkk}}
\def\bO{{\bf O}}\def\rr{{\bf r}} \def\bk{{\bf k}}  \def\bp{{\bf p}}
 \def\bP{{\bf P}}\def\bl{{\bf l}} 
\def\ss{{\underline \sigma}}\def\oo{{\underline \omega}}

\def\PPP{{\cal P}}\def\EE{{\cal E}}\def\cF{{\cal F}}
\def\mF{{\mathfrak{F}}}
\def\MM{{\cal M}} \def\VV{{\cal V}}\def\cB{{\cal B}}\def\cA{{\cal A}}\def\cI{{\cal I}}
\def\bV{{\bf V}_n}
\def\cE{{\cal E}}\def\si{{\sigma}}\def\ep{{\epsilon}} \def\cD{{\cal D}}\def\cG{{\cal G}}
\def\CC{{\cal C}}\def\FF{{\cal F}} \def\FFF{{\cal F}}\def\cJ{{\cal J}}
\def\cF{{\cal F}} \def\cT{{\cal T}}\def\cS{{\cal S}}\def\cQ{{\cal Q}}
\def\HHH{{\cal H}}\def\WW{{\cal W}}\def\cP{{\cal P}}
\def\TT{{\cal T}}\def\NN{{\cal N}} \def\BBB{{\cal B}}\def\III{{\cal I}}
\def\RR{{\cal R}}\def\cL{{\cal L}} \def\JJ{{\cal J}} \def\OO{{\cal O}}
\def\DD{{\cal D}}\def\AAA{{\cal A}}\def\GG{{\cal G}} \def\SS{{\cal S}}
\def\KK{{\cal K}}\def\UU{{\cal U}} \def\QQ{{\cal Q}} \def\XXX{{\cal X}}

\def\hh{{\bf h}} \def\HH{{\bf H}} \def\AA{{\bf A}} \def\qq{{\bf q}}
\def\bG{{\bf G}}
\def\BB{{\bf B}} \def\XX{{\bf X}} \def\PP{{\bf P}} \def\bP{{\bf P}} 
\def\pp{{\bf p}}
\def\vv{{\bf v}} \def\xx{{\bf x}} \def\yy{{\bf y}} \def\zz{{\bf z}}
\def\dd{{\bf d}}
\def\aaa{{\bf a}}\def\bbb{{\bf b}}\def\hhh{{\bf h}}\def\II{{\bf I}}
\def\ii{{\bf i}}\def\jj{{\bf j}}\def\kk{{\bf k}}\def\bS{{\bf S}}
\def\mm{{\bf m}}\def\Vn{{\bf n}}\def\uu{{\bf u}}\def\tt{{\bf t}}
\def\B{\hbox{\msytw B}}
\def\RRR{\hbox{\msytw R}} \def\rrrr{\hbox{\msytww R}}
\def\rrr{\hbox{\msytwww R}} \def\CCC{\hbox{\msytw C}}\def\EEE{\hbox{\msytw E}}
\def\cccc{\hbox{\msytww C}} \def\ccc{\hbox{\msytwww C}}
\def\MMM{\hbox{\euftw M}}\font\euftw=eufm10 scaled\magstep1%
\def\NNN{\hbox{\msytw N}} \def\nnnn{\hbox{\msytww N}}
\def\nnn{\hbox{\msytwww N}} \def\ZZZ{\hbox{\msytw Z}}\def\QQQ{\hbox{\msytw Q}}
\def\zzzz{\hbox{\msytww Z}} \def\zzz{\hbox{\msytwww Z}}
\def\SSS{{\bf S}}
\def\SSSS{\hbox{\euftwww S}}
\def\1{\hbox{\msytw 1}}
\newcommand{\mR}{{\msytw R}}
\def\virg{\quad,\quad}


\def\\{\hfill\break}
\def\={:=}
\let\io=\infty
\let\0=\noindent\def\pagina{{\vfill\eject}}
\def\media#1{{\langle#1\rangle}}
\let\dpr=\partial
\def\sign{{\rm sign}}
\def\const{{\rm const}}
\def\tende#1{\,\vtop{\ialign{##\crcr\rightarrowfill\crcr\noalign{\kern-1pt
    \nointerlineskip} \hskip3.pt${\scriptstyle #1}$\hskip3.pt\crcr}}\,}
\def\otto{\,{\kern-1.truept\leftarrow\kern-5.truept\to\kern-1.truept}\,}
\def\defin{{\buildrel def\over=}}
\def\wt{\widetilde}
\def\wh{\widehat}
\def\to{\rightarrow}
\def\ra{\right\rangle}
\def\qed{\hfill\raise1pt\hbox{\vrule height5pt width5pt depth0pt}}
\def\Val{{\rm Val}}
\def\ul#1{{\underline#1}}
\def\lis{\overline}
\def\V#1{{\bf#1}}
\def\be{\begin{equation}}
\def\ee{\end{equation}}
\def\bea{\begin{eqnarray}}
\def\eea{\end{eqnarray}}
\def\bd{\begin{definition}}
\def\ed{\end{definition}}

\def\nn{\nonumber}
\def\pref#1{(\ref{#1})}
\def\ie{{\it i.e.}}
\def\cC{{\cal C}}
\def\lb{\label}
\def\eg{{\it e.g.}}
\def\sl{{\displaystyle{\not}}}
\def\Tr{\mathrm{Tr}}
\def\BBBB{\hbox{\msytw B}}
\def\bbb{\hbox{\msytww B}}
\def\TTT{\hbox{\msytw T}}
\def\d{\delta}
\def\bT{{\bf T}}
\def\mod{{\rm mod}}
\def\der{{\rm d}}
\def\bs{\backslash}
\newtheorem{corollary}{Corollary}[section]
\newtheorem{lemma}{Lemma}[section]
\newtheorem{example}{Example}[section]
\newtheorem{remark}{Remark}[section]
\newtheorem{definition}{Definition}[section]
\newtheorem{theorem}{Theorem}[section]
\newtheorem{proposition}{Proposition}[section]
\newtheorem{oss}{Remark}

\title{{\bf Non-Fermi Liquid Behaviors in the Hubbard model on the Honeycomb lattice}}
\author{Zhituo Wang\\ Institute for Advanced Study in Mathematics, \\Harbin Institute of Technology\\
Email: wzht@hit.edu.cn}

\maketitle

\begin{abstract}
In this paper we study the low temperature ($T\rightarrow 0$) behaviors of the weakly interacting Hubbard model on the honeycomb lattice with a bare chemical potential, which takes values in a small neighborhood of the renormalized chemical potential $\mu$, which is fixed to be $1$. We prove that
the two-point Schwinger's function is an analytic function of the coupling constant $\lambda$ in the domain $|\lambda|\cdot|\log T|^2<1$. But the ground state of this model is not a Fermi liquid, because analytic properties of self-energy don't satisfy the Salmhofer's criterion on the Fermi liquids at finite temperature. 
\end{abstract}

\tableofcontents
\renewcommand{\thesection}{\arabic{section}}
\newpage

\section{Introduction}
Landau Fermi liquid theory is one of the cornerstones of quantum many-body theory. It essentially states that, at temperature zero, the single particle excitations of a non-interacting Fermi gas becomes quasi-particles in a Fermi liquid. The quasi-particle spectrum has almost the same structure
as the non-interacting single particle excitation spectrum and the quasi-particle density function has a jump at the Fermi surface. Fermi liquid theory has successfully explained the experimental fact that a metal formed by interacting electrons behaves almost like a free fermion system.

It is well known that the Fermi liquid state may transit to other states when the temperature of the interacting Fermi gas is low enough. The various mechanisms for such phase transitions are also called the quantum instabilities. The most celebrated one is the BCS instability for the formation of Cooper pairs \cite{BCS}, which lead to superconductivity in $2$ and $3$ dimensions. This instability may happen to any time-reversal invariant system. However, despite its discovery more than $60$ years ago, we still don't have a fully, mathematical understanding of the BCS theory.
We believe that an important step toward constructing a rigorous BCS theory is to control rigorously what occurs in various interacting Fermion system, before the phase transition happens. This is one important motivation of our work. 

Another instability is the Luttinger instability \cite{Lu} in certain one-dimensional solvable models. The ground state is called a Luttinger liquid. One important difference between a Fermi liquid and a Luttinger liquid is that, while the density function for the latter is continuous across the Fermi surface, it has infinite slope there. For a rigorous treatment of Luttinger liquids in one dimension, see \cite{BG1} and the references therein.

Fermi liquids properties at positive temperature have been proved to exist in various models, including the 2d jellium model \cite{DR1}, in which the Fermi surface is a circle; a general $2$-d model with symmetric Fermi surfaces \cite{BGM1}, and the doped Hubbard model that is far from half-filling \cite{BGM2}. In a large series of papers the construction of two-dimensional Fermi liquids at zero temperature for a wide class of systems with non-symmetric Fermi surfaces has been completed by Feldman, Kn\"orrer and Trubowitz \cite{FKT}. 

The Hubbard model \cite{hubb} is one of the most important model for describing interacting electrons in a solid, taking into account the quantum mechanical hopping of the electrons on the lattice and the (simplified) repulsive coulomb interactions between electrons, through a density-density interaction.  Despite its simple definition, it exhibits various interesting phenomena, including the Fermi liquid behavior, the ferromagnetism, the antiferromagnetism, the Tomonaga-Luttinger liquid behaviors and the superconductivity \cite{lieb}. The nature of the ground state of the Hubbard model depends crucially on the spatial lattice and the doping. The Hubbard model on the square lattice that is far from half filling exhibit Fermi liquid behaviors at sufficiently low temperature \cite{BGM2} while it becomes Luttinger liquid \cite{Lu, To} at half filling \cite{Riv}. The Hubbard model on the honeycomb lattice at half filling, which is an important model for studying the various behaviors of {\it Graphene} \cite{review1, HCM, N, W}, displays also the Fermi liquid behavior \cite{GM}. In this paper we will consider the Hubbard model on the honeycomb lattice in which the value of bare chemical potential is in a neighborhood of $\mu=1$, which is the value of the renormalized chemical potential, and is fixed to be $1$. We shall prove that the ground state is not a Fermi liquid but a Luttinger liquid, when the temperature is low enough, according to Salmhofer's criterion for the finite temperature Fermi liquids (see \cite{salm} and Theorem \ref{salmc}). This confirms a conjecture of Anderson \cite{and}, which states that a two-dimensional Fermi gas should exhibit behaviors similar to the $1$-D Luttinger liquid.

One common difficulty in the many-Fermions problem is that the Fermi surfaces change their shapes due to the interactions. Let the non-interacting band structure be $E$, which is given by $E(\bp)=\e(\bp)-\mu$. The set $\cF_0=\{\bp: E(\bp)=0\}$ defines the non-interacting Fermi surface. It is well known that the perturbation series of the Green functions are eventually divergent when the spatial momentum $\bp$ approaches the Fermi surface at zero temperature. When the interaction $\lambda V$ is turned on, a self-energy term $\cE$ will be generated and the band structure is deformed to be $e=E+\cE$. The interacting Fermi surface is defined by $\cF=\{\bp: e(\bp)=0\}$, which is deviated from the non-interacting one. It is not static but is a function of the the momentum, and is changing its shape in the course of the multi-slice analysis. This is the case for our model, which is not at half-filling for our choice of chemical potential and the chemical potential receives renormalization due to the presence of the tadpole terms. In order to solve the problem of moving surface, we introduce a counter-term $\delta\mu$ to the interaction such that the renormalized chemical potential is fixed be $\mu=1$ (see also \cite{DR1}). Then we prove that the flow of the counter-terms is bounded by a very small number (see Theorem \ref{flowmu}). Remark that this method is equivalent to the decomposition of a bare chemical potential into a fixed, renormalized one and the counter-term \cite{FST1}-\cite{FST4}, \cite{BGM1}. It is also possible to solve the moving Fermi surface problem without the introduction of the counter-term but to renormalize the Gaussian measure at each renormalization step \cite{BGM2}, so that the propagators may still have the desired decay property at each renormalization scale, as long as the self-energy can be properly bounded. Since the main goal of this paper is to study the analyticity and regularity of the self-energy, we don't follow this construction.

%

  

Another common difficulty in Fermi liquid theory is how to treat properly the constraints from the conservation of momentum on the Fermi surfaces (cf. \cite{FMRT}). The Fermi surface in our model 
at $\mu=1$ is a union of perfect triangles, which is highly non-isometric, so is the integration domain in a small neighborhood of the Fermi surface. In order to implement the conservation of momentum near the Fermi surface and to obtain better decay properties for the propagator, we introduce the highly non-isometric sectors, which are the support of the non-interacting propagators. A sector counting lemma, which states the possible relations among the sector indices for these momentum, is proved. Then we apply the renormalization group analysis \cite{B, G, M2, rivbook}. In order to obtain the optimal bounds for the self-energy, we introduce an auxiliary expansion, called the the multi-arch expansion, which is a convergent expansion that permits us to obtain the 2PI graphs from tree graphs.


This paper is organized as follows. In Section $2$ we introduce the model and present the main results. In Section $3$ we study the scaling behaviors of the propagator and introduce the {\it sectors}. In Section $4$ we introduce the multi-slice expansions and prove the power-counting theorem. We also study the convergence and analytic properties of the Schwinger functions with $4$ and more external fields. In Section $5$ we consider the renormalization of the $2$-point Schwinger functions and study the convergence and analytic properties of the self-energy, for which we introduce the multi-slice expansions. We conclude in Section $6$ with discussions about possible future work and the Appendix is about the geometry of the Fermi surface.

\section{The model and Main results}
\subsection{The Hubbard model on the honeycomb lattice}
Let $\tilde\Lambda=\{\xx\ \vert\ \xx=n_1\bl_1+n_2\bl_2, n_1, n_2\in\ZZZ\}\subset\RRR^2$ be the infinite triangle lattice generated by the two basis vectors 
${\bl}_1=\frac12(3,\sqrt{3})$, ${\bl_2}=\frac12(3,-\sqrt{3})$. Given $L\in\NNN_+$, let the torus $\tilde\Lambda_L=\tilde\Lambda/L\tilde\Lambda$ be the finite triangle lattice of side $L$ with periodic boundary conditions. The honeycomb lattice $\Lambda_L=\Lambda^A_L\cup\Lambda^B_L$ 
is the superposition of two torus 
$\Lambda^A_L=\tilde\Lambda_L$ and $\Lambda^B_L=\tilde\Lambda_L+\dd_i$, $i=1, 2, 3$,
 where 
\be {\dd_1}=(1,0)\;,\quad {\dd_2}=\frac12
(-1,\sqrt{3})\;,\quad{\dd_3}=\frac12(-1,-\sqrt{3})\;.\label{delta}
\ee
The metric on $\Lambda_L$ is defined by the Euclidean distance on the torus, denoted by 
$\vert\xx-\yy\vert=\min_{(n_1, n_2)\in\ZZZ^2}\vert \xx-\yy+n_1L\bl_1+n_2L\bl_2\vert$. 

\begin{figure}[htp]
\centering
\includegraphics[width=.7\textwidth]{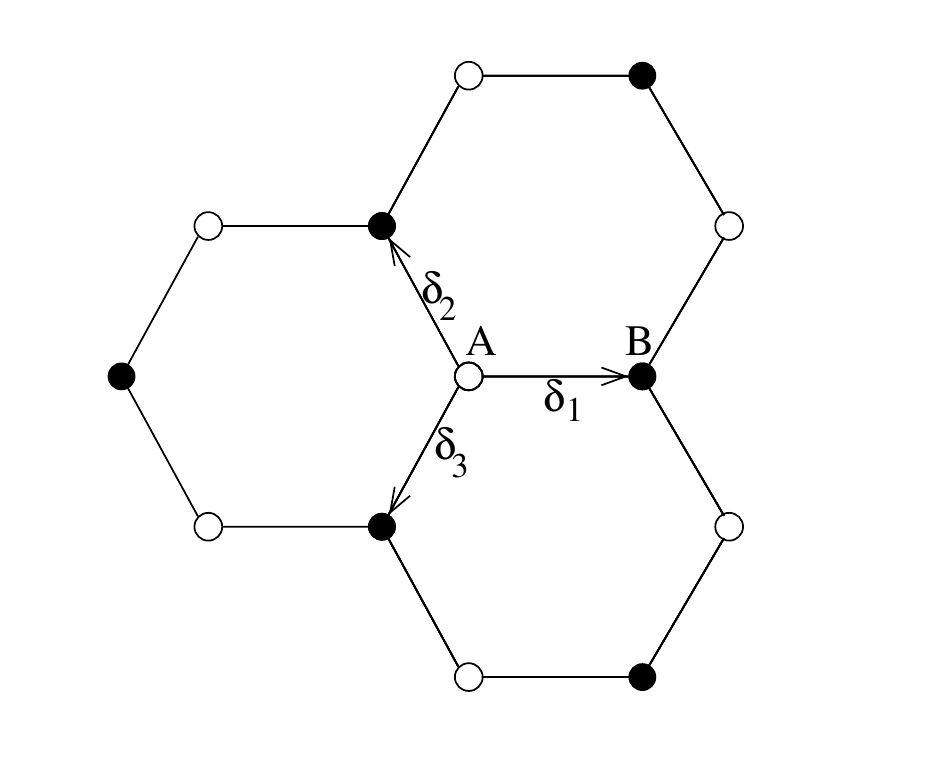}
\caption{\label{lattice}
A portion of the honeycomb lattice $\L$. The white and black dots correspond 
to the sites of the $\L_A$ and $\L_B$ triangular sub-lattices, respectively. These two 
sub-lattices are one the translate of the other. They are connected by nearest neighbor vectors
$\dd_1,\dd_2,\dd_3$ that, in our units, are of unit length.}
\end{figure}


The single particle wave function is denoted by $\Psi_{\xx, \a,\t}$, where $\t\in\{\uparrow,\downarrow\}$ are the spin variables and $\a\in\{A,B\}$ labels the two kinds of lattice points. The wave function is an element of the one-particle Hilbert space $\HHH_L=\CCC^{L^2}\otimes\CCC^4$ and satisfies the usual normalization condition
$\Vert\Psi\Vert^2_2=\sum_{\xx,\t,\a}\vert\Psi_{\xx, \a,\t}\vert^2=1$. The dynamics of the particle is generated by the Hamiltonian $H:\HHH_L\rightarrow\HHH_L$.

Let $\bigwedge^N \HHH_L$ be the $N$-th anti-symmetric tensor product Hilbert spaces $\HHH_L$. A system of $N\ge2$ particles is described by the normalized, antisymmetric wave function $\Psi^{N}:=\Psi^N(\xi_1,\ \cdots, \xi_N)\in\bigwedge^N\HHH_L$, for which we have $\Psi^N(\cdots,\xi_i, \xi_{i+1};\cdots)=-\Psi^N(\cdots,\xi_{i+1},\xi_i,\cdots),$ where $\xi_i=(x_i,\a_i,\t_i)$ is the coordinate of the $i$-th particle. Let $H$ be the single particle Hamiltonian on $\HHH_L$, the many-body Hamiltonian $H_N:\bigwedge^N \HHH_L\rightarrow\bigwedge^N  \HHH_L$ is defined as
\be
H_N=\sum_{i=1}^N H^{(i)}+\lambda\sum_{1\le i<j\le N}V_{ij},
\ee
where $H^{(i)}:=\mathbbm{1}^{\otimes(i-1)}\otimes H\otimes\mathbbm{1}^{\otimes(N-i)}$, $\mathbbm{1}$ the identity operator, $V_{ij}$ is the pair interaction between the $ij$ particles and $\lambda$ is the coupling constant.

It is convenient to define the model in the second quantization formalism. Define the Fermionic Fock space $\FFF_L$ as:
\be
\FFF_L=\CCC\oplus\bigoplus_{N=1}^{4L^2}\FFF_L^{(N)},\quad \FFF_L^{(N)}=\bigwedge^N \HHH_L.
\ee
$\FFF_L$ is a finite-dimensional Hilbert space when $L<\infty$, due to the Pauli's exclusion principal. At a given site $\xx\in\Lambda_L$ we introduce the Fermionic creation and annihilation operators ${\bf a}^\pm_{\xx,\a,\t}$, with $\a\in\{A, B\}$ and $\t\in\{\uparrow,\downarrow\}$, acting on the Fermionic Fock space as follows. For any element $\Psi=(\Psi^{(0)}, \Psi^{(1)},\cdots,\Psi^{(N)},\cdots)\in\FFF_L$, we define:
\bea
&&({\bf a}^+_{\zz,\a,\t}\Psi)^{(N)}(\xi_1,\cdots, \xi_N)\nn\\
&&\quad\quad\quad:=\frac{1}{\sqrt{N}}\sum_{j=1}^N(-1)^j\delta_{\zz,\xx_j}\delta_{\a,\a_j}\delta_{\t,\t_j} \Psi^{(N-1)}(\xi_1,\cdots ,\xi_{j-1},\xi_{j+1},\cdots,\xi_{N}),\\
&&({\bf a}^-_{\zz,\a,\t}\Psi)^{(N)}(\xi_1,\cdots, \xi_N):= \sqrt{N+1}\ \Psi^{(N+1)}(\zz,\a,\t;\ \xi_1,\cdots,\xi_{n}),
\eea 
where $\xi_i=(\xx_i,\a_i,\t_i)$ is the coordinate of the $i$-th particle, $\delta_{.,.}$ is the Kronecker delta function. These Fermionic operators satisfy the canonical anti-commutation relations (CAR): 
\be\{{\bf a}^+_{\xx,\a,\t}, {\bf a}^-_{\xx',\a',\t'}\}=\delta_{\xx,\xx'}\delta_{\a,\a'}\delta_{\t,\t'},\quad \{{\bf a}^+_{\xx,\a,\t}, {\bf a}^+_{\xx',\a',\t'}\}=\{{\bf a}^-_{\xx,\a,\t}, 
{\bf a}^-_{\xx',\a',\t'}\}=0,\ee and are subjected to the periodic boundary conditions ${\bf a}^\pm_{\xx+n_1L+n_2L,r,\t}={\bf a}^+_{\xx,r,\t},\ \forall \xx\in\Lambda_L.$

The dual lattice ${\tilde\Lambda}^*$ to $\tilde\Lambda$ is the triangular lattice generated by the basis vectors $\bG_1=\frac{2\pi}{3}(1,\sqrt{3})$, $\bG_2=\frac{2\pi}{3}(1,-\sqrt{3})$ obeying the reciprocal relation $\bG_i\cdot\bl_j=2\pi\delta_{ij}$. The first Brillouin zone is defined as the torus $\DD_L:=\RRR^2/\tilde\Lambda^*=\big\{\bk\in\RRR^2\ \vert\ \bk=\frac{n_1}{L}\bG_1+\frac{n_2}{L}\bG_2, n_{1,2}\in[-L, L-1]\big\}$.
Define the Fourier transform of the Fermionic creation and annihilation operators as 
\be
{\bf a}^\pm_{\xx,\a,\t}=\frac{1}{L^2}\sum_{\bk\in\BBB_L}e^{\pm i\kk\cdot\xx}\hat {\bf a}^\pm_{\bk,\a,\t}, \quad\forall\ \xx\in\Lambda_L,
\ee 
and the inverse Fourier transform is 
\be\hat {\bf a}^\pm_{\bk,\a,\t}=\sum_{\xx\in\Lambda_L}e^{\mp i\bk\cdot\xx} {\bf a}^\pm_{\xx,\a,\t},\quad\forall\ \bk\in\DD_L.
\ee 
The periodicity of ${\bf a}^\pm_{\xx,\a,\t}$ implies that
$\hat {\bf a}^\pm_{\bk+n_1\bG_1+n_2\bG_2,\a,\t}=\hat {\bf a}^\pm_{\bk,\a,\t}.$
Moreover, let $|\L_L|$ be the volume of the lattice $\L_L$, these operators obey the following anti-commutation relations: 
\bea&&\quad\{{\bf a}^+_{\bk,\a,\t}, {\bf a}^-_{\bk',\a',\t'}\}=|\L_L|\delta_{\bk,\bk'}\ \delta_{\a,\a'}\ \delta_{\t,\t'},\nn\\
&&\quad \{{\bf a}^+_{\bk,\a,\t}, {\bf a}^+_{\bk',\a',\t'}\}=\{{\bf a}^-_{\bk,\a,\t}, {\bf a}^-_{\bk',\a',\t'}\}=0.
\eea

The second quantized Hamiltonian $H_{\L}(\lambda)=H^0_{\L_L}+ V_{\lambda,\L_L}$ is the sum of the non-interacting Hamiltonian
$H^0_{\L_L}$ and the many-body interaction $V_{\lambda,\L_L}$. We have
\bea
H^0_{\L_L}&=&-t\sum_{\substack{\xx\in \L_L\\ i=1,2,3}}\sum_{\t=\uparrow\downarrow} \Big(\ 
{\bf a}^+_{\xx,A,\t} {\bf a}^-_{\xx+\dd_i, B,\t} +{\bf a}^+_{\xx+\dd_i,B,\t} {\bf a}^-_{\xx, A,\t}\ \Big)\nn\\
&&\quad\quad\quad-\mu\sum_{\substack{\xx\in \L_L\\ i=1,2,3}}\sum_{\t=\uparrow\downarrow}\Big(\ 
{\bf a}^+_{\xx,A,\t}{\bf a}^-_{\xx,A,\t}+{\bf a}^+_{\xx+ \dd_i,B, \t}{\bf a}^-_{\xx+\dd_i,B, \t}\ \Big),\label{hamil0}
\eea
where ${\dd_i}$ are the shift vectors with $i=1, 2, 3$ (see Formula \eqref{delta}),
$t>0$ is the nearest neighbor hopping parameter and $\mu$ is the {\it renormalized} chemical potential, which controls the average particle density in the Gibbs state. We will fix $t=1$ for the rest of this paper. The many-body interaction $V_{\lambda,\L_L}$ of strength $\lambda$ is given by
\bea  
V_{\L_L}=\sum_{\xx\in \L_L}
\Big(\ {\bf a}^+_{\xx,A,\uparrow}{\bf a}^-_{\xx,A,\uparrow}{\bf a}^+_{\xx,A,\downarrow}{\bf a}^-_{\xx,A,\downarrow}+{\bf a}^+_{\xx,B,\uparrow}{\bf a}^-_{\xx,B,\uparrow}{\bf a}^+_{\xx,B,\downarrow}{\bf a}^-_{\xx,B,\downarrow}\ \Big).
 \label{hamil1}\eea

We introduce also a counter term
\be\label{hamil1c}
\delta V_{\L_L}=\delta\mu\sum_{\substack{\xx\in \L_L\\ i=1,2,3}}\sum_{\t=\uparrow\downarrow}\Big(\ 
{\bf a}^+_{\xx,A,\t}{\bf a}^-_{\xx,A,\t}+{\bf a}^+_{\xx+ \dd_i,B, \t}{\bf a}^-_{\xx+\dd_i,B, \t}\ \Big),
\ee
to the interaction term, where $\delta\mu$ is the bare chemical potential counter-term, which is chosen in such a way that the renormalized chemical potential is fixed to be $\mu=1$. Hence the bare chemical potential in our model is $\mu+\delta\mu$. We will prove in Section \ref{secflow} (see also Theorem \ref{flowmu}) that there exists a positive constant $K$ such that $|\delta\mu|\le K|\lambda|$, where $|\lambda|<1/|\log T|^2$. So we have $|\delta\mu|\ll1$, when $T\ll1$.

We define the new interaction term as the combination of the above two terms, and still call it $V_{\L_L}$:
\bea  
V_{\L_L}&=& \sum_{\xx\in \L_L}
\Big(\ {\bf a}^+_{\xx,A,\uparrow}{\bf a}^-_{\xx,A,\uparrow}{\bf a}^+_{\xx,A,\downarrow}{\bf a}^-_{\xx,A,\downarrow}+{\bf a}^+_{\xx,B,\uparrow}{\bf a}^-_{\xx,B,\uparrow}{\bf a}^+_{\xx,B,\downarrow}{\bf a}^-_{\xx,B,\downarrow}\ \Big)\\
&&\quad+\delta\mu\sum_{\substack{\xx\in \L_L\\ i=1,2,3}}\sum_{\t=\uparrow\downarrow}\Big(\ 
{\bf a}^+_{\xx,A,\t}{\bf a}^-_{\xx,A,\t}+{\bf a}^+_{\xx+ \dd_i,B, \t}{\bf a}^-_{\xx+\dd_i,B, \t}\ \Big),
 \label{hamil1}\eea

We can organize the creation and annihilation operators $\{{\bf a}^\pm_{\ \xx,A,B,\t}\}$
and write the non-interacting Hamiltonian $H^0_{\L_L}$ in the following form:
\be
H^0_{\L_L}=\sum_{\substack{(\xx, \yy)\in \L_L\\ \t,\t'\in \{\uparrow,\downarrow\}}}{\bf a}^+_{\ \xx,\a,\t}\ \Big[H(\xx-\yy)\ \Big]_{\a\a'}\ {\bf a}^-_{\ \yy,\a',\t'}\ ,
\ee
in which the kernel $H(\xx-\yy)$ is a $2\times2$ matrix.
In order to keep tracking of the matrix elements, we set
$(\a,\a')=(A,B):=(1,2)$. Let $\hat H(\bk)$ be the Fourier transform of $H(\xx-\yy)$, we have:
\be
\hat H(\bk)=\begin{pmatrix}\ -\mu&-\Omega^*(\bk)\\-\Omega(\bk)&-\mu\ 
\end{pmatrix},
\ee
where 
\be\label{disp2}
\O({\bk})=\sum_{i=1}^3 
e^{i(\dd_i-\dd_1) \bk}=1+2
e^{-i \frac32 k_1}\cos(\frac{\sqrt{3}}2 k_2)
\ee
is the {\it non-interacting complex dispersion relation} and $\Omega^*(\bk)$ is the complex conjugate. 

The Gibbs state associated with the Hamiltonian $H_{\L_L}$ is:
$
\langle\cdot\rangle=\Tr\ [\ \cdot\ e^{-\beta H_{\L_L}}]/Z_{\beta,\L_L}$, 
where $Z_{\beta,\L_L}=\Tr_{\FFF_L}e^{-\beta H_{\L_L}}$ is the partition function. 
Given any operator $\OO$ on $\FFF_L$ and $x^0\in[-\beta,\beta)$, the imaginary-time evolution of $\OO$ generated by the Hamiltonian $H$ is defined as $\OO_{x^0}:=\OO(-ix^0)=e^{x^0H}\OO e^{-x^0H}$, where $\OO(x^0)$ is the real-time evolution generated by $H$ and $x^0$ is called the imaginary time variable. For example, the imaginary-time evolution of the Fermionic operators ${\bf a}^\pm_{\xx,r,\t}$ are given by ${\bf a}^\pm_{x,r,\t}= e^{Hx^0}{\bf a}^\pm_{\xx,r,\t} e^{-H x^0}$, with $x=(x^0,\xx)$ and ${\bf a}^\pm_{(\xx,x^0),r,\t}={\bf a}^\pm_{\xx,r,\t}(-ix^0)$. 

Let $p=(p_0, \pp)$ be the momentum, where
the discrete-valued component $p_0\in\frac{2\pi}{\beta}(n+\frac12), n\in\ZZZ$ is called as the Matsubara frequency, the space-time Fourier transform of an operator $\OO_x$ is given by
\be
\hat\OO_p=\int_{-\beta}^\beta dx^0\sum_{\xx\in\L_L}e^{-i\pp\cdot\xx+ip_0x_0}\ \OO_x.
\ee

Given $n$ operators $\OO^{(1)}_{x^0_1},\cdots, \OO^{(n)}_{x^0_n}$ acting on $\FFF_L$, each of which can be written as a normal-ordered polynomial in the time-evolved Fermionic operators ${\bf a}^\pm_{\xx,r,\t}$, we define the Euclidean correlation function as 
\be\label{corre1}
\langle\bT\OO^{(1)}_{x^0_1},\cdots, \OO^{(n)}_{x^0_n}\rangle_{\beta,L}=\frac{1}{Z_{\beta,\Lambda_L}}\Tr_{\FFF_L}\ e^{-\beta H_{\L_L}}\bT\{\OO^{(1)}_{x^0_1},\cdots, \OO^{(n)}_{x^0_n}\} ,
\ee
where ${Z_{\beta,\Lambda_L}=\Tr_{\FFF_L}e^{-\beta H_{\L_L}}}$ is the partition function and $\bT$ is the Fermionic time-ordering operator, acting on a product of $n$ Fermionic operator as
\be
\bT\ {\bf a}^{\e_1}_{(\xx_1, {x_1^0}),\a_1,\t_1}\cdots {\bf a}^{\e_n}_{(\xx_n, {x_n^0}),\a_n,\t_n}={\rm sgn} (\pi)\ {\bf a}^{\e_{\pi(1)}}_{(\xx_{\pi(1)}, {x}_{\pi(1)}^0),\a_{\pi(1)},\t_{\pi(2)}}\cdots {\bf a}^{\e_{\pi(n)}}_{(\xx_{\pi(n)}, {x}_{\pi(n)}^0),\a_{\pi(n)},\t_{\pi(n)}},
\ee
and $\pi$ is a permutation on the set $\{1,\cdots, n\}$ with signature ${\rm sgn} (\pi)$ such that
${x}_{\pi(1)}^0\ge{x}_{\pi(2)}^0\ge\cdots\ge{x}_{\pi(n)}^0$. If some operators are evaluated at the same time, the ambiguity is solved by taking the normal-ordering on these operators. 


The $n$-point Schwinger function is defined as
\be\label{nptsch}
S^{\beta}_{n}(x_1,\e_1,\a_1,\t_1;\cdots x_n ,\e_n,\a_n,\t_n):=\lim_{L\rightarrow\infty}
\langle\bT\ {\bf a}^{\e_1}_{x_1,\a_1,\t_1}\cdots {\bf a}^{\e_n}_{x_n,\a_n,\t_n}\rangle_{\beta,L},
\ee
for $x_i\in\Lambda_{\beta, \Lambda_L}:=[-\b,\b[\times\L_L$, $\t_i=\uparrow\downarrow$,
$\e_i=\pm$, $\a=1,2$. 

For $\xx\neq\yy$ and $x-y\neq
(\pm\b,\vec 0)$, the non-interacting two-point Schwinger function (also called the {\it free propagator}) is given by:
\bea
C_{\b}
(x-y) \ := \lim_{L\rightarrow\infty}S_{2,\b,L}(x,\t,-;y,\t,+)\Big|_{\l=0}=\lim_{L\rightarrow\infty} {1\over \b|\L_L|}
\sum_{k\in\DD_{\b,L}}e^{-ik\cdot(x-y)} C(k)\label{free2pt},
\eea
where $\DD_{\b, L}=\DD_\b\times\DD_L$, with $\DD_\b:=\{k_0={2\pi\over \b}(n_0+{1\over 2})\;:\;n_0\in\ZZZ
\}$, the set of the Matsubara frequencies and $\DD_L=\{\bk={n_1\over L}{\bG_1}+ {n_2\over L}{\vec
G_2}\ :\ -L\le n_1,n_2\le L-1\}$, and 
\bea\label{2ptk}
C(k)&=&\begin{pmatrix}-i k_0-\m & -\O^*_0(\bk) \\ -\O_0(\bk) &
-ik_0-\m\end{pmatrix}^{-1}\\
&=&\frac{1}{k_0^2+|\O_0(\bk)|^2-\m^2-2i\mu k_0} \begin{pmatrix}i k_0+\m &-\O^*_0(\bk) \\ -\O_0(\bk) &
ik_0+\m\end{pmatrix},\nn
\eea
is the free propagator in momentum space. 

Remark that the denominator of \eqref{free2pt} is never vanishing 
when the temperature $T=\frac{1}{\beta}$ is strictly positive, which means that the temperature serves as the infrared cutoff for the model.

\subsection{The Fermi surfaces}
The {\it non-interacting band structure} is defined as 
\bea\label{band1}
e(\bk,\mu):=-\det \begin{pmatrix}-\m & -\O^*_0(\bk) \\ -\O_0(\bk) &
-\m\end{pmatrix}=
|\O_0(\bk)|^2-\mu^2,
\eea
which is equal to $$4\cos(3k_1/2)\cos(\sqrt{3} k_2/2)+
4\cos^2(\sqrt{3} k_2/2)+1-\mu^2.$$ The noninteracting Fermi surface (F.S.) $\cF_0$ is defined as the locus of the dispersion relation:
\be{\cal F}_0:=
\{\bk=(k_1, k_2)\in \RRR^2\vert\ e(\bk,\mu)=0\}.\label{freefs}\ee
The Fermi surface $\cF_0$ may have several connected components in general, each of which is called a Fermi curve (F.C.). The geometric properties of the Fermi surface depend crucially on the value of $\mu$.

When $\mu=0$, the solution to the equation $e_0(\bk,\mu)=0$ composes of a set of points, called 
{\it the Fermi points}, among which only $\bk^F_1=(\frac{2\pi}{3},\frac{2\pi}{3\sqrt{3}})$, and $\bk^F_2=(\frac{2\pi}{3}, -\frac{2\pi}{3\sqrt{3}})$ are considered as fundamental. Other Fermi points are just translations of them.  

\begin{figure}[!htb]
\centering
\includegraphics[scale=.6]{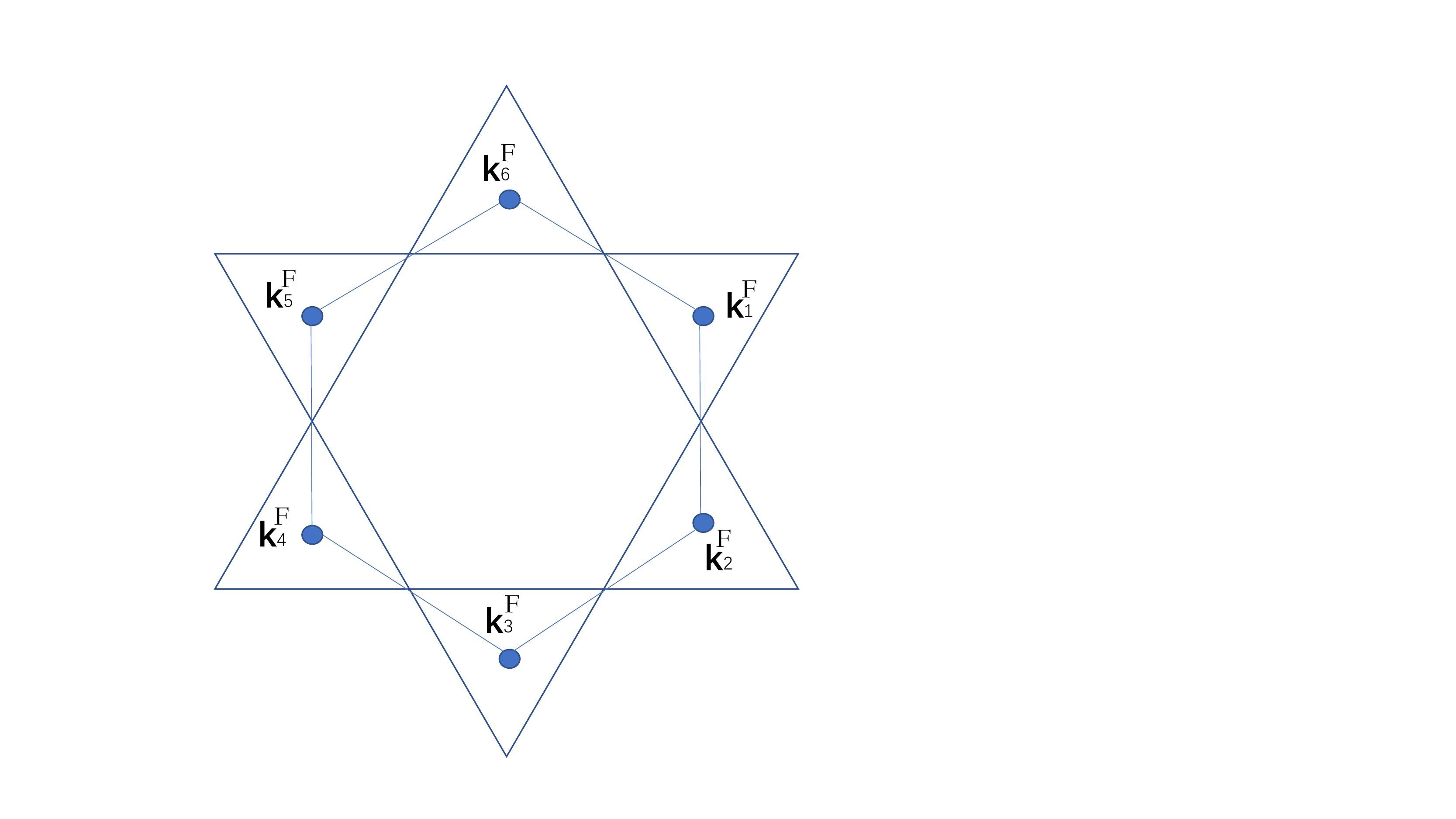}
\caption{The Fermi Triangle at $\mu=1$. The centers of the Fermi triangles are the Fermi points.}\label{fpt}
\end{figure}

When $\mu=1$, the band structure reads:
\bea\label{div0}
e(\bk,1)=[\cos(\sqrt{3} k_2/2)]\cdot[\cos( \frac14(3k_1+\sqrt{3} k_2))]\cdot[\cos( \frac14(3k_1-\sqrt{3} k_2))],
\eea
and the solutions of $e(\bk,1)=0$ are a set of lines:
\be\label{sol0}
k_2=\pm\frac{(2n+1)\pi}{\sqrt3},\  n\in\ZZZ \quad {\rm and}\quad  k_2=\pm {\sqrt{3}} k_1\mp\frac{4n+2}{\sqrt3}\pi,\ 
n\in\ZZZ\ ,
\ee
which form a set of perfect triangles, each of which is called a Fermi triangle and the Fermi surfaces is the union of these perfect triangles. Among all these triangles the two triangles surrounding the Fermi points $\bk^F_1$ and $\bk^F_2$, respectively, are considered as fundamental and all others are considered as translations of them. Figure \ref{fpt} shows the Fermi surfaces composed of $6$ Fermi triangles surrounding the corresponding Fermi points.  

When $0<\mu<1$, the Fermi surfaces are closed convex curves centered around the Fermi points and contained in the Fermi triangles \cite{GMW}. The lines of the Fermi triangles form boarders of these Fermi curves; When $\mu>1$ the Fermi surfaces can be more complicated and can be more degenerate.

\subsection{Main result}
We consider the weakly interacting Hubbard model on the mono-layer honeycomb lattice with a non-vanishing bare chemical potential which takes values in a neighborhood of $1$. Since this model is not at half-filling for this choice of chemical potential, the shape of the Fermi surface is changing due to the interactions. We introduce a chemical potential counter-term to the interaction term so that the renormalized chemical, which decides the non-interacting dispersion relation hence the shape of the Fermi surface, is fixed to be $1$. We will prove that the ground state of this model doesn't displays Fermi liquid behavior but the Luttinger liquid behavior. We also prove that the chemical potential counter-term $\delta\mu$ is bounded by $K|\lambda|$ for certain positive constant $K$. The main theorem of this paper is:
\begin{theorem}
The perturbation series of the self-energy of the Hubbard model on the honeycomb lattice with renormalized chemical potential $\mu=1$ at temperature $T>0$ are convergent for the coupling constant $\lambda<1/|\log T|^2$. The second derivative of the self-energy is not uniformly bounded when $T\rightarrow0$. So the ground state of the system is not Fermi liquid, by Salmhofer's criterion on Finite temperature Fermi liquids (cf. Theorem \ref{salmc} or \cite{salm}). 
\end{theorem}
\section{The Multi-scale Analysis}
\subsection{The functional integrals}
It is mostly convenient to express the partition functions and correlations in terms of Grassmann integrals over the Grassmann algebra ${\bf Gra}$, which is an associative, non-commutative, nilpotent algebra over the set of Grassmann variables 
$\{\hat\psi^\e_{k,\t,\a}\}^{\t=\uparrow\downarrow;\a=1,2;\epsilon=\pm}_{k\in\DD_{\b,L}}$ 
such that 
$\hat\psi^\e_{k,\t,\a}\hat\psi^{\e'}_{k',\t',\a'}=-\hat\psi^{\e'}_{k',\t',\a'}\hat\psi^\e_{k,\t,\a}$, for  $(\e,\t,\a,k)\neq(\e',\t',\a',k',)$ and $(\hat\psi^\e_{k,\t,\a})^2=0$.
The Grassmann differentiation is defined by 
${\partial_{ \hat\psi^\e_{k,\t,\a}}}{ \hat\psi^{\e'}_{k',\t',\a'}}=\delta_{k,k'}\delta_{\t,\t'}\delta_{\a,\a'}\delta_{\e,\e'}$
while the Grassmann integral is defined as follows: 
given a monomial $Q( \hat\psi^-, \hat\psi^+)$ in the
variables 
$\hat\psi_{k,\t,\a}^-, \hat\psi_{k,\t,\a}^+$, let \be D\psi=\prod_{k\in\DD_{\b, L}, \atop
\t=\pm, \a=1,2}d\hat\psi_{k,\t,\a}^+
d\hat\psi_{k,\t, \a}^-,\lb{2.10}\ee be the Grassmann Lebesque measure,
the Grassmann integration $\int D\psi$ is $0$, except in the case $Q( \hat\psi^-, \hat\psi^+)=
\prod_{k,\t,\a} \hat\psi^-_{k,\t,\a} \hat\psi^+_{k,\t,\a}$, up to a
permutation of the variable, in which case its value is $1$. 

The Grassmann Gaussian measure $P(d\psi)$ with covariance $\hat C_k$ is defined by the following
equation:
\be
\lim_{L\rightarrow\infty}\int P(d\psi)\hat\psi^{-}_{k_1,\t_1,\a_1}\hat\psi^{+}_{k_2,\t_2,\a_2}=\delta_{k_1,k_2}\delta_{\t_1,\t_2}[\hat C_{k_1}]_{\a_1,\a_2}.
\ee
We have:
\be
P(d\psi) = N^{-1} D\psi \cdot\;\exp \Bigg\{-{1\over
|\L_L|\b} \sum_{k\in\DD_{\b, L},\t={\uparrow\downarrow}, \a=1, 2 } 
\hat\psi^{+}_{k,\t, \a}{\hat g}_k^{-1}\hat\psi^{-}_{k,\t,\a}\Bigg\}\;,
\label{ggauss}\ee
where
\be
N=\prod_{\kk\in\DD_L,\t={\uparrow\downarrow}}{1\over
\b|\L_L|}
\begin{pmatrix}-i k_0-1 & -\O^*(\bk) \\ -\O(\bk) &
-ik_0-1\end{pmatrix},\label{norma}
\ee
is the normalization factor. 

Define the Grassmann fields in the direct space by
\be
\psi^\pm_{x,\t,\a}=\frac{1}{\b|\Lambda_L|}\sum_{k\in\DD_{\b, L}}
e^{\pm ikx}\hat\psi^\pm_{k,\t,\a},\ \ x\in\Lambda_{\beta,L},
\ee
the interacting potential becomes:
\bea \VV(\psi)&=&
\l\sum_{\a,\a'=1,2}\ \int_{\Lambda_{\beta,L}} d^3x \, \psi^+_{x,\uparrow,\a}
\psi^-_{x,\uparrow,\a'}\psi^+_{x,\downarrow,\a}
\psi^-_{x,\downarrow,\a'}\nn\\
&&\quad\quad+\delta\mu \int_{\Lambda_{\beta,L}} d^3x\ \Big(\ \sum_{\a=1,2}\sum_{\tau=\uparrow, \downarrow} \psi^+_{x,\t,\a}
\psi^-_{x,\t,\a}\ \Big)
\nonumber\\
&=&\frac{\l}{\b|\L_L|}\ \Big[\ \sum_{\a,\a'=1,2}\sum_{\tau,\tau'=\uparrow, \downarrow}\sum_{k,k',p\in\DD_{\beta, L}}\hat\psi^+_{k-p,\tau,\a}\hat\psi^-_{k,\tau,\a'}
\hat\psi^+_{k'+p,\tau',a}\hat\psi^-_{k',\tau',\a'}\nn\\
&&\quad\quad\quad\quad +\delta\mu\ \Big(\ \sum_{k\in\DD_{\beta, L}}\sum_{\a=1,2}\sum_{\tau=\uparrow, \downarrow} \hat\psi^+_{k,\t,\a}
\hat\psi^-_{k,\t,\a}\ \Big)\ \Big]
\label{potx}\ ,\eea
where $\int_{\Lambda_{\beta,L}} d^3x:=\int_{-\beta}^\beta dx_0\ \sum_{\xx\in\L_L}$ is the short-handed notion for the integration and sum, the summation over $k, k'$ and $p$ runs over $\DD_{\b,\L_L}$.

It is easy to prove that both the Grassmann Gaussian measure and the quartic term $V(\psi)$ are invariant under the following transformations (see cf. \cite{GM}):
\begin{itemize}
\item spin exchange: $\hat\psi^\e_{k,\t,\a}\leftrightarrow\hat\psi^\e_{k,-\t,\a}$, where we use the notations $\e=\pm$, $\t=\uparrow,\downarrow$ while $-\t=\downarrow,\uparrow$, $\a=1,2$;

\item global $U(1)$: $\hat\psi^\epsilon_{k,\tau,\a}\rightarrow e^{i\epsilon\theta_1}\hat\psi^\epsilon_{k,\tau,\a}$, with 
$\theta_1\in\RRR$ independent of $k$;

\item global spin $SO(2)$: 
\be
\begin{pmatrix}\hat\psi^\epsilon_{k,\uparrow,\a}\\
\hat\psi^\epsilon_{k,\downarrow,r}\end{pmatrix}\rightarrow\begin{pmatrix}\cos\theta_2&\sin\theta_2\\
-\sin\theta_2&\cos\theta_2\end{pmatrix}\ \begin{pmatrix}\hat\psi^\epsilon_{k,\uparrow,\a}\\
\hat\psi^\epsilon_{k,\downarrow,\a}\end{pmatrix};
\ee

\item discrete spatial rotations: $\hat\psi^\pm_{(k_0,\kk),\t,\a}\rightarrow e^{\mp i(\a-1)\kk\cdot(\dd_3-\dd_1)}\hat\psi^\pm_{(k_0,T_{2\pi/3} \kk),\t,\a}$, where
$T_{2\pi/3} \kk$ means rotation of the vector $\kk$ by $2\pi/3$. Remark that the discrete spatial 
rotation symmetry group is also the isometric group for the Fermi surface, to be discussed later.

\item complex conjugation: $\psi^\epsilon_{(x_0,\xx),\t,\a}\rightarrow \psi^\epsilon_{(-x_0,\xx),\t,\a}$, $c\rightarrow c^\star$ where $c$ is a generic constant appearing in the Grassmann Gaussian measure $p(d\psi)$ or the interaction $\VV(\psi)$; 
\end{itemize}


The Schwinger functions are the moments (only formally, as one needs renormalizations for mathematically well defined Schwinger functions) of the Grassmann measure $P(d\psi)e^{-\VV(\psi)}$
\be
S_n(x_1,\t_1,\e_1,\a_1,\cdots,x_n,\t_n,\e_n,\a_n)=\frac{\int\psi^{\epsilon_1}_{x_1,\t_1,\a_1}\cdots \psi^{\epsilon_n}_{x_n,\t_n,\a_n} P(d\psi)e^{-\VV(\psi)}}{\int P(d\psi)e^{-\VV(\psi)}}\ ,
\ee 
while the free propagator is given by:
\be
[S_0(x-y)]_{\a_1,\a_2}=\lim_{L\rightarrow\infty}\int P(d\psi)\psi^-_{x,\t,\a_1}
\psi^+_{y,\t,\a_2}\ ,
\ee
for $x-y\neq \beta\ZZZ\times\{0\}$. 
\begin{remark}
In the following we will be interested in the $L\rightarrow\infty$ limit of the Schwinger functions and the self-energy. So we will take this infinite volume limit in the future parts of this paper, although we still use $\Lambda_{\beta, L}$ and $\DD_{\beta, L}$ to indicate the domains of the spatial-tempo variables and the momentum variables. 
\end{remark}


\subsection{Scale Analysis}
Since the model is defined on a lattice, the momentum dual to the minimal distance between lattice points plays the role of the ultraviolet momentum cutoff. So the correlation functions are not divergent at short distance or high momentum, but could be singular in the infrared limit, namely, when $|\L_L|\rightarrow\infty$ and $T\rightarrow0$, for which we have $e(\kk,1)\rightarrow0$. So we fix the ultraviolet cutoff to be $1$ and study only the infrared behaviors of the correlation functions. 

Recall the Gevrey class of compacted supported functions $G^s_0$:
\begin{definition}
Given $\Omega\subset\RRR^d$ and $s\ge1$, the Gevrey class $G^s_0(\Omega)$ of index $s$ is defined as the set of smooth functions $f\in\cC^\infty(\Omega)$ such that for every compact subset $K\subset\Omega$ there exist two positive constants $A$ and $L$ satisfying
\bea
\max_{x\in K}|\partial^\alpha f(x)|\le AL^{-|\alpha|}(|\alpha!|)^s,\ \alpha\in\ZZZ^d_+,\ |\alpha|=\alpha_1+\cdots+\alpha_d.
\eea
The Gevrey class of functions with compact support, $G_0^s$, is defined as 
\be
G_0^s(\Omega)=G^s(\Omega)\cap C^\infty_0(\Omega).
\ee
The Fourier transform for any $f\in G_0^s$ satisfies
\be
\max_{k\in\RRR^d}|\hat f(k)|\le Ae^{-s(\frac{L}{\sqrt{d}}|k|)^{1/s}}.
\ee
\end{definition}

In order to implement the infrared analysis, we introduce the smooth, compacted supported functions $\chi\in G^s_0(\RRR)$, which satisfy:
\be
\chi(t)=\chi(-t)=
\begin{cases}
=0\ ,&\quad {\rm for}\quad  |t|>2,\\
\in(0,1)\ ,&\quad {\rm for}\quad  1<|t|\le2,\\
=1,\ &\quad {\rm for}\quad  |t|\le 1. 
\end{cases}\label{support}
\ee
Given any fixed constant $\gamma\ge10$, we can construct a partition of unity
\bea\label{part1}
1&=&\sum_{j=0}^{\infty}\chi_j(t),\ \ \forall t\neq 0;\\
\chi_0(t)&=&1-\chi(t),\nn \\ 
\chi_j(t)&=&\chi(\gamma^{2j-1}t)-\chi(\gamma^{2j}t)\ {\rm for}\ j\ge1.\nn
\eea

Remark that the free propagator \eqref{2ptk} can be written as
\be C(k)=\tilde C(k) A(k),\label{freep2}\ee
where
\be\label{freep3}
A(k)=\begin{pmatrix}i k_0+1&-\O^*_0(\bk) \\ -\O_0(\bk) &
ik_0+1\end{pmatrix}
\ee
is the $2$ by $2$ matrix and
\be\label{redprop}\tilde C(k):=\frac{1}{-2ik_0-e(\kk,1)+k^2_0},\ee
which is non-singular when either $k_0$ or $e(\kk,1)$ is an order one quantity; When both $k_0$ and
$e(\kk,1)$ are approaching zero, the term $k_0^2$ is negligible comparing to the first two terms. 

Following this simple argument, the free propagator is decomposed as :
\bea\label{multiprop}
&&C(k)_{\a\a'}=\sum_{j=0}^\infty \ C^j(k)_{\a\a'},\ \a,\a'=1,2,\\
&&C^j(k)_{\a\a'}=C(k)_{\a\a'}\cdot \chi_j[4k_0^2+e^2(\kk)].\nn
\eea
It is useful to introduce the multi-slice decomposition for $\tilde C(k)$:
\bea\label{multip2}
&&\tilde C(k)=\sum_{j=0}^{j_{max}} \tilde C^j(k),\\
&&\tilde C^j(k)=\tilde C(k)\cdot \chi_j[4k_0^2+e^2(\kk)].\label{ctilde}
\eea
Remark that the support function $\chi_j[4k_0^2+e^2(\kk)]$ vanishes for $j\ge j_{max}:=\EEE(\tilde j_{max})$, where $\tilde j_{max}$ is the solution to $\gamma^{\tilde j_{max}-1}= 1/\sqrt2\pi T$ and
$\EEE(\tilde j_{max})$ is the integer part of $\tilde j_{max}$. So there are only finitely many of indices $j$ to be summed in \eqref{multiprop}.

The support support of $\chi_j$ is noted by $\cD_j$:  
\be\label{multi1}
\cD_j=\Big\{k=(k_0,\kk) \vert\g^{-2j}\le 4k_0^2+e^2(\kk)\le 2\g^{-2j+2}\ \Big\},\ee
for each index $j$.
Since the temperature $T$ is strictly positive, we have $k_0^2\ge\pi^2 T^2>0$, which implies that
\be
-\sqrt2\g^{-j+1}\le e(\kk)\le\sqrt2\g^{-j+1}.
\ee
Define also the topological annulus as:
\be
\cA_j=\{\kk\ | -\sqrt2\g^{-j+1}\le e(\kk)\le\sqrt2\g^{-j+1}\},
\ee
so the non-interacting Fermi surface $\cF_0=\{\kk\vert\ e_j(\kk)=0\}$ is completely covered by the annulus. 


Formula \eqref{multi1} implies also the following lemma for the matrix elements of $A$ (see \eqref{freep3}):
\begin{proposition}\label{mat0}
Let $j\in[0, j_{max}]$ be any scaling index and $\gamma\ge4$ a fixed constant such that $\g^{-2j}\le k_0^2+e^2(\kk)\le 2\g^{-2j+1}$, let
$\vert A_{\a\a'}\vert$ be the absolute value of the matrix element $A_{\a\a'}$,$\a,\a'=1,2$,
then there exist two constants $O(1), O'(1)$ with $0.5<O(1)<1$, $1<O'(1)<10$, such that
\be\label{matele2}
O(1)\le\vert A_{\a\a'}\vert\le O'(1),\  \a,\a'=1,2\ .
\ee 
\end{proposition} 
\begin{proof}
We consider first $\vert A_{11}\vert=\vert A_{22}\vert=\vert ik_0+1\vert=(1+k_0^2)^{1/2}$. Since $0<\gamma^{-j}\le k_0\le2$, we have $1<(1+k_0^2)^{1/2}<3$. So we can choose $O(1)$ and $O'(1)$ to satisfy the inequality \eqref{matele2}. Now we consider the elements $\vert A_{12}\vert=\vert A_{21}\vert=\vert \O_0(\kk)\vert$. By definition (see \eqref{band1}) we have $\vert \O_0(\kk)\vert=(1+e(\kk,1))^{1/2}$. From Formula \eqref{multi1} we can easily deduce that $-\sqrt2/2\le e(\kk,1)\le\sqrt2/2$ and $0.54\le(1+e(\kk,1))^{1/2}\le1.31$. So we proved this lemma.
\end{proof}

\begin{remark}\label{mat1}
Since each of the matrix element $A_{\a\a'}$ in \eqref{freep2} is bounded by an order one constant both from below and from above, the infrared asymptotic behavior of the propagator $C(k)_{\a\a'}, \a=1,2$ is dominated by that of $\tilde C(k)$. So in the future analysis we can simply replace each matrix element of $C^j(k)$ by $\tilde C^j(k)$. 
\end{remark}
Following this remark we have the following lemma:
\begin{lemma}
Let $C^j(k)_{\a\a'}$, $\a,\a'=1,2$, be any matrix element of the momentum space free propagator at slice $j$, we have
\be\label{tad1}
\sup_{k\in \cA^j}\Vert C^j(k)_{\a\a'}\Vert\le O(1)\g^{j},
\ee
where $O(1)$ is certain positive constant of order one.
\end{lemma}
\begin{proof}
From the definition of the support function $\chi_j$ and \eqref{ctilde} it is easily find that
\be
\sup_{k\in A^j}\Vert \tilde C^j(k)_{rr'}\Vert\le O(1)'\g^{j}\ ,
\ee
for certain positive constant $O(1)'$. This lemma can be easily proved when we combining this result with the above remark. 
\end{proof}

The bare chemical potential counter-term can be rewritten as $\delta\mu^1_{j_{max}}$, which is a function of the ultraviolet cutoff $\g^0=1$ and infrared cutoff $\g^{-j_{max}}$. The value of the renormalized chemical potential $\mu$, which is the one appears in the free-propagator $C^j$, is fixed to be $1$. The renormalization condition (BPHZ condition) states that:
\be\label{rncondition}
\delta\mu_{ren}=\delta\mu^{j_{max}}_{j_{max}}=\hat\Sigma_2(k_F)=0,
\ee
where $\hat\Sigma_2$ is the self-energy in the momentum space representation and $k_F$ is the Fermi momentum (cf. Section $2.2$), which means that there is no renormalization to the chemical potential when the energy scale of the model coincides with the infrared cutoff. So this is a very natural condition. 
Let the counter-term at slice $j$ be $\delta\mu^j_{j_{max}}$, BPHZ condition implies that:
\be
\delta\mu^1_{j_{max}}=\sum_{j=0}^{j_{max}}\delta\mu^j_{j_{max}}.
\ee
Theorem \ref{flowmu} (cf. Section $5.5$) states that $\delta\mu^1_{j_{max}}$ is bounded by a small number $c.|\lambda|$, where $|\lambda|<1/|\log T|^2$, $T$ is the temperature and $c$ is certain positive constant. 

\subsection{Sectors and angular analysis}
Notice that Formula \eqref{multi1} implies only that the size of $k_0^2+e^2(\bk)$ is bounded by
$O(1)\gamma^{-2j}$, at scaling index $j$, but does not fix the size of $e^2(\bk)$. 
In order to obtain the optimal decay behavior for the propagator, we further decompose the support of the $\chi_j(k)$, which is $\cA_j$, into {\it sectors}, which is a set of rectangle covering the annulus $\cA_j$. The size of the sectors depends also on the scaling indices. So the sliced propagators are now supported now on the sectors. The conception and techniques for sectors have been introduced for the first time in \cite{FMRT} and have been developed in \cite{DR1,FKT, Riv, AMR1, AMR2, BGM1, BGM2}. Since the Fermi surfaces in our model have flat edges, a curve in the neighborhood of the Fermi surface define by the dispersion relation is highly non-isometric. Hence the sectors introduced in this paper is very similar to that in \cite{Riv}, in which the Fermi surface is a fixed square. The methods of the angular analysis for the Fermi surface in this paper also come from \cite{Riv}.

\begin{figure}[htp]
\centering
\includegraphics[width=1.0\textwidth]{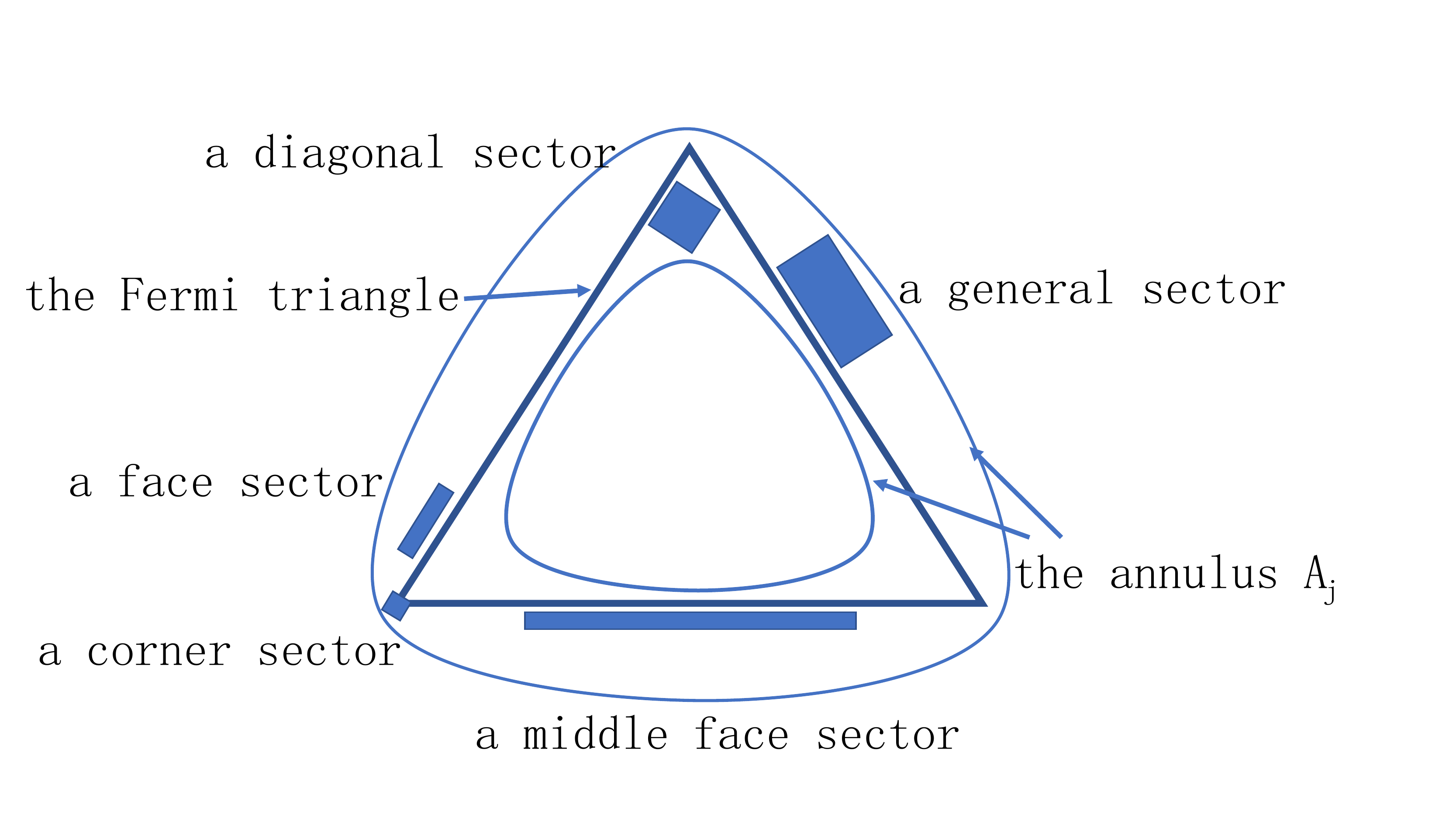}
\caption{\label{figsec}An illustration of the various sectors.
}
\end{figure}

Taking into account the $\ZZZ^3$ symmetry of the Fermi triangle (see Figure \ref{figsec}), it is convenient to introduce a new basis $(e_+, e_-)$ that is neither orthogonal nor normal:
\be
e_+=\frac{\pi}{3}(1, \sqrt3),\quad e_-=\frac{\pi}{3}(-1, {\sqrt3}),
\ee
and let $(k_+,k_-)$ be the coordinates in the new basis, we have
\be
k_1=\frac{\pi}{3}(k_+-k_-),\quad k_2=\frac{\pi}{\sqrt3}(k_++k_-),
\ee
and the band formula \eqref{div0} becomes:
\be\label{band3}
e(\bk,1)=8\cos\frac{\pi(k_++k_-)}{2} \cos\frac{\pi k_+}{2} \cos\frac{\pi k_-}{2}.
\ee

We introduce the following partition of unity:
\bea\label{secf}
1=\sum_{s=0}^{j}v_{s}(t),\ \ \begin{cases} v_0(t)=1-\chi(\g^2t),\\
v_{s}(t)=\chi_{s+1}(t),\\
v_j(t)=\chi(\g^{2j}t),
\end{cases}
\ {\rm for}\ \ 1\le s\le j-1,
\eea
where $t$ stands for the term $\cos^2\frac{\pi(k_++k_-)}{2}$, $\cos^2\frac{\pi k_+}{2}$ or $\cos^2\frac{\pi k_-}{2}$ in $e^2(\kk,1)$ and $s=s_0$, $s_+$, or $s_-$ is the corresponding sector index. Since $\cos^2\frac{\pi(k_++k_-)}{2}$ is a function of $\cos^2\frac{\pi k_+}{2}$ and $\cos^2\frac{\pi k_-}{2}$, $s_0$ is not independent of $s_\pm$. 

Following \cite{Riv}, we have the following names for the sectors with indices $\{s_+,s_-\}$ that have different shapes and cover different regions of the annulus $\cA_j$:
\begin{itemize}
\item the sectors $(0,j)$ and $(j,0)$ are called the middle-face sectors;
\item the sectors $(s,j)$ and $(j,s)$ with $j>s>0$ are called the face sectors;
\item the sector $(j,j)$ is called the corner sector;
\item the sectors $(s,s)$ with $(j-2)/2\le s<j$ are called the diagonal sectors;
\item the other sectors are called the general sectors. 
\end{itemize}


We have the following simple but useful lemma concerning the amplitudes of the three factors in \eqref{secf}.
\begin{lemma}\label{bdcos}
Let $s\ge2$ be a sector scaling index. If $\vert\cos\frac{\pi k_+}{2}|\le\gamma^{-s}$ or $|\cos\frac{\pi k_-}{2}|\le\gamma^{-s}$, 
then there exists an order-one constant $0<O(1)\le1$ such that 
\be
O(1)\le\vert\cos\frac{\pi(k_++k_-)}{2}\vert\le1.
\ee
\end{lemma}
\begin{proof}
By trigonometrical formula we have
\be
\cos\frac{\pi(k_++k_-)}{2}=\cos\frac{\pi k_+}{2} \cos\frac{\pi k_-}{2}-\sin\frac{\pi k_+}{2} \sin\frac{\pi k_-}{2}.
\ee
We consider two possible cases. In the first case let $|\cos\frac{\pi k_-}{2}|=O'(1)$ and $|\cos\frac{\pi k_+}{2}|\sim\gamma^{-s}$, where $O'(1)$ is another constant such that both $O'(1)$ and $(1-{O'(1)}^2)^{1/2}$ are greater than $\gamma^{-1}$ and $A\sim B$ means $A=O''(1)\cdot B$,
where $0<O''(1)\le1$ is another constant. Then we have
\be
|\cos\frac{\pi(k_++k_-)}{2}|\ge \gamma^{-1}-O'(1)\gamma^{-s}-O(\gamma^{-2s})
\ge \gamma^{-1}(1-\gamma^{-s+1}-\gamma^{-2s+1})>0.
\ee
So we can choose $O(1)$ to be $\gamma^{-1}(1-2\gamma^{-1})$, which is strictly bounded away from zero.

Now we consider the second case, in which $|\cos\frac{\pi k_-}{2}|\le\gamma^{-s}$ and $|\cos\frac{\pi k_+}{2}|\le\gamma^{-s}$. We have
\be
|\cos\frac{\pi(k_++k_-)}{2}|\ge 1-2\gamma^{-2s}-O(\gamma^{-4s}).
\ee
So we can choose the constant $O(1)$ to be $1-3\gamma^{-4}$, which is of order one and strictly bounded away from zero. So we proved this lemma.
\end{proof}

\begin{remark}\label{s0}
Remark that since $|\cos^2\frac{\pi(k_++k_-)}{2}|$ is always larger than an order-one constant, there is no need to assign the compact supported function $v_{s_0}$ or we can simply let $s_0=0$. So only two sector indices $s_+$ and $s_-$ are needed to define the sectorized propagators.
\end{remark}
\begin{definition}
Let $\{\sigma=(s_+, s_-)\}$ be a set of two tuples, in which $s_\pm=0,\cdots,j$ are the sector indices. We can further decompose the propagator according to the sectors and define the resulting  propagator (which is also called the {\it sectorized propagator}) as:
\be\label{sec0}
C_j(k)=\sum_{\s=(s_+, s_-)}C_{j,\s}(k),
\ee
where
\be
C_{j,\s}(k)=C_j(k)\cdot v_{s_+}[\cos^2(k_+/2)]\ v_{s_-}[\cos^2(k_-/2)].\label{sec1}
\ee
\end{definition}

Remark that when the scaling index $j$ is a small, say $j\le10$, then both $k_0^2$ and $e^2(\kk,1)$ are strictly bounded away from zero hence the sliced free propagators are not singular. In this case there is indeed no need of introducing the sectors. So it is useful to indicate a scaling index, called the infrared threshold, above which the infrared analysis becomes important:
\begin{definition}[The infrared threshold]
The infrared threshold $j_0$ is defined as a fixed finite positive integer such that the sector decompositions become important when the scaling index $j$ is greater than $j_0$. Remark that since we can always change the energy scales of the model by choosing the hopping parameter $t$, there is 
the ambiguity for choosing $j_0$ and we can always shift $j_0$ by a finite number. The infrared behavior of the model (for $j\rightarrow\infty$) is independent of the exact value of $j_0$. 
\end{definition}
We have the following lemma concerning the various sector indices. 
\begin{proposition}\label{spm}
Let $j$ be the scaling index, then the possible values of the sector indices $s_+$ and $s_-$ must satisfy:
\be
s_++s_-\ge j-2.
\ee
\end{proposition}
\begin{proof}
By Remark \ref{s0} we can always set the sector index $s_0=0$. From the definition of the compact support functions \eqref{secf} we can easily derive that
\be
e^2(\kk,1)\ge\gamma^{-2s_+-2}\gamma^{-2s_--2}\ .
\ee
In order that the sliced propagator is non-vanishing, $e^2(\kk,1)$ must obey Formula \eqref{multi1}, which implies that 
\be
\gamma^{-2s_+-2}\gamma^{-2s_--2}\ge\g^{-2j}\ .
\ee
So we have:
\be
s_++s_-\ge j-2.
\ee
\end{proof}
\begin{corollary}
As a corollary we can easily find that for the choice of the infrared threshold $j_0=5$ so that
for $j\ge 5$, at least one of the sector index,
$s_+$ or $s_-$ is greater than $2$. This choice of $j_0$ is also consistent with Lemma \ref{bdcos}, which assumes that at least one of the sector index is greater or equal to $2$.
\end{corollary}
\begin{definition}[The depth of a sector]
It is useful to define a new scaling index, the $depth$ $l(\s)$ of a sector $\sigma$, to be $l=s_++s_--j+2$. This index describes the distance of an $s$ sector to the Fermi surface. 
\end{definition}

\subsection{Conservation of momentum and the constraints on the sector indices}
By decomposing the support of the correlation functions into sectors, we need to sum over not only the scaling indices $j$, but also the sector indices $\sigma=(s_+,s_-)$. Conservation of momentums at each interaction vertex places constraints on the values of the associated sectors indices. First of all, let us consider the supporting properties for the propagators. In order that $C^j_{\s}(k)\neq0$, the momentum $k=(k_0,\bk)$ must satisfy the following bounds:
\be
\frac{1}{4}\gamma^{-2j}\le k^2_0\le \frac12 \g^{-2j+2},
\ee
\bea\label{supp1}
\begin{cases}
\g^{-1}\le\vert\cos(\pi k_\pm/2)\vert\le 1,\quad\quad\quad\quad\quad\ {\rm for}\quad s_\pm=0,\\
\g^{-s_\pm-1}\le\vert\cos(\pi k_\pm/2)\vert\le\sqrt2 \g^{-s_\pm},\quad {\rm for}\quad  1\le s_\pm\le j-1,\\
\vert\cos(\pi k_\pm/2)\vert\le\sqrt2 \g^{-j},\quad\quad\quad\quad\quad\quad{\rm for}\quad  s_\pm=j.
\end{cases}
\eea
So we have $\vert k_\pm\vert\le1$ in the first Brillouin zone (recall that $|k_{\pm}|=\frac{1}{2\pi}(3k_1+\sqrt3k_2)$, so we have
$|k_{\pm,max}|=\frac{1}{2\pi}(3\cdot\frac{2\pi}{3}+\sqrt3\cdot\frac{2\pi}{\sqrt3})=2$) for any scaling index $j\ge 1$. Let $q_\pm$ be the fractional part of $k_\pm$, defined by
$q_\pm=k_\pm-1$ for $k_\pm>1$ and $q_\pm=k_\pm+1$ for $k_\pm>1$.

So \eqref{supp1} is equivalent to:
\bea
\begin{cases}
2/{\pi \g}\le\vert q_\pm\vert\le 1,\quad\quad\quad\quad\quad\quad\quad\quad {\rm for}\quad s_\pm=0,\\
2{\g^{-s_\pm-1}}/{\pi }\le\vert q_\pm\vert\le\sqrt2 \g^{-s_\pm},\quad\quad\ {\rm for}\quad  1\le s_\pm\le j-1,\\
\vert q_\pm\vert\le\sqrt2 \g^{-j},\quad\quad\quad\quad\quad\quad\quad\quad\quad {\rm for}\quad  s_\pm=i.
\end{cases}\label{supp2}
\eea

Let $v$ be an interaction vertex with four fields, whose momentum $k_1,\dots,k_4$ are supported in the sectors with scaling indices $j_1,\dots, j_4$, and sector indices $\sigma_1,\cdots,\sigma_4$, respectively.  The conservation of momentum and the periodic boundary conditions imply that :
\bea
&&k_{1,0}+k_{2,0}+k_{3,0}+k_{4,0}=0,\label{com0}\\
&&k_{1,+}+k_{2,+}+k_{3,+}+k_{4,+}=2n_+,\label{com1}\\
&&k_{1,-}+k_{2,-}+k_{3,-}+k_{4,-}=2n_-,\label{com2}
\eea
where $n_+$ and $n_-$ are integers with the same parity. Since $q_\pm=k_\pm\pm1$ and since even sums of the numbers $\pm1$ is still an even number, the above formula implies that
\bea
&&q_{1,+}+q_{2,+}+q_{3,+}+q_{4,+}=2m_+,\label{coq1}\\
&&q_{1,-}+q_{2,-}+q_{3,-}+q_{4,-}=2m_-,\label{coq2}
\eea
where $m_=\pm$ are integers. The following lemma states that $m_\pm=0$ except for very special cases.
\begin{lemma}
Let $i=1,\cdots,4$ be the labeling of the four momentum $k_i$ entering or existing the vertex $v$. The integers $m_\pm\neq0$ only when $s_{i,\pm}=0$ for at least two values of $i$.
\end{lemma}
\begin{proof}
We prove first the case for $m_+$, using \eqref{coq1}; the proof for the case of $m_-$ is exactly the same.  Since $\vert q_{i,+}\vert\le1$, with $i=1,\cdots,4$, we have $\vert m_+\vert\le2$. When
$\vert m_+\vert=2$, we have $\vert q_{i,+}\vert=1$ for all $i$, which implies that $s_{i,+}=0$, for all $i$. Now we consider the case of $\vert m_+\vert=1$. Suppose that $s_{i,+}\neq0$ for three labellings the $i=1,2,3$ and $s_{4,+}=0$ . By Formula \eqref{supp2} we have $\vert q_{i,+}|\le \sqrt2\g^{-1}$ for $i=1,2,3$, so that
\be
q_{1,+}+q_{2,+}+q_{3,+}+q_{4,+}\le 3\sqrt2\g^{-1}+1<2,\quad {\rm for}\quad \g> 3\sqrt2.
\ee
So Formula \eqref{coq1} can't hold if more than two of the sector indices $s_{i,+}$ are not equal to zero. So we proved that $m_+\neq0$ only when $s_{i,+}=0$ for at least two values of the labeling index $i$. With exactly the same method we can prove the case for $m_-$. So we proved the Lemma. 
\end{proof}

Now we study in more detail constraints on the sector indices $s_{i,\pm}$, $i=1,\cdots,4$ for $m_\pm=0$.
\begin{proposition}\label{secmain}
Let $q_{i,+}$, $i=1,\cdots,4$ be the momenta enter or exist the vertex $v$ and $j_i$, $s_{i,+}$, $i=1,\cdots,4$, be the scaling indices and sector indices of the support functions for the corresponding propagators. Let
$s_{1,+}, s_{2,+}$ be the two smallest indices with smallest values among $s_{i,+}$, $i=1,\cdots,4$ such that $s_{1,+}\le s_{2,+}$, we have the following two cases for the possible values of $s_{1,+}$ and $s_{2,+}$: either $|s_{1,+}-s_{2,+}|\le1$ or $s_{1,+}=j_1$ and $j_1$ is strictly smaller than $j_2, j_3$ or $j_4$. We have exactly the same results for the sector indices $s_{i,-}$, $i=1,\cdots,4$.
\end{proposition}
\begin{proof}
We can always arrange these indices such that $s_{1,+}\le s_{2,+}\le s_{3,+}\le s_{4,+}$ and $j_{1}\le j_{2}\le j_{3}\le j_{4}$. Then we have either $s_{1,+}<j_1$ or $s_{1,+}=j_1$.
If $s_{1,+}<j_1$, the definition of the support function \eqref{supp2} imply that $q_{i,+}\le\sqrt2\g^{-s_{2,+}}$ for $i=2,3,4$ and $q_{1,+}\ge2{\g^{-s_{1,+}-1}}/{\pi }$.
In order that the equation $q_{1,+}+q_{2,+}+q_{3,+}+q_{4,+}=0$ holds we must have
\be
2{\g^{-s_{1,+}-1}}/{\pi }\le 3\sqrt2\g^{-s_{2,_+}},
\ee
which means
\be
s_{2,+}\le s_{1,+}+1+\log_\g (3\pi\sqrt2).
\ee
Since $0<\log_\g (3\pi\sqrt2)<1$ for $\g>3\pi\sqrt2$, we have $|s_{2,+}-s_{1,+}|\le1$. 

If $s_{1,+}=j_1$, by definition of the support function \eqref{supp2} and the conservation of momentum \eqref{coq1} we have
\be
\sqrt{2}\g^{-j_1}\ge3 \cdot{2\g^{-s_{2,+}}}/\pi\g,
\ee
which means
\be
j_1\le s_{2,+}+1-\log_{\g}(3\sqrt2/\pi)\le j_2-\log_{\g}(3\sqrt2/\pi).
\ee
So we have $j_2\ge j_1+1$. So $j_1$ is strictly smaller than all other scaling indices $j_2,\cdots j_4$.

Following the same arguments we can prove the results for the sectors in the $"-"$ direction.
So we proved this proposition.
\end{proof}

\subsection{Decay property of the propagator}
In this section we study the decay behavior of the free propagators in the direct space. Define the direct space coordinates $(x_+,x_-)$, which are dual to $(k_+,k_-)$, by $x_+=\pi(x_1/3+x_2/\sqrt3)$ and $x_-=\pi(-x_1/3+x_2/\sqrt3)$, with $\xx=(x_1,x_2)\in\Lambda_L$. We have the following lemma.

\begin{lemma}\label{bdx1}
Let $[C_{j,\sigma}(x-y)]_{\a\a'}=(\beta\vert\L\vert)^{-1}
\int d{k}e^{-ik(x-y)} [C_{j,\sigma}(k)]_{\a\a'}$, be the Fourier transform of the sectorized
propagator $[C_{j,\sigma}(k)]_{\a\a'}$ (c.f. Eq. \eqref{sec1}) with scaling index $j$ and sector indices $\s=(s_+, s_-)$. The propagator obey the following scaled decay: 
\be\label{decay1}
\vert\ [C_{j,\sigma}(x-y)]_{\a\a'}\vert\le O(1) \g^{-j-l}\ e^{-c[d_{j,\s}(x,y)]^\a},
\ee
where 
\be
d_{j,\s}(x,y)=\g^{-j}\vert x_0-y_0\vert+\g^{-s_+}\vert x_+-y_+\vert+\g^{-s_-}\vert x_--y_-\vert,
\ee
$\a\in(0,1)$ is the index characterizing the Gevrey class of functions (\cite{DR1}) and $O(1)$ is some order one constant \footnote{The interested readers who are familiar with the sectors for strictly convex Fermi surfaces (cf. \cite{DR1}, \cite{BGM2}, \cite{FKT} are invited to compare the different decaying properties.}.
\end{lemma}
\begin{proof}
From Lemma \ref{mat0} and Remark \ref{mat1} we know that each of the matrix elements $[C_{j,\sigma}(k)]_{\a\a'}$, $\a,\a'=1,2$, can be written as $\tilde C_{j,\sigma}(k)$ ( c.f. Eq. \eqref{multip2} ) times an order one constant and the latter doesn't contribute to the spatial decay of the propagator.
Let 
\be\tilde C_{j,\sigma}(x-y)=\frac{1}{\beta|\L_L|}\int d{k_0}dk_+dk_- \tilde C_{j,\sigma}(k_0,k_+,k_-)e^{ik_0(x_0-y_0)+k_+(x_+-y_+)+ik_-(x_--y_-)}
\ee
be the Fourier transform of $\tilde C_{j,\sigma}(k)$. Recall the explicit formula for $\tilde C_{j,\sigma}(k)$:
\be\label{decay3}
\tilde C_{j,\sigma}(k)=\frac{\chi_j[k_0^2+e^2(\kk,1)]}{-2ik_0-e(\kk,1)+k_0^2}\ v_{s_+}[\cos^2(k_+/2)]\ v_{s_-}[\cos^2(k_-/2)]\ ,
\ee
where $e^2(\kk,1)=64[\cos^2(k_++k_-)/2] [\cos^2(k_+/2)] [\cos^2(k_-/2)]$ (c.f. eq. \eqref{band3}).
Then in order to prove this lemma, it is enough to prove that:
\be\label{decay2}
\vert\tilde C_{j,\sigma}(x-y)\vert\le O(1) \g^{-j-l}\ e^{-c[d_{j,\s}(x,y)]^\a}.
\ee
This is essentially Fourier analysis and integration by parts. The integration $\int dk_0dk_+dk_-$ in the domain of the support function gives a factor $\g^{-j}\cdot \g^{-s_+}\cdot\g^{-s_-}$ while the integrand is bounded by $\frac{1}{\gamma^{-j}}$. These factors give $\g^{-s_+-s_-+j-j}=\g^2\g^{-j-l}$, which gives the pre-factor of \eqref{decay2}. 

Let $\frac{\partial}{\partial k_0}f=(1/2\pi T)[f(k_0+2\pi T)-f(k_0)]$ be the difference operator.  To prove the decay behavior of \eqref{decay2} it is enough to prove that
\bea\label{decay4}
\Vert\frac{\partial^{n_0}}{\partial k_0^{n_0}}\frac{\partial^{n_+}}{\partial k_+^{n_+}}\frac{\partial^{n_-}}{\partial k_-^{n_-}} \tilde C_{j,\s}(k_0, k_+,k_-)\Vert
\le O(1)^n\g^{jn_0}\g^{s_+n_+}\g^{s_-n_-}(n!)^{1/\alpha},
\eea
where $n=n_0+n_++n_-$ and $\Vert\cdot\Vert$ is the sup norm. Indeed, when $\frac{\partial}{\partial k_-}$ acts on $v_{s_-}[\cos^2(k_-/2)]$, using \eqref{supp2} we can easily find that is bounded by 
$O(1)\gamma^{-s_-}\cdot\g^{2s_{-}}=O(1)\g^{s_-}$, where $O(1)$ is a general constant; when it acts on $\chi_j[k_0^2+e^2(\kk,1)]$ it is simply bounded by $\g^{2j-2s_+-s_-}$; when it acts on the denominator $[-2ik_0-e(\kk,1)+k_0^2]^{-1}$ it is bounded by $\g^{j-s_+}$. Using $s_++s_-\ge j-2$ we find that each of the three factors is bounded by $O(1)\g^{s_-}$; when it acts on a factor $\cos(k_-/2)$ generated in the previous derivations, it costs a factor $O(1)\g^{s_-}$.
Similarly each operator $\frac{\partial}{\partial k_+}$ acts on various terms of $\tilde C_{j,\s}(k_0,k_+,k_-)$ is bounded by $O(1)\g^{s_+}$. The factor $(n!)^{1/\alpha}$ comes from derivations on the compact support functions which are Gevrey functions of order $\alpha$. When $j=j_{max}$ there is only the decay in the $x_0$ direction but no decay in the $x_+$ or $x_-$ direction. 
\end{proof}
We have the following lemma concerning the $L_1$ norm for the direct space propagator:
\begin{lemma}
The $L_1$ norm of $[C_{j,\sigma}(x)]_{\a\a'}$, $x\in\Lambda_{\beta,L}$, $\a,\a'=1,2$ is bounded as follows: 
\be\label{tad2}
\Big\Vert\ [C_{j,\sigma}(x)]_{\a\a'}\ \Big\Vert_{L^1}\le O(1)\g^{j}.
\ee
\end{lemma}
\begin{proof}
This lemma can be proved straightforwardly using Lemma \ref{bdx1}. We have
\bea
&&\Big\Vert\ [C_{j,\sigma}(x)]_{\a\a'}\ \Big\Vert_{L^1}=\Big|\ \int_{\Lambda_{\beta, L}} dx_0 dx_+ dx_-  [C_{j,\sigma}(x)]_{\a\a'}\Big|\\\
&&\quad\le O(1)\ \Big|\ \int_{\Lambda_{\beta, L}} dx_0 dx_+ dx_- \ \tilde C_{j,\sigma}(x)\ \Big|\nn\\
&&\quad\le O(1)\g^{-j-l}\g^{(j+s_++s_-)}\le O(1)\g^{j}.
\eea
\end{proof}
\begin{remark}
As a corollary we find that we lose a factor $\g^{2j+l}$ when we take the $L_1$ norm for the sliced propagator $[C_{j,\sigma}(x)]_{\a\a'}$, $\a,\a'=1,2$, comparing to its $L_\infty$ norm. Since the $L_1$ norm of a tree propagator (propagators associated with the tree lines, which be introduced shortly) will play an import role in the power-counting theorem, it is convenient to define a new scaling index in the future analysis, defined as follows.
\end{remark}

\begin{definition}\label{indexr}
We define a new scaling index $r$, which will play the role of the scaling index for the multi scale expansions, by $r=\EEE(j+l/2)$, where $\EEE(\cdot)$ is the nearest integer function. So $r\in [0,r_{max}(T)]$, where $r_{max}(T):=\EEE(1+\frac{3}{2}\ j_{max}(T))$.
Then we have the following decomposition of the propagator into slices $r$:
\be
C=\sum_{j,\sigma}C_{j,\s}=\sum_{r=1}^{r_{max}(T)}C_r,
\ee
where
\be
C_r=\sum_{{\s}\atop j(\s)+l(j,\s)/2=r}\ C_{j,\s} =\sum_{l}C_{r,l},\quad 
C_{r,l}=\sum_{{j,\s}\atop {j+l(\sigma)/2=r}} C_{j,\s}. 
\ee
\end{definition}
Since $|x-\EEE(x)|\le1$, $\forall x\in\RRR$, in the coming sections we shall simply forget the integer part $\EEE(\cdot)$ of the the above definition.

\begin{remark}\label{jrind}
Remark that although we have introduced four different kinds of indices $j$, $s_+$, $s_-$ and $r$, they are not independent but are related by the relation
\be
r=j+\frac l2=\frac{j+s_++s_-}{2}+1.
\ee
With the new index $r$ the constraints for the sector indices
\be
s_++s_-\ge j-2,\ 0\le s_{\pm}\le j,\ 0\le j\le j_{max},
\ee
becomes
\be\label{newr}
s_++s_-\ge r-2,\ 0\le s_{\pm}\le r,\ 0\le r\le r_{max}=3j_{max}/2\ ,
\ee
and the depth index becomes
\be\label{newl}
l=2(s_++s_--r+2).
\ee
\end{remark}

\section{The expansions}
In this section we introduce the multi-slice expansions for the $2p$-point Schwinger functions:
\bea
&&S_{2p}(\lambda, y_1,\t_1,\a_1,\cdots,y_p,\t_p,\a_p;z_1,\t_1,\a_1,\cdots, z_p,\t,\a)\\
&&\quad\quad=\frac{1}{Z}\int d\mu_C(\bar\psi,\psi)\Big[ \prod_{i=1}^p\prod_{\e_i=\pm}\psi^\e_{\tau_i, \a_i}(y_i)\ \Big]\ \Big[
\prod_{i=1}^p\prod_{\e_i=\pm}\psi^\e_{\tau_i,\a_i}(z_i)\Big]e^{-\lambda\int d^3xV(\bar\psi,\psi)},\nn
\eea
where
\be
Z=\int d\mu_C(\bar\psi,\psi)e^{-V(\psi,\bar\psi)}
\ee
is the partition function.


Expanding the exponential as power series and integrating out the Grassmann varaibles w.r.t. the Grassmann Gaussian measure, we have
\bea\label{part1}
&&S_{2p}(\lambda,y_{1,\e_1,\t_1, \a_1}\cdots y_{p,\e_p,\t_p, \a_p};z_{1,\e_1,\t_1, \a_1}\cdots z_{p,\e_p,\t_p, \a_p})\\
&=&\sum_{N}\frac{\l^n}{n!}\frac{(\delta\mu^1_{j_{max}})^{n'}}{n'!}\int_{{(\L_{\beta,L})}^{n}}d^3x_1\cdots d^3x_{n}\nn\\
&&\quad\quad\cdot\sum_{\underline{\t},\underline{\a}}\Bigg\{\begin{matrix}y_{1,\e_1,\t_1, \a_1}\cdots y_{p,\e_p,\t_p, \a_p}x_{1,\e_1,\t_1,\a_1}\cdots x_{ n,\e_{n},\t_{ n},\a_n}\\
z_{1,\e_1,\t_1, \a_1}\cdots z_{p,\e_p,\t_p, \a_p}x_{1,\e_1,\t_1,\a_1}\cdots x_{{n},\e_{n},\t_{n},\a_n}
\end{matrix}\Bigg\},\nn
\eea
where $\underline\t=\{\t_1,\cdots,\t_{p+n}\}$, $\t_i\in\{\uparrow,\downarrow\}$, are spin indices, $\underline{\a}=\{\a_1,\cdots,\a_{p+n}\}$, $\a_i=1,2$, are the indices for the matrix elements of the propagators and $n$ is the number of four-point vertices, each of which is associated with the coupling constant $\lambda$. These four-point vertices are also called {\it the bare vertices}. $n'$ is the number of two-point vertices, each one is associated with a bare chemical potential counter-term $\delta\mu^{1}_{j_{max}}$. These two-point vertices are also called {\it the counter-term vertices}. $N=n+n'$ is the total number of vertices and we have used Cayley's notation ( c.f. \cite{rivbook} ) for the determinants:
\bea
\Bigg\{\begin{matrix}
x_{v,\t,\a}\\ x'_{v',\t',\a'}
\end{matrix}\Bigg\}=\det\Big[\ \delta_{\t\t'}[C_{j,\t}(x_v-x'_{v'})]_{\a,\a'}\ \Big].
\eea


If we fully expand the determinant, the Schwinger functions are given by a sum of $(2N+2p)!$ terms labeled by Feynman graphs. It is well known in constructive renormalization theory that Feynman graphs proliferate very fast (actually $(2N+2p)!\sim C^N (N!)^2$ for $N\rightarrow \infty$, $C$ is a positive constant), so that the perturbation series is not convergent. One needs to reorganize the perturbation series into packages, each of which is labeled by a tree. According to Cayley's theorem, tree graphs doesn't proliferate very fast, so that the reorganized perturbation series is convergent. This convergent perturbation expansion are realized by using the forest formulas, which test links between the interacting vertices and stop as soon as the final connected components are built. The result of the forest formula is a sum over terms labeled by forests graphs and taking the logarithms select a sum over trees. 

In order to make the expansions consistent with the multi-slice analysis, the tree expansions need to be ordered with respect to increasing values of the scaling index $r=j+l/2$, so that
the tree lines with smallest $r$ value (highest momentum) are expanded first. The multi-slice expansions is performed using the BKAR jungle formula (see \cite{BK, AR}):
\begin{theorem}[The BKAR jungle Formula.]\label{ar1}
Let $n\ge1$ be an integer, $I_N=\{1,\cdots, n\}$ be an index set and $\cP_n=\{\ell=(i,j), i, j\in I_n, i\neq j\}\subset I_n\times I_n$ be the set of unordered pairs in $I_n$, in which an element $\ell=(i,j)$ is also called a link. Let $\cF$ be the set of spanning forests over $I_n$ and $\cS$ be the set of smooth functions from $\RRR^{\cP_n}$ to an arbitrary Banach space. Let ${\bf x}=(x_\ell)_{\ell\in\cP_n}$ be an arbitrary element of $\RRR^{\cP_n}$ and $f\in \cS$, we have
\be\label{BKAR}
f({\bf 1})=\sum_{\cJ=(\cF_1\subset\cF_1\cdots\subset\cF_{m})\atop m-jungle}\Big(\int_0^1\prod_{\ell\in\cF_m} dw_\ell\Big)\Bigg(\prod_{k=1}^m\Big(\prod_{\ell\in\cF_k\setminus\cF_{k-1}}\frac{\partial}{\partial {x_\ell}}\ \Big)\Bigg)\ f[X^\cF(w_\ell)],
\ee
where the sum over $\cF$ runs over all forests with $n$ vertices, including the empty forest with no edge, ${\bf 1}\in \RRR^{\cP_n}$ is the vector with every entry equals $1$, $X^\cF(w_\ell)$ is the vector ${(x_\ell)}_{\ell\in \cP_n}$ defined by $x_\ell= x_{ij}^\cF(w_\ell)$, the value at which we evaluate the derivations on $f$ and is defined as follows
\begin{itemize}
\item $x_{ij}^\cF=1$ if $i=j$, 
\item $x_{ij}^\cF=0$ if $i$ and $j$ are not connected by $\cF_k$,
\item $x_{ij}^\cF=\inf_{\ell\in P^{\cF}_{ij}}w_\ell$, if $i$ and $j$ are connected by the forest $\cF_k$ but not $\cF_{k-1}$, where $P^{\cF_k}_{ij}$ is the unique path in the forest that connects $i$ and $j$,
\item $x_{ij}^\cF=1$ if $i$ and $j$ are connected by $\cF_{k-1}$.
\end{itemize}
The elements $(x_{ij}^\cF)$, $i,j=1,\cdots,n$, form an $n\times n$ matrix that is positive. 
\end{theorem}

Remark that in this paper we consider only the amputated $2p$-point Schwinger's function, in which the external propagators (the propagators connecting the external fields $\psi^\e_y$ or $\psi^\e_z$ with $\psi^\e_{x_i}$) are taken away. Using the BKAR jungle formula, we obtain 
\bea\label{rexp1}
&&S_{2p}=\sum_{N=n+n'}\lambda^n (\delta\mu^1_{j_{max}})^{n'} S_{2p,N},\\
&&S_{2p,N}=\frac{1}{n!n'!}\sum_{\{\underline{\tau}\}}\sum_{\cJ}\prod_v\int_{\Lambda_{\beta, L}} d^3x_v\prod_{\ell\in\cF}\int dw_\ell
C_{r,\sigma_\ell}(x_\ell,x'_\ell){\det}_{left}[ C_{\t}(w)]\ ,\nn
\eea
where $\cJ=({\cF}_0\subset{\cF}_1\cdots\subset{\cF}_{r_{max}}={\cF})$ is a layered object of forests $\{\cF_0,\cdots,\cF_{r_{max}}\}$, also called a jungle, in which the last forest $\cF=\cF_{r_{max}}$ is a spanning forest of the fully expanded graph $G$ with $n$ vertices and $2p$ external edges. $\cF_{0}:={\bf V_n}$ is the completely disconnected forest of $n$ connected components, each of which corresponds to the bare vertex $\VV(\psi)$ ( cf. \eqref{potx} ).
Taking the logarithm of the \eqref{rexp1}, we obtain the {\it connected} $2p$-point Schwinger function, noted by $S^c_{2p}$, and we have
\bea\label{rexp2}
&&S^c_{2p}=\sum_{N=n+n'}\lambda^n (\delta\mu^1_{j_{max}})^{n'} S^c_{2p,N},\\
&&S^c_{2p,N}=\frac{1}{n!n'!}\sum_{\{\underline{\tau}\}}\sum_{\cJ'}\prod_v\int dx_v\prod_{\ell\in\cT}\int dw_\ell
C_{r,\sigma_\ell}(x_\ell,y_\ell){\det}_{left}[ C_{r,\sigma}(w)]\ ,\nn
\eea
where the sum is now constrained over the jungles $\cJ'=({\cF}_0\subset{\cF}_1\cdots\subset{\cF}_{r_{max}}={\cT})$ in which final layered forest ${\cF}_{r_{max}}$ is the spanning tree $\cT$ connecting all the $n$ vertices, $\e(\cJ)$ is the product of the $\pm$ signs along the jungle and is again an inessential sign. 
The notation $\det_{left}(C_{r,\sigma}(w))$ means the determinant made of the fields and anti-fields after extracting the tree propagators. 
Since the number of fields not in the tree is equal to $4n+2n'-2(n+n'-1)=2(n+1)$, the determinant is one for a $2(n+1)\times 2(n+1)$ square matrix of the Cayley type similar to \eqref{rexp1}, but with additional multiplicative parameters depending the interpolation parameters $\{w\}$. Let $r_f$ be the index of a field or anti-field $f$, the $(f,g)$ entry of the determinant in \eqref{rexp1} is
\bea\label{intc1}
C_{r}(w)_{f,g}=\delta_{\t(f)\t'(g)}\sum_{v=1}^n\sum_{v'=1}^n\chi(f,v)\chi(g,v') x^{\cF,r_f}_{v,v'}(\{w\})C_{r,\t(f),\s(f)}(x_v,x_{v'}),
\eea
where $[x^{\cF,r_f}_{v,v'}(\{w\})]$ is an $N\times N$ positive matrix whose elements are the weakening parameters, defined in the same way as in \eqref{BKAR}:
\begin{itemize}
\item If the vertices $v$ and $v'$ are not connected by $\cF_r$, then $x^{\cF,r_f}_{v,v'}(\{w\})=0$,
\item If the vertices $v$ and $v'$ are connected by $\cF_{r-1}$, then $x^{\cF,r_f}_{v,v'}(\{w\})=1$,
\item If the vertices $v$ and $v'$ are connected by $\cF_r$ but not $\cF_{r-1}$, then  $x^{\cF,r_f}_{v,v'}(\{w\})$ is equal to the infimum of the $w_\ell$ parameters for $\ell\in\cF_r/\cF_{r-1}$ which is in the unique path connecting the two vertices. The natural convention is that $\cF_{-1}=\emptyset$ and that $x^{\cF,r_f}_{v,v'}(\{w\})=1$.
\end{itemize}

At any level $r$ the forests $\cF_r$ defines a set of $c(r)$ connected components. The lines of the forest $\cF_r$ subject to each connected component forms a tree, noted by $\cT_r^k$.
To each component one associates an object $G^k_r$, $k=1,2,\cdots, c(r)$, which has a well defined set of vertices (called the internal vertices of $G^k_r$), the internal tree lines, which are the set of the lines $\ell(T)\in\cT_r^k$ connecting these vertices and a well defined set of external fields (half-lines) $e(G^k_r)$, whose cardinality $|e(G^k_r)|$ is an even number. By construction, the scaling indices of the external fields $r_f$ greater than the internal lines $r_{\ell(T)}$, which ensures the connectivity of each component.

The graphical structure of a component $G^k_r$ is highly nontrivial: besides a tree structure $\cT_r^k$, it contains also a set of internal fields (which would form loop lines if they are expanded) which still form a determinant. So $G^k_r$ is not a subgraph of the forests. 
Each connected component $G^k_r$ is contained in a unique connected component with lower $r$-index. This inclusion relation has a tree structure, called the Gallavotti-Nicol\`o tree.

\begin{definition} 
A Gallavotti-Nicol\`o tree (GN tree for short \cite{GN}), $\cG$ is an abstract tree graph whose nodes \footnote{The nodes of a Gallavotti-Nicol\`o tree are also called {\it vertices} by some authors. In this paper we use the word {\it vertices} for the vertices in the Feynman graphs.} are the connected components $G^k_r$ introduced above, with $r\in[0,r_{max}]$, $k=1,2,\cdots, c(r)$, such that for each tree line $\ell(T)\in G^k_r$ one has $r_{\ell(T)}\le r$, and whose tree lines are the inclusion relations of the nodes with different $r$-indices. The node $G_{r_{max}}$ of $\cG$, which is unique and corresponds to the fully connected graph $G$, is called the root of $\cG$ and each element of the set ${\bf V_n}=\cF_0$ is called a leave of $\cG$. The cardinality of ${\bf V_N}$, which is $N$, is also called the order of $\cG$. A GN tree of order $N$ is also noted as $\cG_N$.
\end{definition}

\begin{figure}[htp]
\centering
\includegraphics[width=1.0\textwidth]{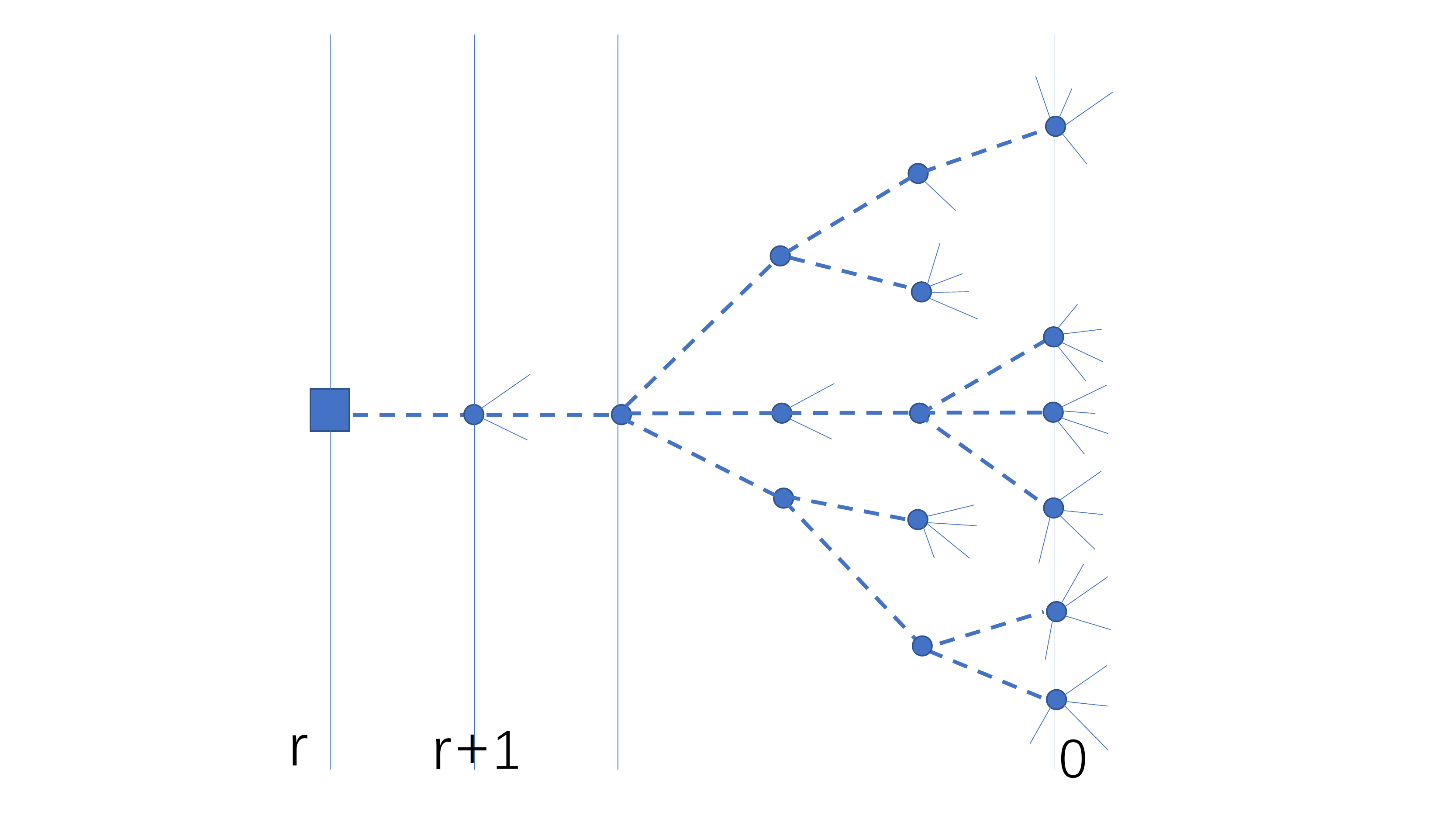}
\caption{\label{gn1}A Gallavotti-Nicol\`o tree with $16$ nodes and $8$ leaves, which are the bare vertices. The round dots
represent the nodes and bare vertices, and the big square represents the root. The dash lines are the inclusion relations between these nodes and the thin lines are the external fields of the nodes. The scaling indices for the nodes are in an increasing order from the root towards the leaves.
}
\end{figure}


\subsection{Power Counting}
Now we consider the power counting theorem for the connected $2p$-point Schwinger functions. Remark that since the chemical potential counter-terms are vertices with only two external fields, they only
contribute the factors $(\delta\mu^1_{j_{max}})^{n'}$ to the power-counting. As will be proved in Theorem \ref{flowmu}, each of the counter-term $\delta\mu^1_{j_{max}}$ is simply bounded by a small number so that the factors $(\delta\mu^1_{j_{max}})^{n'}$ are not divergent. So we will forget the chemical potential counter-terms in this section, just for simplicity. Remark that these counter-terms will be important for the renormalization of the two-point Schwinger functions and the self-energies.

So we rewrite the general $2p$-point Schwinger function as:
\bea
S_{2p} &=&\sum_{n}S_{2p,n}\lambda^n,\\
S_{2p,n}&=&{1 \over n!}\sum_{ \{\underline{\t}\},\cG, \cT} \ 
\sum'_{\cJ}
\epsilon (\cJ) \prod_{j=1}^{n}   \int d^3x_{j}  \delta(x_1)
\prod_{\ell\in \cT} \int_{0}^{1} dw_{\ell}
C_{\t_{\ell},\si_{\ell}} 
(x_{\ell}, \bar x_{\ell}) \nonumber\\
&&\quad \prod_{i=1}^{n} 
\chi_{i}(\si)    
\det\nolimits_{{\rm left}} (C_{j}(w)) \ .\label{form} 
\eea

Note that each matrix element \eqref{intc1} in $\det[C_{r,\sigma}(w)]_{left}$ can be written as the inner product of two vectors in a certain Hilbert space:
\be
C_{r}(w)_{f,g}=( e_\tau\otimes A_f(x_v, ),  e_{\tau'}\otimes B_g(x_{v'},))\ ,
\ee
in which $e_{\uparrow}=(1,0)$, $e_{\downarrow}=(0,1)$ are the unit vectors indicating the spins,
\bea
A_f&=&\frac{1}{\beta|\Lambda_L|}\sum_{k\in\DD_{\beta, L}}\sum_{v=1}^n\chi(f,v)[x^{\cF,r_f}_{v,v'}(\{w\})]^{1/2}e^{-ik\cdot x_v}\cdot\\
&&\quad\quad\quad\cdot\Big[\ \chi_j[k_0^2+e^2(\kk,1)]\cdot
v_{s_+}[\cos^2(k_+/2)]\cdot v_{s_-}[\cos^2(k_-/2)]\ \Big]^{1/2},\nn
\eea
and
\bea
B_g&=&\frac{1}{\beta|\Lambda_L|}\sum_{k\in\DD_{\beta, L}}\sum_{v'=1}^n\chi(g,v')[x^{\cF,r_f}_{v,v'}(\{w\})]^{1/2}e^{-ik\cdot x_{v'}}C(k)\\
&&\quad\quad\quad\cdot\Big[\ \chi_j[k_0^2+e^2(\kk,1)]\cdot
v_{s_+}[\cos^2(k_+/2)]\cdot v_{s_-}[\cos^2(k_-/2)]\ \Big]^{1/2}\nn,
\eea
in which $C(k)$ is the free propagator. (cf. Formulae \eqref{2ptk}.)

By the Gram-Hadamard inequality \cite{DR1, GeM} we have
\be
|\det(A_f, B_g)|\le\prod_{f}||A_f||\cdot ||B_f||\ ,
\ee
in which
\bea
||A_f||\le O(1)\g^{-j_f-s_{f,+}-s_{f,-}},\quad ||B_f||\le O(1)\g^{-s_{f,+}-s_{f,-}}.
\eea
Using $l_f=s_{f,+}+s_{f,-}-j+2$, we have:
\be
||A_f||\cdot||B_f||\le O(1)'\g^{-j_f-l_f/2}.
\ee



Remark that if we use directly the Gram's inequality we would bound all the oscillation terms by one, hence we lose the constraints from the conservation of momentum, which is important for obtaining the correct power-counting theorem. To take into account the conservation of momentum, we introduce 
for each vertex the function 
$\chi_{i}(\si)= \chi (\sigma^1_i, \sigma^2_i, \sigma^3_i, \sigma^4_i)$ which is
1 if the condition of Lemma 4 is satisfied and 0 otherwise.
These insertions are free since the contributions
for which these $\chi $ functions are not 1 are zero.
They must be done before taking the Gram bound, which destroys the Fourier 
oscillations responsible for momentum conservation at each vertex. Now we use Gram's inequality for the determinant $\det\nolimits_{{\rm left}}$ and we have:
\be
\det\nolimits_{{\rm left}}\le c^n\prod_{f\ left}\g^{-(j_f+l_f)/2}= c^n\prod_{f\ left}\g^{-r_f/2-l_f/4}.
\ee 

Integrating over the positions of the vertices save the fixed one
$x_1$ using the Gevrey scaled decay (\ref{decay1}),
we obtain a bound on the $n$-th order of perturbation theory:

\be | S_{2p,n} | \le {c^n \over n!}
\sum_{\{\underline\tau\}, \cG, \cT}\ \sum'_{\{\si \}} 
\prod_{i=1}^{n} \chi_{j}(\si) 
\prod_{\ell \in \cT} \g^{2r_{\ell}}
\prod_{f} \g^{-r_f/2-l_f/4}, \label{absol1}
\ee
where the product over $f$ now runs over all the $4n$ fields
and anti-fields of the theory.

\begin{lemma}\label{indmain}
Let $c(r)$ be the number of connected components at level $r$ in the GN tree, we have the following inductive formulas:
\bea &&\prod_{f} \g^{-r_{f}/2}= \prod_{r=0}^{r_{max}}\ \prod_{k=1}^{c(r)} \g^{-|e(G_r^k)|/2}\ ,
\label{induc1}\\
&&\prod_{\ell \in \cT} \g^{2r_{\ell}}=\g^{-2r_{max}-2}
\prod_{r=0}^{r_{max}}\ \prod_{k=1}^{c(r)}  M^{2}\ .
\label{induc2}
\eea
\end{lemma}
\begin{proof}
We consider \eqref{induc1} first. Let $N_f$ be the set of all fields that form the graph $G$ and $|N_f|$ be the cardinality of $N_f$, obviously we have $|N_f|=4n$ for $|G|=n$. Let $n(r,f)$ be the number of fields $f$ such that $r_f=r$, with $r\in[0, r_{max}]$, then the l.h.s. of \eqref{induc1} is equal to
$
\g^{-\frac12\sum_{f\in N_f} r_{f}}=\g^{-\frac12[\ \sum_{r=0}^{r_{max}}\ n(r,f)\cdot r\ ] }$.
Let $\cup_{k=1}^{c(r)}e(G_r^k)$ be the set of external fields of any connected component of the GN tree at level $r$ with cardinality $|e(G_r)|=\sum_{k=1}^{c(r)}\  |e(G_r^k)|$,
the r.h.s. of \eqref{induc1} is equal to $\g^{-\frac12[\ \sum_{r=0}^{r_{max}}\  |e(G_r)|\ ]}$.
Let $f_e$ be an external field of a connected component in $\cG$ and $r_{f_e}=r\ge1$. So $f_e$ can be an external field of the components $G_{\le r-1}$ but becomes an internal field of $G_{r}$.
So we have $f_e\in e(G_0)\cap\cdots\cap e(G_{r-1})$, which means that $f_e$ is counted exactly $r$ times in the sum $\sum_{r=0}^{r_{max}} |e(G_r)|$. Recall that when $r_{f_e}=0$ then by definition it is an external field to $G_{0}$, which by definition is the set of $n$ bare vertices and $e(G_{0})=N_f$. So we have
$ \sum_{r=0}^{r_{max}} |e(G_r)|= \sum_{r=0}^{r_{max}}\ r\cdot n(f_e,r)$, where $n(f_e,r)$ is the number of external fields at level $r$. Since the set of external field $e(G_{0})$ is identical to the set of fields $N_f$, we have
\be
\prod_{r=0}^{r_{max}}\ \prod_{k=1}^{c(r)} \g^{-|e(G_r^k)|/2}=\g^{-\sum_{r=0}^{r_{max}}\ r\cdot n(f,r)/2}=\prod_{f} \g^{-r_{f}/2}\ .
\ee
So we proved \eqref{induc1}.

Formula \eqref{induc2} can be proved in a similar way. Without losing generality, suppose that the tree $\cT$ has $k$ lines $\ell_1, \ell_2,\cdots, \ell_k$. Let $r_{\ell_i}$ be the scaling index for
the tree line $\ell_i$, $1\le i\le k$. We can always reorganize the set of tree lines such that $r_{\ell_1}\le r_{\ell_2}\cdots\le r_{\ell_k}$, with $r_{\ell_k}\le r_{max}$. Then at the slice $r_{\ell_k}$ the tree $\cT$ is completed so we have $c(r_{\ell_k})=1$. Let and $n(\ell,r)$ be the number of lines in $\cT$ whose scaling indices are equal to $r$, the l.h.s. of \eqref{induc2} is equal to
$\g^{2\sum_{\ell\in\cT}r_\ell}=\g^{2\sum_{r=0}^{r_{max}}\ r\cdot n(l,r)}$, while the r.h.s. of \eqref{induc2} can be written as 
\bea\label{induc2r}
&&\g^{-2r_{max}-2}\ 
\prod_{r=0}^{r_{max}}\ \g^{2c(r)}=\g^{-2r_{max}-2}
\cdot \g^{2 \big[c(0)+c(1)+\cdots c(r_{max})\ \big]}\\
&&\quad\quad=
\g^{2 \sum_{r=0}^{r_{max}}\ \big[c(r)-1\big]}\ .\nn
\eea

Since, by definition, $c(i)-c(i+1)$ is equal to the number of tree lines at scale $r=i+1$ and since $c(r_{max})=1$, we have
\be
c(r)-1=n(\ell,r+1)+n(\ell,r+2)+\cdots n(\ell,r_{max}),
\ee
where $n(\ell,i)=c(i)-c(i+1)$.
Now we sum over all scaling index $r$ for the l.h.s. of the above formula. We can easily the term $n(\ell,r_{\ell_k})$ is summed for exactly $r$ times, so we have 
\be
\sum_{r=0}^{r_{max}}\ \big[c(r)-1\big]=1\cdot n(\ell,1)+2\cdot n(\ell,2)+\cdots +r_{max}\cdot  n(\ell,r_{max}),
\ee
which means
\be
\g^{2 \sum_{r=0}^{r_{max}}\ \big[c(r)-1\big]}=\g^{2 \sum_{r=0}^{r_{max}}\ r\cdot n(\ell,r)}=\prod_{\ell\in\cT}\ \g^{2r_\ell}\ .
\ee

So we have proved Formula \eqref{induc2} hence have completed the proof this lemma.
\end{proof}

\begin{theorem}[The power counting formula]
The connected Schwinger's function is bounded as follows: 
\be\label{pc1} | S_{2p,n} | \le {c^n \over n!}
\sum_{\{\underline\tau\}, \cG, \cT}\ \sum'_{\{\si \}}\prod_{i=1}^{n}\ \big[\chi_{i}(\si)\g^{-(l_i^1 + l_i^2 + l_i^3 + l_i^4)/4}\big]\ 
\prod_{r=0}^{r_{max}}\prod_{k}  \g^{2-e(G_r^k)/2}\ ,
\ee
which means that the two-point functions are linearly divergent, the four point functions are marginal while Schwinger functions with external legs $2p\ge6$ are irrelevant.
\end{theorem}
\begin{remark}
Remark that once the final spanning tree $\cT$ and the Gallavotti tree $\cG$ are determined, the jungle structure $\cJ'$ is completely determined. So we write the sum over $\cJ'$ in \eqref{pc1} simply as sum over $\cT$. This theorem is only the "raw" power counting theorem, as we still haven't summed over the the sectors indices in \eqref{pc1}, which would result in the logarithm corrections to the coupling constants. 
\end{remark}
\begin{proof}
Writing the product $\prod_f\g^{-l_f/4}$ in Formula \eqref{absol1} as $\prod_{i=1}^{n}e^{-(l_i^1 + l_i^2 + l_i^3 + l_i^4)/4}$ and taking into account the conservation of momentum at each vertex $i$, the result follows.
\end{proof}

Now we consider summation over the sector indices associated to the fields hooked to the vertices $i=1,\cdots,n$, taking into account the constraints from the conservation of momentum. We have the following sector counting lemma:
\begin{lemma}[Sector counting lemma for bare vertices]\label{sec1}
Let $f_1,\cdots, f_4$ be the four half fields associated with the vertex $i\in G_r^k$ with scaling indices $j_1,\cdots, j_4$, sector indices $\si_1=(s_{1,+}, s_{1,-}),\cdots,\si_4=(s_{4,+}, s_{4,-})$ and the generalized scaling indices $r_1\cdots, r_4$ such that
$r_{f_1}=r_{f_2}=r_{f_3}=r$ while $r_{f_4}>r$. There exists a positive constant $c$ such that, for fixed $\si_4$, we have
\be\label{ss1}
\sum_{\si_1, \si_2, \si_3} \chi (\si_1, \si_2, \si_3, \si_4) 
M^{-(l_1+l_2+l_3 )/4} \le c.r\ .\ee
\end{lemma}
\begin{proof}
This lemma has been proved in \cite{Riv} for a similar setting. We present a proof here for reader's convenience and for completeness. First of all remark that among the four half fields $f_1,\cdots, f_4$ hooked to vertex $i$ in a node $G_r^k$ with sector indices $\s_1,\cdots,\s_4$ and depth indices $l_i^1,\cdots, l_i^4$, we can always choose one external leg, say, $f_4$, as the {\it root} half line, hence the generalized scaling index $r_4$ greater than all the other indices $r_1=r_2=r_3=r$.  We can always organize the sector indices $\s_1=(s_{1,+},s_{1,-}),\cdots, \s_3=(s_{3,_+}, s_{3,_-})$ such that $s_{1,+}\le s_{2,+}\le s_{3,+}$ and $s_{1,-}\le s_{2,-}\le s_{3,-}$. From the support properties of the propagators we know that, among the three sector indices, either $\s_1$ collapses with $\s_2$ or one has $s_{1,\pm}=j_1$, such that $j_1<\min\{j_2,\cdots,j_4\}$. 
So we can consider the following possibilities:
\begin{itemize}
\item if $\sigma_1\simeq\sigma_2$, then we have $s_{2,\pm}= s_{1,\pm}\pm1$. The depth indices are arranges as $l_1\le l_2\le l_3$. Then the l.h.s. of \eqref{ss1} is equal to 
\be
\sum_{\si_1, \si_3}
\g^{-(2l_1+1+l_3 )/4}\le O(1) \sum_{\si_1}\g^{-l_1/2}\sum_{\si_3}
M^{-l_3 /4}\ .
\ee
Using the fact that $r_k=i_k+l_k/2$ and $l_k=s_{k,+}+s_{k,-}-j_k+2$, we have
$l_k=2(s_{k,+}+s_{k,-}-r_k+2)$, for $k=1,\cdots,3$. 
For fixed $s_1=(s_{1,+}, s_{1,-})$, the sum over $\s_3=(s_{3,+}, s_{3,-})$ can be easily bounded:
\bea
&&\sum_{\si_3=(s_{3,+},s_{3,-})}\g^{-l_3 /4} = \sum_{\si_3=(s_{3,+},s_{3,-})} \g^{-(l_3-l_1) /4}
\g^{-l_1/4}\\
&&\le \sum_{s_{3,+}\ge s_{1,+}}\g^{-(s_{3_+}-s_{1,+})/2}\sum_{s_{3,-}\ge s_{1,-}}\g^{-(s_{3_-}-s_{1,-}) /2}\g^{-l_1/4} \le C\cdot  \g^{-l_1/4}\ .\nn
\eea
Now we consider the sum over $\s_1$. Since $s_{1,+}+s_{1,+}\ge r-2$ (cf. Formula \eqref{newr}), taking into account the factor $\g^{-l_1/4}$ from the above formula, we have:

\bea
\sum_{\si_1}\g^{-l_1/2}\cdot \g^{-l_1/4}&\le&\g^{3r/2} \sum_{s_{1,+}=0}^{r}\g^{-3s_{1_+}/2}\sum_{s_{1,-}=r-2-s_{1,+}}^{r}\g^{-3s_{1_-}/2}\\
&&\le \g^{3r/2}\sum_{s_{1,+}=0}^{r}\g^{-3s_{1_+}/2}\g^{-3r/r+3s_{1_+}/2}
\le C\cdot r\ .\nn
\eea
\item if the sector indices $s_{1,\pm}=j_1$, the smallest scaling index among the four, we have $l_1=j_1+2$.
The sum over $\s_1$ is simply bounded by $\g^{-j_1/4}$ and the sum over $\s_3\ge \s_2$ gives the constant. as in the previous case. The sum over $\s_2$ gives the factor $r$ so the results is bounded by $c\cdot\g^{-j_1/4}r\le c\cdot r$.
\end{itemize}
So we proved this lemma.
\end{proof}



Formula \eqref{pc1} suggests the following: a) the un-renormalized two-point functions are linearly divergent, b) the four point functions are marginal while c) Schwinger functions with external legs $2p\ge6$ are irrelevant. However we still need to sum over the sectors, which results in corrections to the power-counting formula. We call these corrections the {\it logarithmic power-counting}. We shall
study the statements b) and c) in this section and consider a) in the next section, for which we need renormalizations. The statements b) and c) have been proved in a very similar setting in \cite{Riv}, with the multi-slice analysis and the sector counting lemma introduced above. So we just recall the results but omit the proof. The interested is invited to consult \cite{Riv} for more details. Before proceeding it is useful to introduce some graphical notations.
\begin{definition}
Let $\cG$ be a Gallavotti-Nicol\`o tree. Let the set of external fields attached to each node $G_r^k$, $r=0,\cdots, r_{max}$, $k=1,\cdots, c(r)$, be $e(G_r^k)$ whose cardinality is noted by
$|e(G_r^k)|$. A biped $b$ is a node in the GN tree $\cG$ such that $|e(G_r^k)|=2$. Remark that the two external fields of a biped can be hooked to two different vertices or to the same vertex. 
A biped tree $\cG_\cB$ is a subtree of a Gallavotti-Nicol\`o tree in which the set nodes, noted by $V(\cG_\cB)$, composes of the following elements: i) the bare nodes $\VV$ of $\cG$, ii), the set of bipeds
$\cB:=\{G_r^k,\ r=1,\cdots, r_{max}; k=1,\cdots, c(r)\ \big|\ |e(G_r^k)|=2\}$ and iii) the root node of $\cG$ which corresponds to the complete graph $G$. The lines of $\cG_\cB$ are the natural inclusion relations of the nodes $V(\cG_\cB)$. An element $b\in\cB$ is called a biped. To each biped $b$ we also identify its external fields $e_b=\{\bar\psi_b,\psi_b\}$ and define ${\cal EB}:=\big(\cup_{b\in B}\ e_b\big)\setminus e(G)$ as the set of external fields of $\cB$.
\end{definition}

Similarly, we can define the quadruped GN trees, as follows. 
\begin{definition}
A quadruped $q$ is a node of a Gallavotti-Nicol\`o tree $\cG$ which has four external fields. The set of all quadrupeds in a $\cG$ is noted by $\cQ$. A quadruped tree $\cG_\cQ$ is a subtree of $\cG$ whose set of nodes, noted by $V(\cG_\cQ)$, composes of the following elements: the bare nodes of $\cG$, the quadruped $\cQ$ and the root of $\cG$， which is the complete graph $G$. Remark that both the bare vertices and the root can be considered as quadrupeds. The tree lines of $\cG_\cQ$ are the inclusion relations of its nodes. Define also the set of external fields associated to a certain $q$ is by $e_q$ and the set of external fields of $\cQ$ by ${\cal EQ}=(\cup_{q\in\cQ}e_q)\setminus e(G)$. 
\end{definition}
The convergent GN tree is defined as follows.
\begin{definition}
A convergent Gallavotti-Nicol\`o tree $\cG_{\cC}$ is a subtree of a Gallavotti-Nicol\`o tree which doesn't contain the nodes $\cB$ nor $\cQ$. The set of nodes of $\cG_{\cC}$, noted by $V(\cC)$, is the union of the set of leaves ${\bf V}_n$ of $\cG$ with the set convergent nodes $\cC:=\big\{ G_{r}^{k},\ r=1,\cdots, r_{max},\ k=1\cdots c(r)\big\vert\ |e( G_{r}^{k})|\ge 6\big\}$.
\end{definition}
We have the following proposition concerning the relations for the nodes in the various sub-trees of the GN tree defined above:
\begin{proposition}
Let ${ V}(\cG)$, ${ V(\cG_{\cC})}$, ${ V}(\cG_\cQ)$ and ${ V}(\cG_\cB)$ be the set of nodes of the GN trees $\cG$, $\cG_{\cC}$, $\cG_\cQ$ and $\cG_{\cB}$, respectively, we have
\bea
&&{ V(\cG_{\cC})}={\bf V}_n\cup\cC={ V}(\cG)\setminus(\cB\cup\cQ),\ {V(\cG_{Q})}=\bV\cup\cQ={ V}(\cG)\setminus(\cB\cup\cC),\nn\\
&&{ V(\cG_{\cB})}={\bf V}_n\cup\cB={ V}(\cG)\setminus(\cC\cup\cQ),\nn\\
&& {\rm and}\quad { V(\cG)}={\bf V}_n\cup\cB\cup\cQ\cup\cC.
\eea
\end{proposition}
We have the following definitions for the corresponding Schwinger's functions:
\begin{definition}
We can define the following Schwinger functions $S^{\cC}_{2p}$, $S^{\cQ}_{2p}$, $S^{\cC}_{2p}$ corresponding the the GN trees $\cG_{\cC}$, $\cG_\cQ$ and $\cG_{\cB}$ of all orders $n=0,1,\cdots$. 
$S^{\cC}_{2p}$ is called the convergent Schwinger functions, which are the contributions to the Schwinger functions $S_{2p}$ from the convergent graphs, namely all the contributions quadrupeds $\cQ$ and bipeds $\cB$ are taken away. So we have $p\ge3$ for $S^{\cC}_{2p}$. $S^{\cQ}_{2p}$ is called the quadruped Schwinger function, in which all the contributions from the bipeds are taken away, and we have $p\ge2$. Finally $S^{\cB}_{2}$ is the biped Schwinger function, in which the root graph of the GN tree is a biped graph.  
\end{definition}
\begin{remark}
Remark that the quadruped Schwinger functions $S^\cQ$ and the biped Schwinger functions $S^\cB$ are also called the {\bf localized} Schwinger functions.
The localization procedure ( cf. \cite{BGM2} , \cite{DR1}) for our setting corresponds to the selection of $S^\cQ$ and $S^\cB$ from the general Schwinger functions.
\end{remark}
Before considering the localized Schwinger functions, we shall first of all study the analytic properties of the convergent Schwinger function $S^\cC_{2p}$.

\subsection{Convergent contributions}
The convergent Schwinger function $S^\cC_{2p}$ is defined as:
\be
S_{\cC,2p}=\sum_n\l^n S_{\cC,2p,n},\ee
\be\label{conv1}
S_{\cC,2p,n} = {1\over n!}\sum_{{\cal B} = \emptyset,
{\cal Q}=\emptyset \atop  \{\underline\tau\}, \cJ }
\sum_{\{\si \}}' \ep (\cJ)\prod_{v} \int_{\Lambda_{\beta, L}} dx_{v} 
\prod_{\ell\in \cT} \int_{0}^{1} dw_{\ell}
C_{i,\si_{\ell}} (x_{\ell}, y_{\ell})
\det\nolimits_{left} (C(w)).
\ee
We have the following theorem:
\begin{theorem}[The Convergent contributions (see also \cite{Riv}]\label{cth1}
The connected Schwinger functions $S_{\cC, 2p}$, $p\ge3$ whose perturbation series are labeled by
$GN^c$ are analytic in $\lambda$ for $\lambda\log T\le C $hence their radius of convergence $R_T$ at temperature $T$ satisfies 
\be
R_T\le C/|\log T|.
\ee
\end{theorem}
\begin{proof}
The main idea for the proof is from \cite{Riv}, which is for a different model. We try to make our proof more simpler and more pedagogical. The interested readers are also invited to compare the sector sum here with that in \cite{BGM2} or \cite{FKT} for the case of strictly convex Fermi surfaces. 

First of all we rewrite Formula \eqref{conv1} to make the sum over the GN trees more explicit, we have:
\bea\label{conv2}
S_{\cC,2p,n} &=& {1\over n!}\sum_{\{G^k_r, {r=0},\cdots,r_{max}; {k=1},\cdots,c(r)\},\atop{ {\cal B} = \emptyset,
{\cal Q}=\emptyset}}\sum_{\underline\tau}
\sum_{\{\si \}}' \ep (\cJ)\nn\\
&&\quad\quad \prod_{v} \int_{\Lambda_{\beta, L}} dx_{v} 
\prod_{\ell\in \cT} \int_{0}^{1} dw_{\ell}
C_{i,\si_{\ell}} (x_{\ell}, y_{\ell})
\det\nolimits_{left} (C(w)),
\eea

By Lemma \ref{indmain} we have (see also Formula \eqref{pc1})
\bea\label{conv3}
|S_{\cC,2p,n}| &\le& 
{c^n \over n!}
\sum_{\{G^k_r, {r=0},\cdots,r_{max}; {k=1},\cdots,c(r)\},\atop{ {\cal B} = \emptyset,
{\cal Q}=\emptyset}}\sum_{\underline\tau,\cT} \sum'_{\{\si \}}\prod_{i=1}^{n}\ \Big[\ \chi_{i}(\si)e^{-(l_i^1 + l_i^2 + l_i^3 + l_i^4)/4}\nn\\
&&\quad\cdot\prod_{r=0}^{r_{max}}\g^{2-|e(G^k_r)|/2}\ \Big]\label{cpt1}\\
&&\le{c^n \over n!}
\sum_{\{G^k_r, {r=0},\cdots,r_{max}; {k=1},\cdots,c(r)\},\atop{ {\cal B} = \emptyset,
{\cal Q}=\emptyset}}\sum_{\underline\tau,\cT} \sum'_{\{\si \}}\prod_{i=1}^{n}\ \Big[\ \chi_{i}(\si)e^{-(l_i^1 + l_i^2 + l_i^3 + l_i^4)/4}\ \Big]\nn\\ 
&&\quad\cdot\prod_{i=1}^n \g^{-[r_i^1+r_i^2+r_i^3+r_i^4]/6}
\nn,
\eea
where we have used the fact that $2-|e(G^k_r)|/2\le-|e(G^k_r)|/6$ for $|e(G^k_r)|\ge 6$ and the fact that $$\prod_{r=0}^{r_{max}}\g^{-|e(G^k_r)|/6}=\prod_{i=1}^n \g^{-[r_i^1+r_i^2+r_i^3+r_i^4]/6}.$$

Now we consider the term 
\be\label{secsumcov}
\sum'_{\{\si \}}\prod_{i=1}^{n}\ \Big[\ \chi_{i}(\si)e^{-(l_i^1 + l_i^2 + l_i^3 + l_i^4)/4}\ \Big]\prod_{i=1}^n \g^{-[r_i^1+r_i^2+r_i^3+r_i^4]/6}\ee
and sum over the sector indices.

At each slice $r$ we shall sum over the sector indices $s_{\pm}$ for each vertex $i$, using the constraints from conservation of momentum. 
First of all we fix the half field with maximal $r$ index (chosen as $r_4$ for the field $f_4$), which eventually goes to the root. Conservation of momentum implies that either the two smallest sector indices among the four, chosen as $s_{1,\pm}$ and $s_{2,\pm}$, are equal (modulo $\pm1$) or the smallest sector index $s_{1,\pm}=j_1$, the smallest scaling index.
Remark that, since the scaling indices for the four fields are not necessarily the same, we can't directly apply Lemma \ref{sec1}. So besides summing over the sector
indices $s_{1,\pm}$, for which we pay the factor $\max \{r_i^1, r_i^2\}$, we also need to sum over the sectors $s_{3,\pm}$, for which we have (see the proof of Lemma \ref{sec1})
\be
\sum_{s_{3,+},s_{3,-}}\g^{-l_i^3/4}\le c_1.r_i^3,
\ee
for a certain positive constant $c_1$. So at each vertex we lose a factor $c \bar r\cdot r_i^3$, where $\bar r=\max\{r_i^1, r_i^2\}$. So at each vertex $i=1,\cdots, n$, the sum over sectors other than the root sector in \eqref{secsumcov} is bounded by
\be
\prod_{i=1}^n\sum_{r_i^1,\cdots, r_i^{4}=0}^{r_{max}}[\bar r \g^{-r_i^1/6}\g^{-r_i^2/6}\cdot r_i^3\g^{-r_i^3/6}\g^{-r_i^4/6}]\le c_3.
\ee 

Finally, when all the scaling indices are summed up and we arrive at the final slice
$r_{max}$, we still have to sum over the sector index for the root half lines, one for each vertex $i$, which is easily bounded by $r_{max}=3j_{max}/2=3|\log T|/2$. Summing over all the GN trees and
spanning trees cost a factor $c^n n!$, where $c$ is certain positive constant (see \cite{Riv} for the detailed proof of this combinatorial result.). So the convergent Schwinger function is bounded by
\be
|S_{\cC,2p,n}|\le \sum_{n=0}^\infty O(1)^n|\lambda|^n|\log T|^n,
\ee
which means that the series is convergent for $O(1)|\lambda| |\log T|<1$, $|\lambda|<c/ |\log T|$.
So we proved this theorem.
\end{proof}

\subsection{The quadruped Schwinger's functions}
In this section we study the analytic properties of the quadruped Schwinger functions $S_\cQ$, for which the corresponding Gallavotti-Nicol\`o trees are quadruped trees $\cG_\cQ$. 

It is convenient to introduce the conservation of momentum not only to all the bare vertices but also to the quadrupeds $q\in\cQ$. We have
\bea\label{conv2}
S_{\cQ}&=&\sum_{n=0}^\infty \lambda^n S_{\cQ,n},\\
S_{\cQ,n} &=& {1\over n!}\sum_{\cG_\cQ,\cal{EQ}}\sum_{\underline\tau}
\sum_{\{\si \}}' \ep (\cJ)\prod_{i=1}^n\chi_i(\{\sigma\})\prod_{q\in\cQ}\chi_q(\{\sigma\})\nn\\
&&\quad\quad \prod_{v} \int_{\Lambda_{\beta, L}} dx_{v} 
\prod_{\ell\in \cT} \int_{0}^{1} dw_{\ell}
C_{r_\ell,\si_{\ell}} (x_{\ell}, y_{\ell})
\det\nolimits_{left} (C(w)).
\eea

We have the following theorem:
\begin{theorem}\label{mqua}
Let $T$ be a fixed positive number, which is the temperature of the model. Let $\RR^\cQ_T=\{\lambda\ \vert\ |\lambda\log^2T|<1\}$ be a set of values of the coupling constant $\lambda$, the quadruped Schwinger functions $S_{\cQ}(\lambda)$ are analytic functions of $\lambda$ for $\lambda\in\RR_T$, and the radius of convergence $R^\cQ_T$ for the perturbation series satisfies
\be  R^{\cQ}_T\ge c / | \log^{2} T| \ ,
\ee
where $c$ is a positive constant.
\end{theorem}
First of all let us introduce the following definitions.
\begin{definition}
Let $q\in\cG_\cQ$ be any quadruped. Following the tree lines in $\cG_\cQ$ from the root to the leaves, $q$ is linked directly to a set of quadrupeds $\{q'_1,\cdots,q'_{d_q}\}$, $d_q\ge1$. These quadrupeds, which could be either bare vertices or general quadrupeds, are called the maximal sub-quadrupeds of $q$. The number $d_q$ of the maximal sub-quadrupeds contained in $q$ is equal to the number of links in $\cG_\cQ$ starting at $q$. 
\end{definition}
\begin{remark}
Remark that we always follow the root-to-leaves direction in $\cG_\cQ$ when we consider the links start at any quadruped $q$. A maximal sub-quadruped $q'$ of $q$ may still contain some sub-quadrupeds $q''_1,\cdots, q''_{d(q')}$. By definition, the inclusion relation between a quadruped $q$ and a sub-quadruped $q''$ is not a link.
\end{remark}

First of all we have the following lemma.
\begin{lemma}[Sector sum for a single quadruped]\label{secqua}
Let $q$ be any quadruped in $\cG_\cQ$ of scaling indices $r$, which is linked to $d_q$ maximal sub-quadrupeds ${q'_1,\cdots,q'_{d_q}}$. Let the external fields of $q$ be ${f_q^1,\cdots,f_q^4}$, with scaling indices ${r_q^1,\cdots,r_q^4 }$ and sector indices ${\s_q^1,\cdots,\s_q^4}$, respectively. Let the external fields of $q'_v$, $v=1,\cdots q_d$, be ${f_v^1,\cdots,f_v^4 }$, with scaling indices ${r_v^1,\cdots,r_v^4 }$ and sector indices ${\s_v^1=(s_{v,+}^1,s_{v,-}^1),\cdots,\s_v^4=(s_{v,+}^4,s_{v,-}^4 ) }$, respectively. Let $\chi_v(\{\sigma_v\})$, $v=1,\cdots, d_q$, $\{\sigma_v\}:=\{\s_v^1,\cdots,\s_v^4 \}$, be the constraints indicating the conservation of momentum at $q'_v$, we have
\bea
\sum_{\{\sigma_1\},\cdots,\{\sigma_{d_q}\}} \prod_{v=1}^{d_q}\chi_v(\{\sigma_v\})\chi_q(\{\sigma\})e^{-[l_v^1+l_v^2+l_v^3+l_v^4]/4}\le c^{c_q} {r}^{c_q-1}.
\eea
\end{lemma}

\begin{proof}
Let $\cT$ be the spanning tree in the root node $G$ of $\cG_\cQ$ and $\cT_q=\cT\cap q$ be the set of tree lines in $q$. Among all the internal fields contained in $q$ we fix the one $f_{r_q}$ with the maximal scaling index, which we call it the root field of $q$, whose scaling index $r_q$ is called the scaling index of the root. Remark that the root field may belong to any maximal sub-quadruped of $q$. We also fix a root field in each of the other $d_q-1$ quadrupeds, as the one whose scaling index is the maximal among the four external fields. Define the external vertices of $q$ as the external maximal sub-quadrupeds $q'$ to which the external fields of $q$ are hooked. So there are two of them, which are called $e$ and $e'$.
We also call a maximal sub-quadruped of $q$ a vertex. We start summing over the sector indices from the external vertex $e$ towards $e'$, following the tree lines $\cT_q$. 

First of all we consider the constraints on the four sector indices at the external vertex $e$. Since the external fields are fixed, their sector indices are also fixed. Since the root field is also fixed, the sector index for the last field, which belongs to the tree line which connects the next vertex, which is called $q'_1$., is also fixed. Now we consider sector indices of $q'_1$. The root sector indices are always fixed. By conservation of momentum, there could be two kinds of constraints (cf. Proposition \ref{secmain}) on the sector indices: case a) the smallest two sector indices of $q'_1$ collapse and case b) the smallest one is equal to the smallest scaling index $j$ of $q'_1$. In the former case, both of the two fields belong to the tree lines, one connects with the external vertex and one connects to the next vertex, and their sector indices are all fixed. Then the sector indices for the last field of $q'_1$ are also fixed. Continuing this analysis until we arrive at the other external vertex, we find that, once the sector indices for all the root fields are fixed, all sector indices in $q$ are fixed.

In the latter case, when the sector indices $s_{\pm}$ of a field, which is called $f$, is equal to its smallest scaling index $j_f$, then the scaling index is equal to $r_f=3j_f/2$, which is completely determined. $r_f$ must be the minimal scaling index among the four fields and $f$ must belong to the tree line connecting $f$ with the starting external vertex. Hence the root sector indices of the external vertex is also {\it determined}, by conservation of momentum. However there must be a second external field in $q'_1$ that belongs to the tree line connecting the next vertex $q'_2$. Let's call this field $f'$. Then the sector indices for $f'$ is undetermined and should be fixed by hand. 

Now we consider the vertex $q'_2$. If the sector indices of the external fields of $q'_2$ collapse, then we go back to case a) and continue the analysis. If we have the constraints of case b) then the sector indices of $f'$ in the previous vertex are {\it determined} but we need to fix new sector indices for a field belongs to the tree line, in addition to the root ones. Finally we arrive at the vertex adjacent to $e'$, for which the newly fixed sectors for the field that belongs to the tree line connecting with $e'$ is determined by the root sectors of $e'$. 

In conclusion, in order to sum up all the sector indices for $q$, we just need to sum up all the root sector indices. What's more, after fixing root field $f_{r_q}$, there must by another fields contracts with $f_{r_q}$ and we have one pair of sector indices less to sum. Since summing over each root sector is bounded by (cf. Lemma \ref{sec1} )
\be
\sum_{(s_{v,+},s_{v,-})}\g^{-l/4 }\le c.r_v,
\ee
where $r_v$ is the scaling index for the root field at vertex $v$, for which we have
$r_v\le r_q\le r$, finally we have
\bea
\sum_{\{\sigma_1\},\cdots,\{\sigma_{d_q}\}} \prod_{v=1}^{d_q}\chi_v(\{\sigma_v\})\chi_q(\{\sigma\})e^{-[l_v^1+l_v^2+l_v^3+l_v^4]/4}\le c^{c_q} {r}^{c_q-1}.
\eea

So we have proved this lemma.
\end{proof}


\begin{proof}[Proof of Theorem \ref{mqua}]
The idea and main techniques for proving this lemma are from \cite{Riv}. Remark that, since summing over all the quadruped trees is indeed a multi-slice analysis, it is useful to write down explicitly the scaling indices $r$ of the nodes $q\in\cG_\cQ$. We note a quadruped $q$ by $q^k_r$, $r\in[0,r_{max}]$, $k\in[1,c(r)]$, and write a quadruped tree as $\cG_\cQ=\{q_r^k, r=0,\cdots, r_{max}, k=1,\cdots, c(r)\}$. The quadruped Schwinger's functions satisfies the following bound (cf. Formula \eqref{cpt1}):
\bea
|S_{\cQ,n}| &\le& 
{c^n \over n!}
\sum_{\{q^k_r, {r=0},\cdots,r_{max}; {k=1},\cdots,c(r)\},\atop{ {\cal B} = \emptyset}}\sum_{\underline\tau,\cT} \sum'_{\{\si \}}\prod_{q\in\cQ}\chi_q(\{\sigma\})\prod_{i=1}^{n}\ \Big[\chi_{i}(\si)e^{-[l_i^1 + l_i^2 + l_i^3 + l_i^4]/4}\Big]\nn\\
&&\quad\cdot\prod_{r=0}^{r_{max}}\g^{2-|e(G^k_r)|/2}\ .\label{convq}
\eea

One major difference from the convergence case (cf. Formula \eqref{cpt1}) is that, for any quadruped $q_r^k$ we have the $2-|e(q_r^k)|/2=0$. So we couldn't gain any convergent factor from the power-counting. But since we also introduce the conservation of momentum on the external legs of every quadruped, we have additional constraints on the summation over the sector indices, which have been introduced in detail in Lemma \ref{secqua}. Once we have Lemma \ref{secqua}, the conclusion easily follows.

We start the summation procedure over the sector indices from the leaves to the roots, following the quadruped tree. The first quadruped $q_1$ at certain scaling index $r$ contains the bare vertices as the maximal sub-quadrupeds. The next one $q_2$ may contain $q_1$, some other non-trivial quadrupeds at the same scaling index as $q_1$ and some other bare vertices. For each quadruped $q$ we apply  Lemma \ref{secqua}, until we arrive at the root node of $\cG_\cQ$. So we have:
\bea
|S_{\cQ,n}| &\le& \prod_{q\in\cQ}\Big[c_1^{c_q} \sum_{r=0}^{r_{max}} {r}^{c_q-1}\Big]
\le \prod_{q\in\cQ}c_2^{c_q}|\log T|^{c_q},
\label{convq2}
\eea
in which $c_1$ is a positive constant, the sum over scaling indices in $[\cdots]$ means that we sum over the root scaling indices for each quadruped $q$, from $0$ to $r_{max}=3|\log T|/2$, and we have used the fact that the number of all possible quadruped trees is bounded by $c^n n!$ (see \cite{Riv}), and $c_2=3 c_1\cdot c/2$.

Recall the following well-known inductive formula 
\be\label{ind4p}\sum_{q\in \cQ} d_q=|\cQ|+n-1\le 2n-2,
\ee 
where $| \cQ|$ is the cardinality of the set of quadrupeds $\cQ$, for which we have $| \cQ|\le n-1$. Formula \eqref{ind4p} can be easily proved by induction (see also \cite{BGM1}, \cite{AMR1}).

Finally we have the following bound for the quadruped Schwinger function:
\be
|S_{\cQ,n}|\le\prod_{q\in  \cQ}O(1)^{d_q}|\log T|^{d_q}\le O(1)^{2n-2}|\log T|^{2n-2},
\ee
and
\be
|S_{\cQ}(\lambda)|\le\sum_{n=0}^\infty O(1)^{2n-2}\cdot\lambda^n\cdot|\log T|^{2n-2},
\ee

Define 
\be\label{adoq}
\RR^\cQ_T:=\{\lambda\ \vert |\lambda\log^2T|<1 \},
\ee
then the Schwinger function $S_{\cQ}(\lambda)$ is an analytic function for $\lambda\in\RR^\cQ_T$.
So we proved this theorem.

\end{proof}
\begin{remark}\label{rmtad}
Remark that, taking into account the contributions from the quadruped graphs, the analytic domain $\RR^{\cQ}_T$ of of the Schwinger's functions becomes much smaller than $\RR^{c}_T$. 
This fact also set a constraint to the analytic domains for the biped Schwinger functions. Let the analytic domain for the whole Schwinger's functions by $\RR_T$ and the analytic domain for the bipeds  
be $\RR^b_T$. Then we have 
\be
\RR_T=\RR^b_T\cap\RR^{c}_T\cap \RR^{\cQ}_T\subseteq\RR^{\cQ}_T,
\ee 
no matter how big $\RR^b_T$ is.
So the coupling constant $\lambda$ is always bounded by 
\be
|\lambda|<1/{j^2_{max}}=1/{\log^2 T}.
\ee
\end{remark}

\section{The 2-point functions}
In this section we study the convergent and analytic properties of the self-energy of the model.
One important question is that if the system is a Fermi liquid when the temperature is low enough.
This can be checked if they meet the Salmhofer's criterion, which is a set of conditions concerning the convergence and regularity properties of the 1PI 2-point Schwinger functions in the momentum space, at finite temperature, i.e., the self-energy in the momentum space at finite temperature:
\begin{theorem}[Salmhofer's criterion on Fermi liquid at $T>0$ \cite{salm}]\label{salmc}
A $2$-dimensional many-fermion system with dispersion relation $\Omega$ and interaction
$V$ is a Fermi liquid if the thermodynamic limit of the momentum space Green's functions exists for $|\lambda|<\lambda_0(T)$ and if there are constants $C_0, C_1, C_2>0$ independent of $T$
and $\lambda$ such that the following holds. The perturbation expansion for the momentum space self-energy $\Sigma(\lambda, k)$ converges for all $(\lambda,T)$ with $|\lambda\log T|<C_0$, and the self-energy $\Sigma(k, \lambda)$ satisfies the following regularity conditions:
\begin{itemize}
\item $\Sigma(k, \lambda)$ is twice differentiable in $k$ and
\be
\max_{\beta=2}\Vert\partial^\beta\Sigma(k)\Vert_{\infty}\le C_1.
\ee
\item the restriction to the Fermi surface $\Sigma(k, \lambda)_{\{0\}\times\cF }\in C^{\beta_0}(\cF,\RRR)$ and

\be
\max_{\beta=\beta_0}\Vert\partial^\beta\Sigma(k, \lambda)\Vert_{\infty}\le C_1,
\ee
where $\beta_0>2$ is the degree of differentiability of the dispersion relation $\Omega$.

\end{itemize}
\end{theorem}

Remark the original form of Salmfofer's criterion is concerning the regularity properties of the skeleton graphs. But since the self-energy doesn't satisfy these regularities, we don't need to consider further the skeleton graphs. 

We study first the connected $2$-point Schwinger functions $S_2(y,z)$, where $y$ and $z$ are the coordinates for the two external fields. It is useful to expand the Schwinger functions to the order of $n+2$ and identify the two vertices which contract with the two external fields as the external vertices for the graphs of the Schwinger functions. Remark that a tadpole graph has only one external vertex..
Using the BKAR jungle formula we have:
\bea\label{consch}
S^c_2(y,z)&=&\sum_{n=0}^\infty\frac{\lambda^{n+2}}{n!}\int_{({\Lambda_{\beta, L}})^n} d^3x_1\cdots d^3x_n\sum_{\cG}\sum_{\cG_\cB}\sum_{{\cal EB}}\sum_{\{\sigma\}}\sum_{\cT, \underline\tau}\Big(\prod_{\ell\in\cT}\int_{0}^1 dw_\ell\Big)\nn\\
&&\quad\quad\cdot \Big[\prod_{\ell\in\cT}C(f_\ell,g_\ell)\Big]\cdot\det\Big(C(f,g，\{w_\ell\})\Big)_{left},
\eea
where $\cG_\cB$ is the set of all possible Gallavotti-Nicol\`o trees of bipeds that are compatible with the general GN trees $\cG$ and $\cT$ is the set of spanning trees in the roots of $\cG_\cB$ which are connecting all the $n+2$ vertices. 

Remark that the connected Schwinger functions can only tell us the information of the connectedness of the graphs (i.e. information about the tree lines). However, Salmhofer's criterion are concerning about the analytical properties of the self-energies, whose corresponding Feynman graphs have the one particle irreducible structures (1PI for short), which means that the graph can not be disconnected by deleting one line. So we need to further expand the resulting determinant in \eqref{consch} to extract additional lines to ensure the 1PI property. 
As have been noted by many experts in constructive renormalizaton theory, this is in contradiction with the spirit of constructive physics, as expanding loop lines may generate unbounded combinatorial factors which deteriorate the convergence of the perturbation series. However there is one way to perform this auxiliary expansion and generate the 1PI graphs constructively, which is called the multi-arch expansions \cite{DR1}. 

There are two equivalent ways to define the 1PI Schwinger functions $\Sigma_2(y,z)$ from the connected one $S^c_2(y,z)$: by introducing the source terms coupled to the external fields and using Legendre transform, and using the graphical method, namely replacing all the connected graphs (trees) in Formula \eqref{consch} by one particle irreducible graphs. We shall follow the second method in this paper. Let $\{\Gamma\}$ be the set of 1PI graphs over $n+2$ vertices, we can formally "define" the self-energy as:
\bea\label{selfeng}
&&\Sigma_2(y,z)=\sum_{n=0}^\infty\frac{\lambda^{n+2}}{n!}\int_{({\Lambda_{\beta, L}})^n} d^3x_1\cdots d^3x_n\sum_{\cG}\sum_{\cG_\cB}\sum_{{\cal EB}}\sum_{\{\sigma\}, \underline\tau}\sum_{\{\cT\}}\sum_{\{\Gamma\}}\\
&&\quad \Big(\prod_{\ell\in\cT}\int_{0}^1 dw_\ell\Big)\cdot \Big[\prod_{\ell\in\cT}C(f_\ell,g_\ell)\Big]\Big[\prod_{\ell\in\Gamma\setminus\cT}C(f_\ell,g_\ell)\Big]
\cdot\det\Big(C(f,g，\{w_\ell\})\Big)_{left,\Gamma}.\nn
\eea 
Remark that the above expression for the self-energy is only a formal one, as we have to be very careful
about the meaning of $1PI$ graphs over $n+2$ vertices. Expanding all 1PI graphs simply makes the above expansion divergent. A mathematically meaningful definition will be given only after we introduce the multi-arch expansions.

\subsection{The renormalization of the two-point function}
Now we discuss the renormalizations for the two point-functions.
Let $\Sigma_r(y,z)$ be the amplitude of a biped $b$ of scaling index $r$, whose external momentum are $k_{e_1}$ and $k_{e_{2}}$ and are fixed. Let the external fields of $b$ be $\phi_{e_1}$ and $\phi_{e_2}$ with scaling index $r_e$. So we have $r< r_e$, by definition. Without losing generality, we can choose the external field $\phi_{e_1}$ to be a smooth function localized at the point $y$:
$\phi_{e_1}(y')=\eta(y')\delta(y'-y)$, where $\eta\in G^s_0$, and choose the field $\phi_{e_2}$ as
$$\phi_{e_2}(z)=c.\eta(z)\g^{-r_e/4}e^{-\frac12 [|y_0-z_0|\gamma^{-j_0}+|y_+-z_+|\g^{-s_+}+|y_--z_-|\g^{-s_-}]^{\alpha}}.$$ 
Then the amplitude of biped writes:
\bea\label{rn2pt}
&&\int dy' dz\ \phi_{e_1}(y')\Sigma_r(y,z)\phi_{e_2}(z)\\
&&=\int dy [\int dz\ \Sigma_r(y,z)]\phi_{e_1}(y)\phi_{e_2}(y)+\int dy dz\ \Sigma_r(y,z)\phi_{e_1}(y)[\phi_{e_2}(z)-\phi_{e_1}(y)].\nn\\
&&:=\int dy dz\ \phi_{e_1}(y)[\tau+(1-\tau)]\Sigma_r(y,z)\phi_{e_2}(z),\nn
\eea
where $\tau$ is the localization operator and $(1-\tau)\Sigma$ is the remainder term. The first term is called a local term, because both of the external fields are hooked at the vertex $x$. Now define $\delta\mu^r(y)=-[\int dz\ \Sigma_r(y,z)]$, which will be canceled by the counter-term at scale $r$. In other words, for any scaling index $r\in[0, r_{max}]$, we add the corresponding term $\int dy\delta\mu^r\phi_{e_1}(y)\phi_{e_2}(y)$ to the bare counter-term $\delta\mu^1_{r_{max}}\int dy\phi_{e_1}(y)\phi_{e_2}(y)$, and we have:
\be
\delta\mu^r+\delta\mu^r_{r_{max}}=\delta\mu^{r-1}_{r_{max}},
\ee
which defines the renormalized counter-term at scale $r-1$. The BPHZ condition states that 
\be
\sum_{r=0}^{r_{max}}\delta\mu^r+\delta\mu^1_{r_{max}}=0.
\ee

It is important to estimate the amplitude of the counter-terms at any scale $r$, in order to prove the boundedness of $\delta\mu^1_{r_{max}}$. We have the following theorem:
\begin{theorem}\label{flowmu}
There exists a positive constant $K$ such that 
\be\label{bdbare}
|\delta\mu^1_{j_{max}}|\le K|\lambda|,
\ee
where $\lambda\in\RR_T:=\{\lambda\ |\ |\lambda\log^2T|<1 \}$.
\end{theorem}
We will proved this theorem in Section $5.5$. Remark that since $|\delta\mu^1_{j_{max}}|$ cancels exactly with the sum of the localized terms at any order $n$, we can prove this theorem by proving that the amplitudes of the localized terms of any order $n$ are bounded. In the future sections just assume this theorem to be correct, until we prove it in the last section. So we forget the counter-terms $\delta\mu$ for the moment. 

The rest terms in \eqref{rn2pt} are called the remainder terms. We consider first the term:
\bea\label{interp1}
&&\int d^3y d^3z\ \phi_{e_1}(y)\Sigma_r(y,z)[\phi_{e_2}(z)-\phi_{e_1}(y)]\nn\\
&&=
\int d^3z d^3y\ \phi_{e_1}(y)\Sigma_r(y,z)\Big[\ (z-y)\cdot\big[\partial_z\phi_{e_2}(z)\ |_{z=y}\ \big]+\frac{1}{2}(z-y)^2\cdot\big[\partial^2_z\phi_{e_2}(z)\ |_{z=y}\  \big]\nn\\
&&\quad+\frac{1}{2}\int_0^1 dt(1-t)\partial^2_z\phi_{e_2}(tz+(1-t)y)\ \Big].
\eea

Remark that the first term on the R.H.S. of \eqref{interp1} is vanishing due to symmetry, except for the term
\be\label{x0bd1}
\int d^3y d^3z\ \phi_{e_1}(y)(z_0-y_0)\Sigma_r(y,z)\partial_{y_0}\phi_{e_2}(y).
\ee
The idea for bounding this term is to use the decaying property of the tree propagators and use the fact that the scaling indices of the internal lines
in $\Sigma$ is lower than that of the external lines $\phi_e$. 
Let 
\bea C_{r_0}(y',z')=O(1)\g^{-r_0-l/2}e^{-d^\alpha_{r_0}(y',z')},\nn\eea
be a tree propagator with scaling index $r_0$ inside the biped, where
\bea d_{r_0}(y',z')=[\gamma^{-i_{r_0}}(y'_0-z'_0)+\gamma^{-s_{+}}(y'_+-z'_+)+\gamma^{-s_{-}}(y'_--z'_-)],\nn
\eea 
then we have
\be\label{x0bd}
|y'_0-z'_0|e^{-d^\alpha_{r_0}(y'-z')}|
\le\g^{i_{r_0}}e^{-d^\alpha_{r_0}(y'-z')},
\ee
while
\be
\partial_{x_0}\phi_{e_2}(y)\le O(1)\g^{-i_e}\phi_{e_2}(y).
\ee
We can always choose the internal propagators such that $i_e>i_{r_0}$ so that we gain the
convergent factor $\g^{-(i_e-i_{r_0})}$. 


Similarly, the second term of \eqref{interp1} can be written as:
\bea\label{se2}
&&\frac{1}{2}\int dy dz\ \phi_{e_1}(y)\Sigma_r(y,z)\ \Big[\ (z_0-y_0)^2\partial^2_{y_0}\phi_{e_2}(y)\nn\\
&&\quad\quad+(z'_+-y'_+)(z'_--y'_-)\partial_{y_+}\partial_{y_-}\phi_{e_2}(y)\ \Big].
\eea
While the first term can be treated similarly in \eqref{x0bd1}, there are more subtleties for the second term and the third term, since 
\be\label{x1bd}
|y'_+-z'_+|z'_--y'_-||e^{-d^\alpha_{r_0}(y'-z')}|
\le\g^{s_{r_0,+}+s_{r_0,-}}e^{-d^\alpha_{r_0}(y'-z')},
\ee
while
\be
\partial_{y_+}\partial_{y_-}\phi_{e_2}(y)\le O(1)\g^{-(s_{+,e}+s_{-,e})}\phi_{e_2}(y).
\ee
As proved in Section $5.4$ (see also \cite{AMR1} for more details), we can always choose the optimal internal propagators such that $s_{+,e}+s_{-,e}\ge s_{r_0,+}+s_{r_0,-}$ and we gain the convergent factor $\g^{-[(s_{+,e}+s_{-,e})-(s_{r_0,+}+s_{r_0,-})]}$. The last term in \eqref{interp1} is the remainder term of the 2nd order Taylor expansion and is bounded in the same way as \eqref{se2}.

If the biped $b$ is a connected component in a Gallavotti-Nicol\`o tree, we have to replace the external fields $\phi_{e_1}$ and $\phi_{e_2}$ by external propagators, and the renormalization procedure are almost the same (cf. Section $5.5$).


\subsection{The Multi-Arch expansions for 1-PI graphs}
In this part we shall study the analytic properties for the self-energies, whose underlying Feynman graphs are $1$-particle irreducible. We consider first the self-energy without the contributions of the general tadpoles. So it is more convenient to distinguish the two external vertices to which the two external fields are hooked and expand the Schwinger functions to the order of $n+2$. Remark that such two-point Schwinger functions receive contributions not only from the bipeds but also the quadruped graphs and the nodes with more external fields. In order to take into account all these contributions to the self-energies, we consider not only the biped trees $\cG_\cB$ but also the full Gallavotti-Nicol\`o tree structure for the Schwinger functions. Fortunately, the contributions from the quadruped graphs only set constraints to the analytic domain of the coupling constant as $\RR_\cQ=\{\lambda\ |\ \lambda|\log T|^2<1\}$, but don't contain any other divergent terms, and the convergent part are really finite if the values of the coupling constant are constrained in $\RR_\cQ$ (cf. Remark $4.4$.).

As have been discussed in the previous paragraphs, the connected Schwinger functions (cf. Formula \eqref{consch}) are not 1PI in general. So they don't contribute to the self-energy. In order to derive the 1PI structure of the Feynman graphs and the corresponding Schwinger function $\Sigma^b$, we need to introduce a new expansion, called the multi-arch expansion, in addition to the BKAR forest expansion. Remark that although the generalized tadpole graphs are automatically one-particle irreducible, they may contain subgraphs that are general bipeds. So we still need the multi-arch expansions for them. 

In this subsection we shall consider the amplitudes for general 1PI biped graphs and introduce the multi-arch expansions in detail. The general tadpoles will be studied in the next sections.  

The multi-arch expansion is an auxiliary expansion for the integrand of the connected two-point function:
\be\label{main0}
F(\{C_\ell\}_\cT, \{C(f_i,g_j)\})=\Big[\prod_{\ell\in\cT}C_{\s(\ell)}(f(\ell),g(\ell))\Big]\ \det(\{C(f_i,g_j)\})_{left,\cT},
\ee
which further expands the loop determinant in the above formula in such a way that the resulting two point function is 1PI (the corresponding Feynman graphs are 1PI) and no unbounded combinatorial factors can be generated. So it is a convergent expansion.

We write the determinant for the loop lines as $\det(\{C(f_i,g_j)\})_{left,\cT}$ to emphasize that the spanning tree is fixed when we perform the multi-arch expansion. Let $\cT$ be a spanning tree over $n+2$ vertices $\{y,z,x_1\cdots,x_n\}$ and $P(y,z,\cT)$ be the unique path in $\cT$ connecting $y$ and $z$, which are the two ends of the path.
Suppose that there are $p+1$ vertices in the path $P(y,z,\cT)$, with $p\le n+1$, then we can label each vertex in the path with an integer starting with $0$ for $y$ and increasing towards $p+1$ for $z$. Let $\mathfrak{B}_i$ be a branch in $\cT$ at the vertex labeled by $i\in[1,p+1]$, which is defined as the
subtree in $\cT$ whose root is the vertex in $P(y,z,\cT)$ labeled by $i$. We fix two half lines in the path $P(y,z,\cT)$, called the external legs, each attached to one end vertex, so there are $2(n+2)$ half lines left, which are the fields and anti-fields, to be contracted to form the determinant $\det_{left}$. We also call these fields and anti-fields the remaining fields and denote the set of the remaining fields by $\mathfrak{F}_{left}$. Define a packet $\mathfrak{F}_i$ as the set of the remaining fields restricted to the branch $\mathfrak{B}_i$. Obviously, there are $p+1$ packets in total and $\mathfrak{F}_i\cap\mathfrak{F}_j=\emptyset$ for $i\neq j$. In other words, $\mathfrak{F}_{left,\cT}$ can be written as the disjoint union of the $p$ packets: $\mathfrak{F}_{left,\cT}=\mathfrak{F}_1\cup\cdots\cup\mathfrak{F}_p$. Among all possible contractions of the pairs of fields and anti-fields to form the determinant $\det_{left}$, we select through an explicit Taylor expansion with an interpolating parameter $s_1$ those which have a contraction between an element of $\mF_1$ and an element belongs to $\cup_{k=2}^{p}\mF_k$, as follows. Let $\{C(f_i,g_j)\}$ be the elements in the remaining determinant for any loop lines $\{(f_i,g_j)\}$, define \bea
C(f_i,g_j)(s_1)&:=&s_1C(f_i,g_j)\quad {\rm if}\ f_{1}\in \mF_1, g_{1}\notin \mF_1,\\
&:=&C(f_i,g_j)\quad\quad {\rm otherwise},
\eea
we have
\bea
&&\det(\{C(f_i,g_j)\})_{left,\cT}=\det(\{C(f_i,g_j)(s_1)\})_{left,\cT}\ \big|_{s_1=1}\\
&&=\det(\{C(f_i,g_j)(s_1)\})_{left,\cT}\ \big|_{s_1=0}
+\int_0^1 ds_1\frac{d}{ds_1}\det(\{C(f_i,g_j)(s_1)\})_{left,\cT}\nn
\eea
The first term on the second line at $s_1=0$ means that there is no loop line connecting $\mF_1$ to its complements and the graph remains one particle reducible in the $y-z$ channel. The second term means that there is a contraction joins a half-line, which we call it $f_1$, in $\mF_1$ with a half line which is called $g_1$, in $\mF_{k_1}$, $1\le k_1\le p$. Then graphically we add to $\cT$ an explicit line $\ell_1=(f_1, g_1)$ joining the packet $\mF_1$ to $\mF_{k_1}$. This newly added line $\ell_1$ is also called a loop-line or an arch. We also call $\mF_1$ the starting packet of the contraction and index of $\mF_1$, which is $1$, is called the starting index of $\ell_1$. Similarly $\mF_{k_1}$ is called the arriving packet of the contraction and $k_1$ the arriving index of $\ell_1$. These definitions are to be generalized to an arbitrary contraction between a pair of packets and the associated arch. The new graph $\cT\cup\{\ell_1\}$ becomes 1-PI between the vertices $y$ and $x_{k_1}$. If $k_1=p$, then we are done. Otherwise we test whether there is a contraction between an element of $\cup_{k=1}^{k_1}\mF_k$ and its complement, by introducing a second interpolation parameter $s_2$:
\bea\label{mul1}
C(f_i,g_j)(s_1, s_2)&:=&s_2C(f_i,g_j)(s_1)\quad {\rm if}\quad f_{i}\in\cup_1^{k_1} \mF_{k},\  g_{j}\in\cup_{k_1+1}^{p}\mF_k\ ,\\
&:=&C(f_i,g_j)(s_1)\quad\quad {\rm otherwise}\ .
\eea
Then we have:
\bea
&&\det(\{C(f_i,g_j)(s_1)\})_{1,left,\cT}=\det(\{C(f_i,g_j)(s_1,s_2)\})_{1,left,\cT}\})|_{s_2=1}\nn\\
&&=\det(\{C(f_i,g_j)(s_1,s_2)\})_{1,left,\cT}\})|_{s_2=0}\nn\\
&&\quad\quad\quad+\int_0^1 ds_2\frac{d}{ds_2}\det(\{C(f_i,g_j)(s_1,s_2)\})_{1,left,\cT}\}).
\eea
%
%
Again the first term at $s_2=0$ means that the block $\cup_{i=1}^{k_1}\mF_{k}$ is not linked to its complement by any loop line, so that it is a generalized tadpole graph. The second term can be written as
\bea\label{link1}
&&\int_0^1 ds_2\frac{d}{ds_2}\det(\{C(f_i,g_j)(s_1,s_2)\})_{1,left,\cT}\})\\
&&\quad=
\int_0^1 ds_2\ \frac{\partial}{\partial s_2}C(f_{i_1},g_{j_1})(s_1,s_2)\cdot \frac{\partial}{\partial C(f_{i_1},g_{j_1})}\det(\{C(f_i,g_j)(s_1,s_2)\})_{1,left,\cT}\}),\nn
\eea 
with $f_{i_1}\in \cup_1^{k_1} \mF_{k}$, $g_{j_1}\in\cup_{k_1+1}^{p}\mF_k$ and we have
\bea\label{rprop0}
\frac{\partial}{\partial s_2}C(f_{i_1},g_{j_1})(s_1,s_2)&=&C(f_{i_1},g_{j_1})\quad\quad{\rm if}\quad f_{i_2}\in\cup_{i=2}^{k_1}\mF_{k}\nn\\
&=&s_1C(f_{i_1},g_{j_1})\quad{\rm if}\quad f_{i_2}\in\mF_{1}.
\eea
Graphically, formula \eqref{link1} means that we have a contraction between the loop field $f_{i_1}$ and $g_{i_1}$ so we add to $\cT$ another line $\ell_2=(f_{i_1},g_{i_1})$ joining the two packets. Now the new graph $\cT\cup\{\ell_1,\ell_2\}$ becomes 1-PI between the vertices $y$ and $x_{k_2}$. Similarly
for an arch ends at the packet $\mF_{k_r}$, the corresponding interpolated propagator is defined as
\be\label{rprop}
C(f_{r}, g_r)(s_1,\cdots, s_r)=s_rC(f_r,g_r)(s_1,\cdots, s_{r-1}).
\ee

We continue the interpolation process until the whole graph becomes 1-PI. Suppose that we stop the process at step $m\le p$, which means that we add $m$ lines to a $\cT$ to form a 1-PI biped graph, the set of newly added loop lines $\{\ell_1,\cdots,\ell_m\}$ is called an m-arch system:
\bea
&&\Big\{ \ell_1=(f_1,g_1),\cdots,\ell_m=(f_m,g_m)\ \Big|\ f_1\in\mF_1, g_1\in\mF_{k_1};\ f_2\in\cup_{r=1}^{k_1}\mF_r, g_2\in\cup_{k_{1}+1}^{\mF_{k_2}};\nn\\
&&\quad\quad\quad \cdots ;\ 
f_m\in\cup_{r=1}^{k_{m-1}}\mF_r,\ g_m\in\mF_{k_m}=\mF_{p};\  k_1\le\cdots\le k_m;\ m\le p\ \Big\}.
\eea
We have the following expression for the loop determinant:
\bea\label{main1}
&&\det\big(\{C(f_i,g_j)\}\big)_{left,\cT}=\sum_{\substack{{\rm m-arch-systems}\\(f_1,g_1),\cdots, (f_m,g_m)\\ {\rm with}\ m\le p}}\ \int_0^1 ds_1\cdots \int_0^1 ds_m  \nn\\
&&\Bigg[\frac{\partial}{\partial{s_1}}C(f_1,g_1)(s_1)\cdot\frac{\partial}{\partial{s_2}}C(f_2,g_2)(s_1,s_2)\cdots\frac{\partial}{\partial{s_m}}C(f_m,g_m)(s_1,s_2,\cdots,s_m)\cdot\nn\\
&&\quad\quad\cdot\frac{\partial^m det_{left,\cT}}{\prod_{r=1}^m \partial C(f_r, g_r)}\Big(\{s_r\}\Big)\Bigg]\ ,
\eea
where the sum is over all $m$-arch systems with $p$ vertices. It is useful to have a more explicit expression for the second line of the above formula. We have:
\begin{proposition}\label{prodpg}
Let $\ell_r$ be a loop line in an $m$-arch system introduced above. Let $q_r$ be the number of loop lines that fly over $\ell_r$, namely those loop lines whose starting indices are smaller than or equal to that of $\ell_r$ while whose arriving indices are greater than that of $\ell_r$. Let
$\prod_{r=1}^m C(f_r,g_r)(s_1,\cdots, s_{r-1})$ be the compact form of the second line of formula 
\eqref{main1}, then we have
\bea\label{indu0}
&&\prod_{r=1}^m C(f_r,g_r)(s_1,\cdots, s_{r-1}):=\prod_{r=1}^m\partial_{s_r}C(f_r,g_r)(s_1,\cdots,s_r)\nn\\
&&\quad=\big[\ \prod_{r=1}^m C(f_r,g_r)\ \big]\cdot \big[\ \prod_{r=1}^m
s_r^{q_r}\ \big]\ .
\eea
\end{proposition}

\begin{proof}
This proposition can be easily proved by induction, using the definition of the interpolated propagators (c.f. \eqref{rprop}, \eqref{rprop0}). Indeed if the starting indices of the successive loop lines in a $m$-arch system is also in the strictly increasing order, then there is no interpolation parameter left in the product
$\prod_{r=1}^m\partial_{s_r}C(f_r,g_r)(s_1,\cdots,s_r)$; There can be an factor
$s_k^{i_k}$ left when there are loop lines which completely fly over $\ell_k$, whose number is exactly equal to the power $i_k$, which is equal to $q_k$, by definition. So we proved this proposition.
\end{proof}

\begin{remark}
Remark that the remaining determinant on the third line of \eqref{main1} satisfies all properties of the initial determinant, especially the positivity property, because all the $s_{r}$ interpolations are always performed between a subset of packets and 
its complement. So the final covariance, as function of the parameters $s_{r}$, is a convex combination with positive coefficients of block-diagonal covariances. So these $s_{r}$ parameters don't alter Gram's bound on the remaining determinant.
\end{remark}

We still have to check the fact that no factorials can be generated in the multi-arch expansions, which is highly not trivial: by construction, while the arriving index for a successive arch is increasing, the departure index has not to be so. Potentially a factorial might be generated for choosing the departure field once the arriving field is fixed. We will show that the possible factorials are damped by the integration over the interpolation parameters. This has been proved in detail in \cite{DR1,AMR1}. Here we collect some basic notions and results of \cite{DR1,AMR1}, for completeness. 

In order to formulate this more explicitly, let's introduce more notations concerning the multi-arch graphs.

\begin{definition}
Let $\cL_n=\{\ell_1,\cdots,\ell_n\}$ be a set of $n$ loop lines, $n\le m$, in an $m$-arch system such that the arriving indices of these lines are put in the increasing order. The set $\cL_n$ is said to form a nesting system if the starting index of the last line $\ell_n$ is the smallest one among all the starting indices of the loop lines in $\cL_n$. In this case the loop lines $\{\ell_1,\cdots,\ell_{n-1}\}$ are said to be useless in that the graph remains 
1PI if these loop lines are all deleted. 
\end{definition}
\begin{remark}
Remark that a nesting system may contain a subset of loop lines that still form a nesting system.
The nesting systems are indeed the source of the combinatorial factors.
\end{remark}

From Proposition \ref{prodpg} we know that only the nesting loop lines contribute to the interpolating factors in \eqref{indu0}. So the sum over all $m$-arch systems over $p$ vertices, which may result in combinatorial factors, should be weighted by the integral $\Big(\prod_{r=1}^m \int_0^1 s_r\Big)\ 
\prod_{r=1}^m s_r^{q_{r}}$. The result is stated in the following lemma ( see also \cite{AMR1}):
\begin{lemma}
Let $n\ge1$ be the number of vertices of the 1-PI graph formed in the m-arch expansions. There exist some numerical constants $c$ and $K$ such that:
\be
\sum_{m=1}^p\sum_{\substack{{\rm m-arch-systems}\\(f_1,g_1),\cdots, (f_m,g_m)\\ {\rm with}\ m\le p}}\ \Big(\prod_{r=1}^m \int_0^1 s_r\Big)\ 
\prod_{r=1}^m s_r^{q_{r}}\le c K^n.
\ee
\end{lemma}
{\it Proof}:
This lemma for a similar setting has been proved in the appendix of \cite{DR1}. Since this lemma is in the combinatorial nature and applicable to our case without any new techniques, we just omit the proof and ask the interested reader to look at \cite{DR1} for details ( see also \cite{AMR1} for graphical explanations of the multi-arch expansions ). 
\qed

\begin{figure}[htp]
\centering
\includegraphics[width=.8\textwidth]{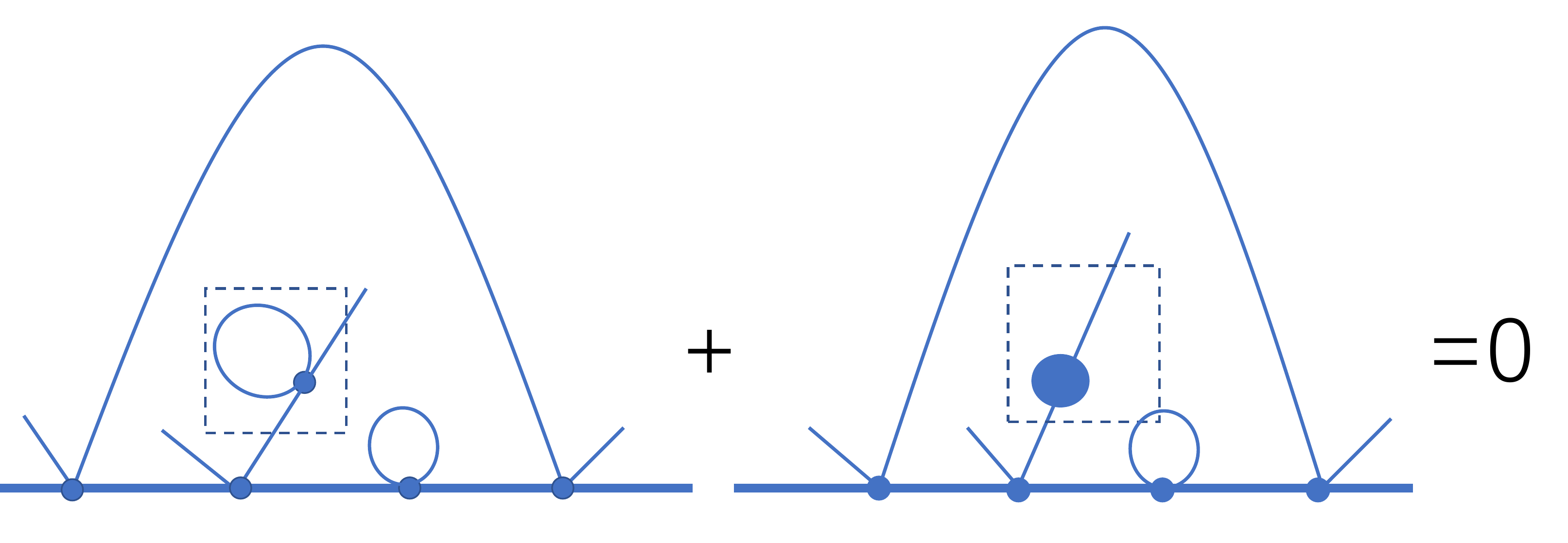}
\caption{\label{rtad}
Cancellation of a tadpole term with the corresponding counter-term. The small dots are the bare interaction vertices and the big dot is the counter-term vertex. Deleting the vertex contained in the dashed square in each of the two graphs one obtains the graph for $F'_{2,n-1}$.}
\end{figure}

We have the following lemma concerning the amplitude of the 1PI Schwinger functions:
\begin{lemma}
The amplitude of the 1PI biped graphs is given by:
\bea\label{sfe}
\Sigma^b (y,z)& = &\sum_{n=0}^\infty \frac{\lambda^{n+2}}{n!} \int_{\Lambda^n} d^3x_1 ... d^3x_n\sum_{\{ \underline\t \}}\sum_{\cG_\cB} \sum_{\text{external fields} \atop \mathcal{EB}}  \sum_{\text{spanning trees}\ \mathcal{T} } \nn\\
&&\sum_{\{ \sigma \}}
\sum_{{m-{\rm arch\ systems} \atop 
\big( (f_1,g_1,...,(f_m,g_m))\big) } \atop
{\rm with} \  m \leq p }
\left( \prod_{\ell \in \mathcal{T}} \int_0^1 dw_\ell \right) \left( \prod_{r = 1}^m \int_0^1 ds_r  \right)
\left( \prod_{\ell \in \mathcal{T}} C_{\sigma(\ell)} (f_\ell,g_\ell)\right)\nn
\\
&&\left( \prod_{r=1}^m C(f_r,g_r) (s_1,...,s_{r-1})\right)
 \frac{\partial^m \det_{\text{left}, \mathcal{T}}}{\prod_{r=1}^m \partial C(f_r,g_r)} 
\big( \{ w_\ell\}, \{ s_r\}\big) \ .
\eea
\end{lemma}
In order to get better bounds for the general biped Schwinger functions, we introduce a second auxiliary BKAR tree expansion, called the 2-particle-irreducible multi-arch expansions, which completes the 1PI graphs into the 2PI (graphs that remain connected after deleting two lines) and one-vertex irreducible(graphs that remain connected after deleting one vertex) graphs. As we will see later, 2PI graphs give us more constraints for summing over the sector indices. We have the following lemma for the second multi-arch expansions on the Schwinger functions:
\begin{lemma}
The amplitude for the 2PI biped graphs, obtained from the two-level multi-arch expansions is
\bea
&&\Sigma^b (y,z)_{2PI}=\sum_{n=0}^\infty \frac{\lambda^{n+2}}{n!} \int_{\Lambda^n} d^3x_1 ... d^3x_n 
\sum_{\{ \underline\t  \}} \sum_{\text{biped structures} \atop \mathcal{B}}\sum_{\text{external fields} \atop \mathcal{EB}}\nn\\
&&\quad\quad\sum_{\cG_\cB} \sum_{\text{spanning trees} \mathcal{T}} \sum_{\{\sigma \}}\sum_{ m-{\rm arch\ systems} \atop \bigl( (f_1,g_1), ... (f_m,g_m) \bigr)}
\sum_{m'-{\rm arch\ systems} \atop \bigl( (f'_1,g'_1), ... (f'_{m'},g'_{m'}) \bigr)}\nn\\
&&\quad\quad\left( \prod_{\ell \in \mathcal{T}} \int_0^1 dw_\ell \right) \left( \prod_{\ell \in \mathcal{T}} C_{\sigma(\ell)} (f_\ell,g_\ell)\right)\left( \prod_{r=1}^m \int_0^1 ds_r \right)\left( \prod_{r'=1}^{m' }\int_0^1 ds'_{r'} \right)\nn \\
&&\quad\quad\left( \prod_{r=1}^m C(f_r,g_r) (s_1,...,s_{r-1})\right) \left( \prod_{r' = 1}^{m'} C({f'}_{r'},{g'}_{r'}) (s'_1,...,s'_{r'-1})\right)\nn\\
&&\quad\quad\frac{\partial^{m+m'} \det_{\text{left}, \mathcal{T}}}{\prod_{r=1}^m \partial C(f_r,g_r)\prod_{r'=1}^{m'} \partial C(f'_{r'},g'_{r'}) }\big( \{ w_\ell\}, \{ s_r\} ,  \{ s'_{r'} \}  \big),
\eea
where we have summed over all the first multi-arch systems with $m$ loop lines and
the second multi-arch systems with $m'$ loop lines. The underlying graphs are two-line irreducible as well as one vertex irreducible.
\end{lemma}
\begin{proof}
This lemma simply states the result of a second multi-arch expansion for the $2$-point Schwinger functions, on top of the multi-arch expansion introduced above, which permits us to obtain the $2$-PI Schwinger functions from the 1PIs. Recall that a $2$PI Schwinger function is the one whose underlying graphs are not disconnected by deleting two lines in the graph. Since this construction has 
been present in great detail in \cite{AMR1}, we don't repeat it here but invite the interested reader to look at that paper for details. Remark that the second multi-arch expansion respects again positivity of the interpolated propagator at any stage, and remains constructive.
\end{proof}
\vskip.3cm
Before considering the amplitudes of general 1PI graphs, let us consider first a tadpole graph, which is local at one vertex. 
Let $F_{2,n}=F'_{2,n-1}||T^r||$ be the amplitude of a biped graph with $n$ vertices which contains a tadpole $T^r$ at scaling index $r$, whose amplitude is noted by $||T^r||$. Remark that $F'_{2,n-1}$ may contain other tadpoles terms. Let $F'_{2,n}=F'_{2,n-1}\delta\mu^r$ be the amplitude of the graph which is formed by contraction of the fields in $F'_{2,n-1}$ with the fields in a counter-term $\delta\mu^r\int d^3x\psi^2(x)$, such that the counter-term located at the same position as the tadpole. To form the graph for $F'_{2,n-1}\delta\mu^r$ we just replace the tadpole vertex by the counter-term vertex. We can always choose $\delta\mu^r=-||T^r||$. So we have $F_{2,n}+F'_{2,n}=0$.
See Figure \ref{rtad} for an illustration of the cancellation of a tadpole graph with the corresponding counter-term. Remark that the graph corresponds to the term $F'_{2,n-1}\delta\mu^r$ is unique.

Since the simple tadpole terms are related to the counter-term, it is important to calculate the amplitude for a single tadpole. We have
\begin{lemma}\label{tadmain1}
Let $T$ be a tadpole graph and $T^j$ be its amplitude at slice $j$, let $\lambda\in\RR_T\subseteq\RR^q_T$ (cf. Formula \eqref{adoq} and Lemma \ref{rmtad}) be the coupling constant. There exist two positive constants $O(1)$ and $O(2)$ such that:
\be\label{tad01}
O(1)(|\lambda|j)\g^{-j}\le\Vert T^j\Vert\le O(2)(|\lambda|j)\g^{-j}.
\ee
\end{lemma}

\begin{proof}
For any scaling index $0\le j\le j_{max}$, we have:
\be
\Vert T^j\Vert=|\lambda|\sum_{\s=(s_+,s_-)}\ \Big|\int dk_0 dk_+dk_-\tilde C_{j,\sigma}(k_0,k_+,k_-)\ \Big|\le O(1) |\lambda| \sum_{(s_+,s_-)}\g^{-s_+-s_-}.
\ee
Since $0\le s_\pm\le j$ and $s_++s_-\ge j-2$, we have
\be
\sum_{(s_+,s_-)}\g^{-s_+-s_-}=\Big(\sum_{s_+=0}^{j}\g^{-s_+}\Big)\ \Big(\sum_{s_-=j-2-s_+}^{j} \g^{-s_-}\Big).
\ee
Now using the fact that
\be
\g^{-j+2+s_+}\le\sum_{s_-=j-2-s_+}^{j} \g^{-s_-}\le \g^{-j+2+s_+}\frac{1-\g^{-(2+s_+)}}{1-\g^{-1}},
\ee
we have
\bea
O(1)(|\lambda|j)\g^{-j}\le\Vert T^j\Vert\le O(2)(|\lambda|j)\g^{-j}.
\eea
So we proved this lemma.
\end{proof}

Since both the upper bound and the lower bound for the amplitude of a tadpole $T^j$ have been obtained, we have the following corollary:
\begin{corollary}\label{tad03}
Let $T^j(p)$ be the amplitude of a tadpole at scaling index $j$, there exist two constants $O''(1)$ and $O''(2)$ such that 
\be
O''(1)\g^j \le ||\frac{\partial^2}{\partial {p^2}} T^j(p)||\le O''(2)\g^j,
\ee
which is divergent for $j=j_{max}$ and $j_{max}\rightarrow\infty$. So that the second derivative of a tadpole amplitude is not uniformly bounded. This is an important evidence that the current model does not show Fermi liquid behavior when the temperature is low enough. 
\end{corollary}
\begin{proof}
The momentum $p$ is bounded by $\gamma^{-j}$ at scale $j$, so that for each derivation w.r.t. the momentum we get a scaling factor $\gamma^{j}$. So we proved this lemma.
\end{proof}
Finally we can easily prove the following lemma for the amplitude of the full tadpole.
\begin{lemma}\label{tad05}
Let $T=\sum_{j=0}^{j_{max}}T^j$ be the full tadpole and $||T||$ be its amplitude. There exist two positive constants $O(1)'$ and $O(2)'$ such that:
\be
O(1)'|\lambda|\le\Vert T\Vert\le O(2)'|\lambda|.
\ee
\end{lemma}
\begin{proof}
Because $||T||=\sum_{j=0}^{j_{max}}||T^j||$, we can prove this lemma directly by summing over the indices $j$, using the explicit expression for  $||T^j||$.
\end{proof}


Since the cancellation between a tadpole term with a counter-term is exact, tadpoles don't contribute to the self-energy. So the graphs we will consider in the future sections are all tadpole free.   
\subsection{Summing over the sectors indices.}
In this part we consider the summation over the sectors indices for biped graphs, which is similar to that for the quadrupeds graphs. We have the following lemma:
\begin{lemma}\label{secbiped}
Let $b_r$ be a 2PI biped with root scaling index $r$ and contains $n$ bare vertices. There exists a positive constant $C$ such that the summation over all the sector indices for the internal fields of $b$ subject to conservation of momentum is bounded by $C^nr^{n-1}$. 
\end{lemma}
\begin{proof}
Consider such a 2PI biped $b_r$. Let $\cT_b=b_{r}\cap\cT$ be the tree lines in $b_r$. Conservation of momentum on each vertex and on the external fields of the biped $b$ implies that the two external fields have identical sector indices. We choose a root field $f_r$ for each vertex and among all the root fields we select the one with maximal scaling index, as the root index for the whole biped. Since the sectors for the external fields are fixed, it is easy to find that all the other sector indices are determined. What's more, there must be another field which contract with the root field $f_r$. So have at most $n-1$ independent sectors to be summed. Since summing over each pair of sector indices is bounded by $r$, the result is the bound $C^n r^{n-1}$.
%
\end{proof}
Remark that this bound is not optimal. The number of pairs of sector may be further reduced when the 2PI biped is formed by more arches that are not useless. But this bound is enough for our purpose.

Since to each vertex is associated a coupling constant $\lambda$, which is bounded by $(|\log T|^2)^{-1}$ in the analytic domain $\RR_T$, and since we expand the self-energy to the order or $n+2$, the summation over all sector indices for the 2PI biped graph with $n+2$ vertices is bounded by:
\be
C^n |\lambda|^{n+2} r^{n+1}\le C^n (\lambda^2 r^2) |\lambda\log^2 T|^n \big(\frac {r}{|\log T|^2}\big)^n\le C'^n(\lambda^2 r^2),
\ee
where $C'$ is a fixed positive constant.

\subsection{The ring sectors and the optimal power-counting formula for the self-energy}
Let $\cT$ be the fixed spanning tree of the graph and $\cL$ be the set of loop lines generated by the two-level multi-arch expansions, such that $\cL\cup\cT$ is a 2PI graph. Menger's theorems ensure that any such 2PI multi-arch graph has three line-disjoint independent paths and two internally vertex-disjoint paths joining the two external vertices \cite{AMR1} of the graph. Among the three paths in $\cL\cup\cT$ we can always choose a ring structure $R$ obtain the optimal bounds for the self-energy. 
\begin{definition}
A ring $R$ is a set of two paths, noted by $P_{R,1}$, $P_{R,2}$, in $\cL\cup\cT$, which connect the two vertices $y$ and $z$ and satisfy the following conditions. Firstly, the two paths in $R$ don't have any intersection on the paths or on the vertices, except on the two external vertices $y$ and $z$. Secondly, the lines in the ring are chosen as optimal: let $b$ be any node in the biped tree $\cG_\cB$, then at least two external fields of $b$ are not contained in the ring. 
\end{definition}
We have:
\begin{lemma}[\cite{AMR1}]
We can always choose two paths in $\cL\cup\cT$ which satisfy all the conditions in the above definition, hence form a ring $R$. 
\end{lemma}
\begin{proof}
This lemma of combinatorial nature has already been proved in \cite{AMR1}, so we don't repeat it here but ask interested reader to look at that paper for details.
\end{proof}

The definition of the ring lines implies that the propagators in the ring have smaller scaling indices (hence higher momentum) than the ones that are not contained in the ring. Now we explain why it is useful to have two explicit line-and-vertex disjoint paths. 
Let $P_1(y,z)=(y,x_1,\cdots,x_{p_1},z)$ be an integration path in the ring in which $y,z$ are fixed external vertices and $(x_1,\cdots, x_{p_1})$ are the internal vertices. There are $p_1+1$ propagators in the path but we only perform integration for the first $p$ propagators, with respect to the corresponding end vertices. So there is one propagator $C_{x_{p_1},z}$ that we don't integrate out and we gain the convergent factor $\g^{-r_1-l_1/2}$, which is the pre-factor of that propagator. If we have two such paths, there is a second propagator in the second path that is not integrated out w.r.t. its end vertex and we gain another convergent factor $\g^{-r_2-l_2/2}$.

The conservation of momentum also sets constraints to the sector indices for the ring propagators.
So we need to regroup some of the sectors so that the ring propagators have minimal $r$ indices and $s_\pm$ indices. The newly adjusted sectors are called the ring sectors. The construction of ring sectors has been introduced in \cite{AMR1}, Section $VII.1-VII.2$. Since this construction is rather 
straightforward and has been discussed in great detail in \cite{AMR1}, we don't repeat it here but ask the interested reader to consult \cite{AMR1} for details.

After all these preparations, we are ready prove the main theorem concerning the self-energy:
\begin{theorem}\label{maina}
Let $\Sigma_2(y,z)^r$ be the biped self-energy with root scaling index $r$, let the corresponding Gallavotti-Nicol\`o tree is noted by $\cG_r$. Let $\lambda\in\RR_T$ and let $\sigma_{\cG_r}$ be the set of all possible sector indices that is compatible with the Gallavotti-Nicol\`o tree $\cG_r$. Then there exists a constant $O(1)$ such that:
\be\label{x2pta}
|\Sigma_2(y,z)^{ r}|\le \lambda^2 r^2\sup_{\sigma\in\sigma_\cG} O(1)M^{-3r(\sigma)}e^{-cd^\alpha_\sigma(y,z)},
\ee
\be\label{x2pt}
|y_+-z_+||y_--z_-||\Sigma_2(y,z)^{ r}|\le \lambda^2 r^2\sup_{\sigma\in\sigma_\cG} O(1)M^{-2r(\sigma)}e^{-cd^\alpha_\sigma(y,z)},
\ee
\be\label{x2ptc}
|y_0-z_0||\Sigma_2(y,z)^{ r}|\le \lambda^2 r^2\sup_{\sigma\in\sigma_\cG} O(1)M^{-2r(\sigma)}e^{-cd^\alpha_\sigma(y,z)}.
\ee
\end{theorem}
\begin{proof}
A similar result for a different setting has been proved in \cite{AMR1}, sections VIII to X, and can be applied to our case without introducing any new techniques. So we don't repeat the proof here but ask the interested readers to consult \cite{AMR1} for details.
\end{proof}

Summing over the scaling indices $r$ from $0$ to $r_{max}$, and using the fact that $r_{max}=3j_{max}/2=3|\log T|/2$, we have:
\begin{corollary}
There exist three positive constants $c_1$, $c_2$ and $c_3$ such that
\be
|\Sigma_2(y,z)^{\le r_{max}}|\le c_1 \lambda^2 |\log T|^3\ ,
\ee
\be
|y_+-z_+||y_--z_-||\Sigma_2(y,z)^{\le r_{max}}|\le c_2 \lambda^2 |\log T|^3\ ,
\ee
and
\be\label{p2pt}
|y_0-z_0||\Sigma_2(y,z)^{ r}|\le c_3 \lambda^2 |\log T|^3.
\ee
\end{corollary}

We can formulate Theorem \ref{maina} in the momentum space representation and we have:
\begin{theorem}[The bound for the self-energy in the momentum space.]\label{mainb}
Let $\hat\Si^{r}(\lambda,k)$ be the self-energy for a biped of scaling index $r$ in the momentum space representation and $\lambda\in\RR_T$. There exists a positive constant $O(1)$ such that:
\be |\hat\Si^{r}(\lambda,k)|\le O(1)\l^2 r^2\g^{-r},\label{spa}
\ee
\be \vert| \frac{\partial}{\partial k_\mu } \hat\Si^{ r} (\lambda,k) \vert|\le O(1)\lambda^2 r^2,\label{spb} 
\ee
\be\label{spc} | \frac{\partial^2}{\partial k_\mu \partial k_\nu}  \hat\Si^{r} (\lambda,k) | \le O(1) \lambda^2 r^2  M^{r}.
\ee
\end{theorem}
\begin{remark}
This theorem states that the self energy is uniformly $C^1$ for $|\lambda|<1/|\log T|^2$, which is 
smaller than the "required" domain of analyticity $|\lambda|<c/|\log T|$ in Slamhofer's criterion.
The second derivative for the self-energy is growing with respect to $r$, which is unbounded as $T$ tends to zero. So that the system lose the Fermi liquid behavior and becomes a Luttinger liquid.
\end{remark}

\begin{proof}[Proof of Theorem \ref{mainb}]
Remark that the expressions for the self-energy in \eqref{x2pta} and \eqref{spa} are simply Fourier dual to each other. So they are equivalent. Here we just check 
that formula \eqref{spa}-\eqref{spc} can be derived from \eqref{x2pta}.
Let $\Sigma(y,z)$ be the Fourier transform of $\hat\Sigma(p)$. The left hand side of \eqref{x2pta} reads
\be
|\Sigma^r(y,z)|=|\int dp\ \hat\Sigma(p) e^{ip(y-z)}|\ge \gamma^{-2r}|\hat\Sigma^r(p)|,
\ee
while 
\be
|\Sigma^r(y,z)|\le \lambda^2 r^2 O(1)M^{-3r},
\ee
so we have 
\be
|\hat\Sigma^r(p)|\le \lambda^2 r^2 O(1)M^{-r}.
\ee
So we have obtained \eqref{spa}. Since the momentum $p$ is bounded by $\g^{-r}$ at scaling index $r$, first order differentiation w.r.t. $p$ on the r.h.s. term of \eqref{spa} gives \eqref{spb} while the second order differentiation gives \eqref{spc}.
\end{proof}

We have the following theorem for the lower bound for the sunshine graph, which is the simplest biped graph with two vertices connected by three propagators.
\begin{theorem}\label{sun1}
Let $\hat\Si^r_{S}$ be the self-energy of the sunshine graph of scaling index $r$. Let $K$ and $K'$ be two positive constant. We have the following lower bound for $\hat\Si^r_{S}$ and its second .
\be \left| \hat\Si^r_{S} (k) \right|
 \ge  K \l^2 |\log T|^{2}  M^{-r} \ , \label{lower}
\ee
\be \left| \frac{\partial^2}{\partial k_\mu \partial k_\nu}  \hat\Si^r_{S} (k) \right|
 \ge  K' \l^2 |\log T|^{2}  M^{+r} \ , \label{lower}
\ee
in the special case of $\mu, \nu$ in the $(+,+)$ 
direction and incoming momentum $(k_0 = \pi T, k_+ = 1, k_- =0)$.
\end{theorem}

\begin{proof}
This theorem can be proved straightforwardly with the techniques introduced in \cite{AMR2}. So we don't repeat it here.
\end{proof}
\begin{figure}[htp]
\centering
\includegraphics[width=.5\textwidth]{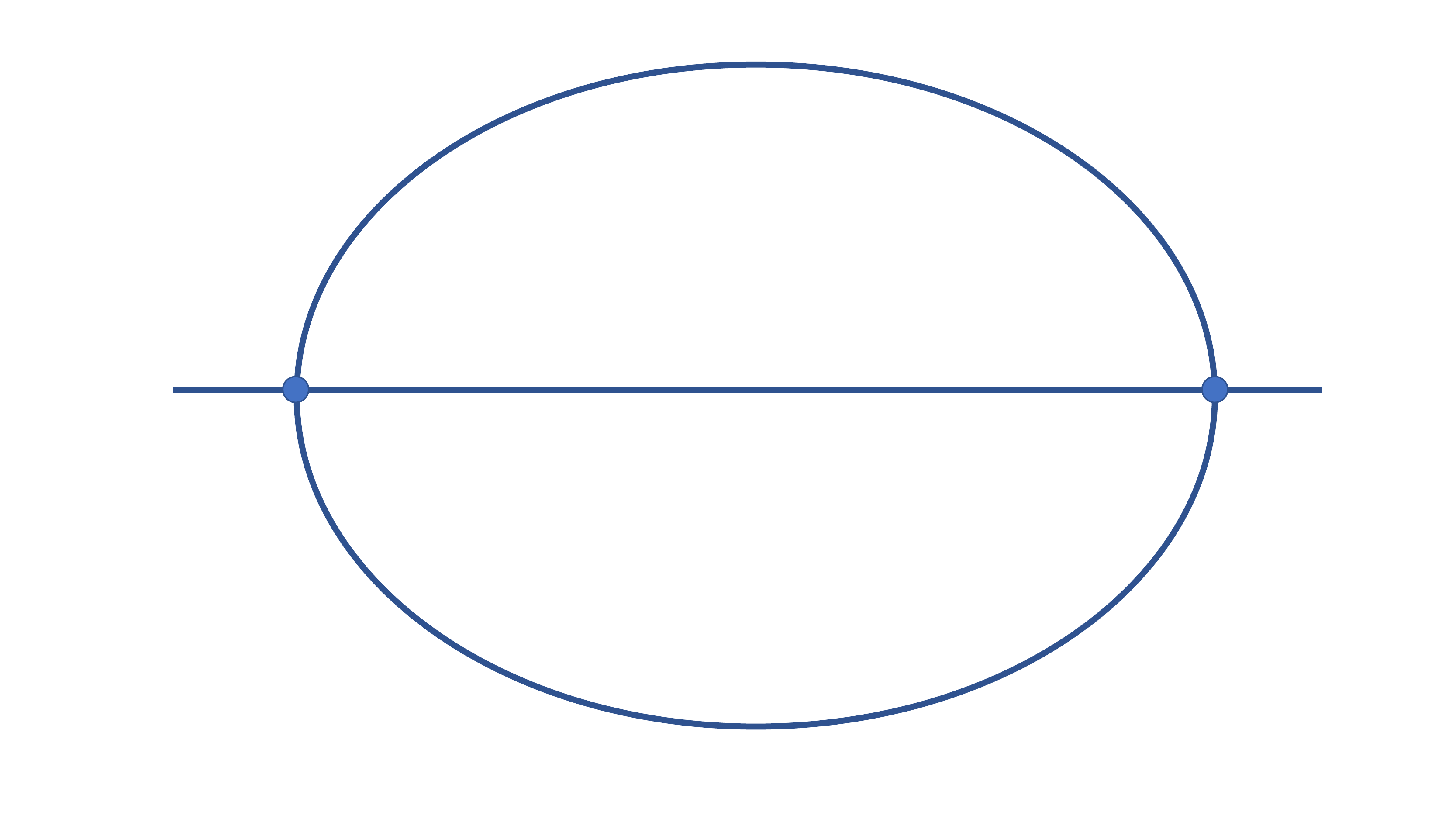}
\caption{\label{gtad1}
A sunshine graph.}
\end{figure}

\subsection{Proof of Theorem \ref{flowmu}.}\label{secflow}
In this section we will study the flow of the counter-term $\delta\mu$ and prove Theorem \ref{flowmu}. By definition, the counter-term is introduced to cancel the amplitude of the local terms and the amplitude of the counter-term is equal to the {\it minus} of the amplitudes of these terms. In order to estimate the bound for the counter-term it is enough to estimate the bounds for the summation of all local terms.

Remark that there are three kinds of local terms: the tadpole term, the localized term: 
\be
\Sigma^{\le r_{max}}(x)=\sum_{r=0}^{r_{max}}\int d^3x\ \Sigma^r(x, y),
\ee
in which all the tadpoles are canceled by renormalization, and the generalized tadpole terms, which is a tadpole whose internal lines are decorated by 1PI bipeds. See Figure \ref{gtad1} for an example of a generalized tadpole.

\begin{figure}[htp]
\centering
\includegraphics[width=.7\textwidth]{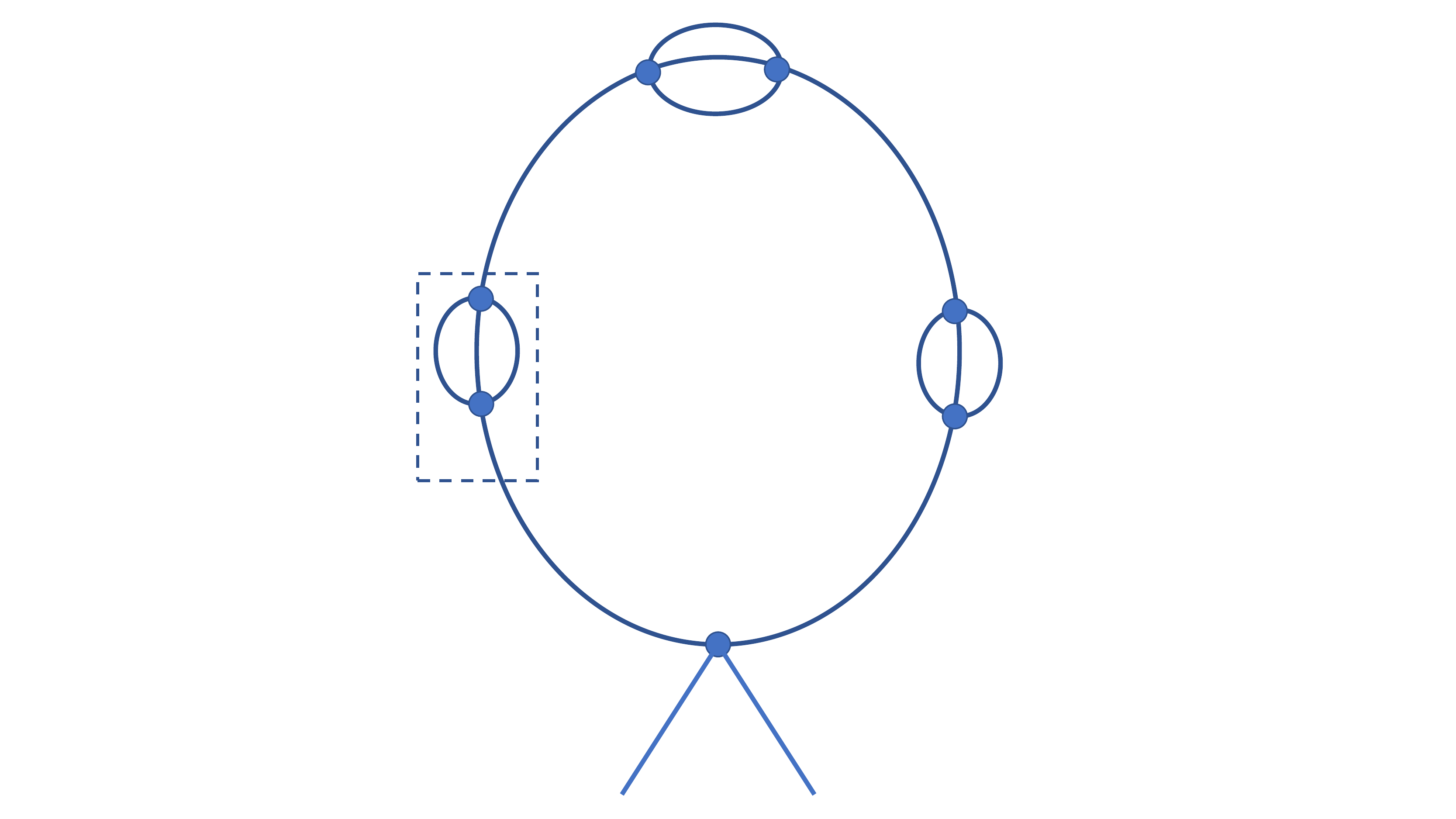}
\caption{\label{gtad1}
A generalized tadpole.}
\end{figure}

Recall that Lemma \ref{tad05} states that the amplitude of a tadpole is bounded by:
\be
||T||\le K_1|\lambda|,
\ee
where $K_1$ is a positive constant and $|\lambda|<1/|\log T|^2$. So the contributions from the tadpoles are corrected bounded.

Now we consider the bound for the localized term. We have:
\be\label{localmu1}
||\Sigma_2^{\le r_{max}}(x)||\le \sum_{r=0}^{r_{max}}\int\ d^3x\ |\Sigma_2^r(x, y)|.
\ee
Since the integrand is bounded by (cf. Theorem \ref{maina}, Formula \eqref{x2pta} )
\be
|\Sigma^r_2(y,z)|\le O(1)|\lambda|^2 r^2 \sup_{\sigma\in\sigma_{\cG}}\g^{-3r(\sigma)}e^{-cd^\alpha_\sigma(y,z)},
\ee
we can perform the integration in \eqref{localmu1} along the spanning tree in the 1PI graph, for which the spatial integration is bounded by 
\be|\int d^3x e^{-cd^\alpha_\sigma(y,z)}|\le c'.\gamma^{2r_\cT},
\ee
where $r_\cT$ is the maximal scaling index among the tree propagators. Combine the above two terms, we have:
\be\label{local3}
\sum_r\int\ dx_0 dx_+dx_-\ |\Sigma_2^r(x, y)|
\le O'(1)\sum_r|\lambda|^2 r^2 \g^{-r}\le K_2 |\lambda|^2,
\ee
where $K_2$ is another positive constant. Since $|\lambda|^2\ll|\lambda|$ for $\lambda\in\RR_T$, the amplitudes of the localized term is also bounded by $K_2|\lambda|$.

Now we consider the amplitude for a generalized tadpoles, which is formed by contracting a chain of bipeds with a bare vertex. Let $T^g_n$ be a generalized tadpole which contain $n$ irreducible and renormalized bipeds and $n+1$ propagators connecting these bipeds. Let the scaling index of an external propagators be $r^e$ and the lowest scaling indices in the biped by $r^i$. Obviously all external propagators have the same scaling indices, due to conservation of momentum.

We can factorize $T^g_n$ as products of more elementary terms 
\be T^\Sigma(r^e,r^i,\lambda,x)= \int d{x_1}\int dy_1 C^{r^e}(x,x_1)(1-\tau)\Sigma^{r^i}(x_1,y_1),
\ee
each of which corresponds to the part contained in the dashed square in Figure \ref{gtad1}.
To remember that (cf. Section $5.1$) each of the above form is bounded by
\be
\int dx_1 dy_1|x_{1,0}-y_{1,0}|\cdot|\Sigma^{r^i}(x_1,y_1)|\cdot|\frac{\partial}{\partial{x_{1,0}}} C^{r^e}(x,x_1)|,
\ee
or
\be
\int dx_1 dy_1|x_{1,+}-y_{1,+}||x_{1,-}-y_{1,-}||\Sigma^{r^i}(x_1,y_1)|\cdot|
\frac{\partial}{\partial{x_{1,+}}}\frac{\partial}{\partial{x_{1,-}}} C^{r^e}(x,x_1)|.
\ee

Using the face that $j^e+s_+^e+s_-^e=2r^e$ and $s_+^e+s_-^e\ge r^e$, we have
 $$|\frac{\partial}{\partial{x_{1,0}}} C^{r^e}(x,x_1)|\le O(1)\g^{-2r^e}e^{-d(x,x_1)},$$ and
$$|\frac{\partial}{\partial{x_{1,+}}}\frac{\partial}{\partial{x_{1,-}}} C^{r^e}(x,x_1)|\le O(1)\g^{-2r^e}e^{-d(x,x_1)},$$

By Theorem \ref{maina}, Formula \eqref{x2pt} and \eqref{x2ptc}, which states that
$$|x_{1,0}-y_{1,0}||\Sigma^{r^i}(x_1,y_1)|\le O(1)\g^{-2r^i}e^{-d^\alpha(x_1,y_1)},$$
and
$$|x_{1,+}-y_{1,+}||x_{1,-}-y_{1,-}||\Sigma^{r^i}(x_1,y_1)|\le O(1)\g^{-2r^i}e^{-d^\alpha(x_1,y_1)},$$
we can easily find that, after performing the integration, each of the above term is bounded by 
$$O(1)\g^{2r^e}\g^{2r^i}\g^{-2r^e}\g^{-2r^i}\le O(1) .$$

Since there is a propagator in $T^g_n$ whose position coordinates are not integrated,
we have
\be
||T^g_n||\le\prod_{i=1}^n||\sum_{r^i}  T^\Sigma_i(r^e,r^i,\lambda,x)||\cdot||C^{r^e}(x_n,x)||_{L^\infty}\le O(1)^n |\lambda|^{2n+1} |\log T|^n\g^{-j^e},
\ee
which is simply bounded by the amplitude of a simple tadpole of scaling index $r^e$. 
Summing over the indices $r^e$ and using the fact that $|\lambda\log^2 T|<1$, then there exists
a positive constant $K_3$ such that
we have
\be
\sum_{n=0}^\infty |T^g_n|\le 2 K_3 |\lambda|.
\ee

Summing up all the local terms and let $K=K_1+K_2+2K_3$ we prove that the amplitudes for all the local terms are bounded by $K|\lambda|$. Hence we proved Theorem \ref{flowmu}. 


\section{Conclusions and Perspectives}
In this paper we proved that the doped Hubbard model on the honeycomb lattice is not a Fermi liquid but a Luttinger liquid if the value of the bare chemical potential is close to $1$. This is very different from the Fermi liquid behavior of this model but at half-filling, which corresponds to $\mu=0$ \cite{GM}. The reason for the non-Fermi liquid behavior may partially due to the fact that the non-interacting Fermi surface is no more convex. So it will be interesting to study the same model but for $0<\mu<1$, which may exhibit crossovers between the Fermi liquids and the Luttinger liquids behaviors. This would be the subject of a future work \cite{GMW}.


\section{Appendix. The geometric properties of the Fermi surfaces}
In this appendix we study the geometric properties of a curve in the neighborhood of the non-interacting Fermi surface.
Let $\cF^h(\kk)$ be the boundary curve of the annulus $\cA_h$, which is in the $O(\gamma^{-2h})$ neighborhood of the Fermi triangle:
\bea\label{tri1}
&&\cF^h(\kk)=\Big\{\ \kk=(k_1,k_2)\ \vert\ \\
&&\quad\quad[\cos(\sqrt{3} k_2/2)]\cdot[\cos( \frac14(3k_1+\sqrt{3} k_2))]\cdot[\cos( \frac14(3k_1-\sqrt{3} k_2))]=\phi(\gamma^{-2h})\ \Big\},\nn
\eea
where $\phi(\gamma^{-2h})$ is a function whose leading term is $\gamma^{-2h}$. 

Remark that, since the size of the Fermi triangle is of order one, any part of $\cF^h(\kk)$ can't be close to all three edges of the Fermi triangle: for any $\kk\in\cF^h(\kk)$ that is in the $\gamma^{-2h}$ vicinity of both $\ell_1$ and $\ell_2$, its distance to $\ell_3$ must be of order one. So in order to consider the geometrical properties of the curve, like the curvature, around the Fermi triangle, we just need to consider the part of the curve $F(\kk)$ that is close to $\ell_1\cup\ell_2$, in which
\bea
&&\ell_1=\{(k_1,k_2)\vert k_2=\frac{\pi}{\sqrt3}\},\\
&& \ell_2=\{(k_1,k_2)\ \vert
k_2=\sqrt3 k_1-\frac{2\pi}{\sqrt3};\ k_1\in[\frac{\pi}{3},\pi], \ k_2\in[0,\frac{\pi}{\sqrt3}]\}.
\eea
By the $\ZZZ_3$ rotational symmetry we obtain the geometric properties of the entire curve. 
So instead of studying the curve defined by \eqref{tri1}, we only need to consider the (partial) curve:
\bea\label{tri2}
F(k_1, k_2)&=&\Big\{(k_1,k_2)\ \vert\ k_1\in[\frac{\pi}{3},\pi], \ k_2\in[0,\frac{\pi}{\sqrt3}],\\
&&\quad\quad\quad[\cos(\sqrt{3} k_2/2)]\cdot[\cos( \frac14(3k_1-\sqrt{3} k_2))]=\gamma^{-2h}\Big\},\nn
\eea

It is useful to shift the $k_1$ coordinate as $k_1=k_1'+\frac{2\pi}{3}$ and the equations for the two lines become
\bea
\ell_1=\{k_2=\frac{\pi}{\sqrt3}\}, \  
\ell'_2=\{k_2=\pm\sqrt3 k_1'\};\ k_1'\in[-\frac{\pi}{3},\frac{\pi}{3}].
\eea

The curvature of the Fermi curve is given by:
\be
R=\frac{(F_{k_1}^2+F_{k_2}^2)^{\frac{3}{2}}}{-F_{k_2}^2F_{{k_1}{k_1}}+2F_{k_1}F_{k_2}F_{{k_1}{k_2}}-F_{k_1}^2F_{{k_2}{k_2}}}\ ,
\ee
where 
$F_{k_1}=\frac{\partial F({k_1},{k_2})}{\partial {k_1}}$, $F_{{k_1}{k_1}}=\frac{\partial^2 F({k_1},{k_1})}{\partial {k_1}^2}$, $F_{{k_2}{k_2}}=\frac{\partial^2 F({k_2},{k_2})}{\partial {k_2}^2}$
and $F_{{k_1},{k_2}}=\frac{\partial^2 F({k_1},{k_2})}{\partial {k_1}\partial {k_2}}$. 
After some straightforward calculations, we have:
\be
R(k_1,k_2)=\frac{2}{27}\ \frac{[9\sin^2\frac{3}{2}k_1+3\sin^2\frac{\sqrt3}{2}k_2]^{\frac32}}{\cos\frac{3}{2}k_1\sin^2\frac{\sqrt3}{2}k_2+\sin^2\frac{3}{2}k_1\cos\frac{\sqrt3}{2}k_2}.
\ee
In the $\gamma^{-2h}$ vicinity of $\ell_1\cup\ell_2$ we have
\be
R(k_1)\simeq\frac{(3k_1/2)^3+\g^{-3h/2}}{\g^{-h}},
\ee
where for two functions $f\simeq g$ means that in the first Brillouin zone we have
$cf\le g\le df$ for some constants $c$ and $d$.
Another important quantity is the width $w(k_1)$ of the band:
\be
w(k_1)\simeq \frac{\g^{-h}}{\g^{-h /2}+3k_1/2},
\ee

Following \cite{Riv}, we can also define the anisotropic length for the sectors
\be
l(k_1)=\sqrt{w(k_1)R(k_1)}\simeq\g^{-h/2}+3k_1/2,
\ee
which ranges from $\g^{-h/2}$ to an order 1 constant and justifies the introduction of the various
sectors (cf. Figure \ref{figsec}).


%
%
%
%

\medskip
\noindent{\bf Acknowledgments}
The author is very grateful to Prof. Vincent Rivasseau for explaining to him many details of the papers \cite{Riv, AMR1} and for many useful discussions during the author's visit to LPT Orsay, Univ. Paris 11. The author is also very grateful to Prof. Horst Kn\"orrer for useful discussions and encouragements, and to Prof. Alessandro Giuliani and Vieri Mastropietro for discussions about the Hubbard model with small chemical potentials. Part of this work has been finished during the author's visit to the Institute of Mathematics, University of Zurich. The author is also very grateful to Prof. Benjamin Schlein for his invitation and hospitality. The author is supported
by NSFC (11701121) and the HIT Young talent program.


\thebibliography{0}


\bibitem{AMR1} S. Afchain, J. Magnen and V. Rivasseau: {\it
Renormalization of the 2-Point Function of the Hubbard Model at Half-Filling},
Ann. Henri Poincar\'e {\bf 6}, 399-448 (2005).

\bibitem{AMR2} S. Afchain, J. Magnen and V. Rivasseau: {\it
The Two Dimensional Hubbard Model at Half-Filling, part III: The Lower Bound on the Self-Energy},
Ann. Henri Poincar\'e {\bf 6}, 449-483 (2005).

\bibitem{and} P. W. Anderson: {\it "Luttinger-Liquid" behavior of the normal metallic state of the 2D Hubbard model}, Phys. Rev. Lett. {\bf 64}, 1839-1841 (1990).

\bibitem{AR}
   A.~Abdesselam and V.~Rivasseau,
  ``Trees, forests and jungles: A botanical garden for cluster expansions,''  in Constructive Physics, Lecture Notes in Physics 446, Springer Verlag, 1995,

\bibitem{BCS} T. Bardeen, L. N. Cooper and J. R. Schrieffer: {\it Theory of
Superconductivity}, Phys. Rev. {\bf 108}, 1175-1204 (1957).

\bibitem{BG} G. Benfatto and G. Gallavotti: {\it
Perturbation theory of the Fermi surface in a quantum liquid.
A general quasiparticle formalism and one dimensional systems},
Jour. Stat. Phys. {\bf  59}, 541-664 (1990).

\bibitem{BG1} G. Benfatto and G. Gallavotti: {\it
Renormalization Group},
Physics Notes, Vol. 1, Princeton University Press (1995).

\bibitem{BGM1} G. Benfatto, A. Giuliani and V. Mastropietro: {\it
Low Temperature Analysis of Two-Dimensional Fermi Systems with Symmetric Fermi Surface},
Ann. Henri Poincar\'e {\bf 4}, 137-193 (2003).

\bibitem{BGM2} G. Benfatto, A. Giuliani and V. Mastropietro: {\it
Fermi liquid behavior in the 2D Hubbard model at low temperatures},
Ann. Henri Poincar\'e {\bf 7}, 809-898 (2006).


\bibitem{B} D. C. Brydges: {|it A short course on cluster expansions}, Ph\'enom\`enes critiques,
syst\`emes al\'eatoires, th\'eories de jauge, Part I, II (Les Houches, 1984), 129, 183, North-Holland, Amsterdam, 1986.

\bibitem{BrF} D. C. Brydges and P. Federbush: {\it A new form of the Meyer expansion
in classical statistical mechanics}, Jour. Math. Phys. {\bf 19}, 2064-2067 (1978).

\bibitem{BK} D. Brydges and T. Kennedy,
{\it Mayer expansions and the Hamilton-Jacobi equation}, Journal of
Statistical Physics, {\bf 48}, 19 (1987).

\bibitem{lieb} D. Baeriswyl,  D. Campbell, J. Carmelo, F. Guinea, E. Louis, 
{\it The Hubbard Model, Its Physics and Mathematical Physics}, Nato ASI series, V. 343, 
Springer Science+Business Media New York, 1995

\bibitem{review1} A. H. Castro Neto, F. Guinea, N. M. R. Peres, K. S. Novoselov,
A. K. Geim: {\it The electronic properties of graphene}, Rev. Mod. Phys. {\bf 81}, 109-162 (2009)

\bibitem{DR1} M. Disertori and V. Rivasseau: {\it Interacting Fermi liquid in
two dimensions at finite temperature, Part I - Convergent attributions} and
{\it Part II - Renormalization},
Comm. Math. Phys. {\bf 215},  251-290 (2000) and
291-341 (2000).


\bibitem{FMRT} J. Feldman, J. Magnen, V. Rivasseau and E. Trubowitz:
{\it An infinite volume expansion for many fermions Freen functions},
Helv. Phys. Acta {\bf 65}, 679-721 (1992).

\bibitem{FKT} J. Feldman, H. Kn\"orrer and E. Trubowitz: {\it
A Two Dimensional Fermi Liquid},
Comm. Math. Phys {\bf 247}, 1-319 (2004).

\bibitem{FST1} J. Feldman, M. Salmhofer and E. Trubowitz: {\it
Perturbation Theory Around Nonnested Fermi Surfaces.
I. Keeping the Fermi Surface Fixed}, Journal of
Statistical Physics, {\bf 84}, 1209-1336 (1996).
 
\bibitem{FST2} J. Feldman, M. Salmhofer and E. Trubowitz:
{\it Perturbation Theory around Non-nested Fermi Surfaces II. Regularity of the Moving Fermi Surface: RPA Contributions }
Comm. Pure Appl. Math. 51 (1998), 1133-1246.

\bibitem{FST3} J. Feldman, M. Salmhofer and E. Trubowitz:
 {\it Regularity of the Moving Fermi
Surface: The Full Selfenergy,} Comm. Pure Appl. Math. 52
(1999), 273-324.

\bibitem{FST4} J. Feldman, M. Salmhofer and E. Trubowitz:
{\it An inversion theorem in Fermi surface theory} Comm. Pure Appl. Math. 53 (2000), 1350-1384.

\bibitem{G} G. Gallavotti: {\it
Renormalization group and ultraviolet stability
for scalar fields via renormalization group methods},
Rev. Mod. Phys. {\bf 57}, 471-562 (1985).

\bibitem{GN} G. Gallavotti and F. Nicol\`o: {\it
Renormalization theory for four dimensional scalar fields. Part I} and
{\it II}, Comm. Math. Phys. {\bf 100}, 545-590 (1985) and
{\bf 101}, 471-562 (1985).

\bibitem{GK} K. Gawedzki and A. Kupiainen: {\it Gross-Neveu model through
convergent perturbation expansions}, Comm. Math. Phys. {\bf 102}, 1-30 (1985).

\bibitem{GeM} G. Gentile and V. Mastropietro: {\it Renormalization group for
one-dimensional fermions. A review on mathematical results}. In:
Renormalization group theory in the new millennium, III,
Phys. Rep. {\bf 352}, 273-437 (2001).

\bibitem{GM} A. Giuliani and V. Mastropietro: {\it The two-dimensional
Hubbard model on the honeycomb lattice}, Comm. Math. Phys. {\bf 293}, 301-346
(2010).

\bibitem{GMprb} A. Giuliani and V. Mastropietro: {\it
Rigorous construction of ground state correlations in graphene: renormalization of the velocities and Ward Identities},
Phys. Rev. B {\bf 79}, 201403(R) (2009); Erratum, ibid {\bf 82}, 199901 (2010).

\bibitem{hubb} J. Hubbard, {\it Electron correlations in narrow energy bands}, Proc. Roy. Soc. (London), {\bf A276}, 238-257 (1963).

\bibitem{HCM} C.-Y. Hou, C. Chamon and C. Mudry: {\it Electron Fractionalization in
Two-Dimensional Graphenelike Structures}, Phys. Rev. Lett. {\bf 98}, 186809 (2007).

\bibitem{Le} A. Lesniewski: {\it Effective action
for the Yukawa$_2$ quantum field theory}, Comm. Math. Phys. {\bf 108},
437-467 (1987).

\bibitem{Lu} J. M. Luttinger: {\it An exactly soluble model of a many fermions
system}, J. Math. Phys. {\bf 4}, 1154-1162 (1963).

\bibitem{M2} V. Mastropietro: {\it Non-Perturbative
Renormalization}, World Scientific (2008).

\bibitem{N}  K. S. Novoselov, A. K. Geim, S. V. Morozov,
D. Jiang, Y. Zhang, S. V. Dubonos, I. V. Grigorieva and
 A. A. Firsov: {\it Electric Field Effect in Atomically Thin Carbon Films},
Science {\bf 306}, 666 (2004).



\bibitem{Riv} V. Rivasseau: {\it The Two Dimensional Hubbard Model at Half-Filling. I. Convergent Contributions}, J. Statistical Phys. {\bf 106}, 693-722 (2002).


\bibitem{rivbook} V. Rivasseau: {\it From Perturbative Renormalization to Constructive Renormalization}, Princeton university press


\bibitem{salm} M. Salmhofer: {\it Continuous Renormalization for Fermions and Fermi Liquid Theory}, Comm. Math. Phys. {\bf 194}, 249-295 (1998).


\bibitem{Sa} M. Salmhofer: {\it Renormalization: An Introduction},
Springer (1999).

\bibitem{To} S. Tomonaga: {\it Remarks on Bloch's methods of sound waves
applied to many fermion systems}, Progr. Theo. Phys. {\bf 5}, 544-569 (1950).

\bibitem{W} P. R. Wallace: {\it The Band Theory of Graphite}, Phys. Rev. {\bf 71},
622-634 (1947).

\bibitem{GMW} A. Giuliani, V. Mastropietro, Z. Wang: {\it Doped Hubbard model on the honeycomb lattice with small chemical potential}, Work in progress.
\endthebibliography
\end{document}